\pdfoutput=1
\documentclass[aps,prl,notitlepage,twocolumn,superscriptaddress]{revtex4-2}

\makeatletter
\def\@hangfrom@section#1#2#3{\normalsize\@hangfrom{#1#2}#3}%
\def\@hangfroms@section#1#2{\normalsize#1#2}%
\makeatother

\RequirePackage[usenames, dvipsnames, x11names]{xcolor}
\usepackage{amsmath,amsfonts,amssymb,bm}
\usepackage{graphicx,color,soul}
\usepackage{dsfont}
\usepackage{enumerate}
\usepackage[colorlinks=true, allcolors=blue]{hyperref}

\usepackage{multirow}

\definecolor{marineblue}{HTML}{01386A}
\definecolor{lightblue}{HTML}{4A90D9}

\hypersetup{
    colorlinks=true,
    citecolor=Green,
    linkcolor=Red,
    urlcolor=NavyBlue
}

\newcommand{\nc}{\newcommand}
\nc{\ket}[1]{|#1\rangle}
\nc{\bra}[1]{\langle#1|}
\nc{\ketbra}[2]{|#1\rangle\!\langle#2|}
\nc{\braket}[2]{\langle#1|#2\rangle}
\nc{\braoprket}[3]{\langle#1|#2|#3\rangle}
\nc{\opr}[1]{\operatorname{#1}}
\nc{\avg}[1]{\langle#1\rangle}
\nc{\ketbrasame}[1]{|#1\rangle\!\langle#1|}
\nc{\tr}{\opr{tr}}
\nc{\E}{\mathbb{E}}
\nc{\var}{\opr{Var}}
\nc{\up}{\uparrow}
\nc{\dn}{\downarrow}
\nc{\bd}[1]{\boldsymbol{#1}}
\nc{\zjx}[1]{\textcolor{lightblue}{\textbf{#1}}}
\nc{\jx}[1]{\textcolor{lightblue}{\textbf{[JX: #1]}}}
\nc{\lb}[1]{\textcolor{brown}{\textbf{[LB: #1]}}}
\nc{\ww}[1]{\textcolor{LimeGreen}{\textbf{#1}}}
\nc{\ls}[1]{\textcolor{blue}{\textbf{[LS: #1]}}}

\nc{\red}[1]{\textcolor{red}{#1}}

\nc{\todo}[1]{\textcolor{blue}{\textbf{[#1]}}}

\nc{\rev}[1]{\textcolor{lightblue}{#1}}
\nc{\fable}[1]{\textcolor{BurntOrange}{\textbf{[#1]}}}
\nc{\orange}[1]{\textcolor{BurntOrange}{\textbf{\boldmath #1}}}
\nc{\orangem}[1]{{\color{BurntOrange}\bm{#1}}}
\nc{\del}[1]{\textcolor{red}{\sout{#1}}}

\nc{\sgn}{\operatorname{sgn}}

\usepackage[normalem]{ulem}

\newcommand{\diag}[2]{\vcenter{\hbox{\includegraphics[height=#2]{#1}}}}
\newcommand{\figref}[1]{Fig.\,\ref{#1}}
\newcommand{\figsref}[1]{Figs.\,\ref{#1}}
\newcommand{\secref}[1]{Sec.\,\ref{#1}}
\newcommand{\eqnref}[1]{Eq.\,\eqref{#1}}

\newcommand{\appref}[1]{SM~Sec.~\ref{#1}}

\begin{document}

\graphicspath{{Figures/}}

\title{Chiral Superconductivity Indifferent to Parent Topology in the Square-Lattice Hofstadter-Hubbard Model}

\author{Jia-Xin Zhang}
\thanks{These two authors contributed equally to this work.}
\affiliation{French American Center for Theoretical Science, CNRS, KITP, Santa Barbara, California 93106-4030, USA}
\affiliation{Kavli Institute for Theoretical Physics, University of California, Santa Barbara, California 93106-4030, USA}

\author{Wen O. Wang}
\thanks{These two authors contributed equally to this work.}
\affiliation{Kavli Institute for Theoretical Physics, University of California, Santa Barbara, California 93106-4030, USA}

\author{Leon Balents}
\affiliation{Kavli Institute for Theoretical Physics, University of California, Santa Barbara, California 93106-4030, USA}
\affiliation{Canadian Institute for Advanced Research, Toronto, Ontario, Canada}
\affiliation{French American Center for Theoretical Science, CNRS, KITP, Santa Barbara, California 93106-4030, USA}

\author{Lucile Savary}
\affiliation{French American Center for Theoretical Science, CNRS, KITP, Santa Barbara, California 93106-4030, USA}
\affiliation{Kavli Institute for Theoretical Physics, University of California, Santa Barbara, California 93106-4030, USA}

\author{D. N. Sheng}
\affiliation{Department of Physics and Astronomy, California State University, Northridge, California 91330, USA}

\date{\today}

\begin{abstract}
We study a $\pi$-flux square-lattice $t$-$t'$ Hofstadter-Hubbard model proposed to be realizable in twisted moir\'e bilayers.
By tuning $U$ and $t'$ at half filling, ground-state density-matrix renormalization group simulations (DMRG) and a parton analysis identify an integer quantum Hall (IQH) phase, a chiral spin liquid (CSL), and an antiferromagnet (AFM) in the phase diagram, while charge and spin fluctuations from finite-temperature determinant quantum Monte Carlo (DQMC) suggest a closing charge gap but an open spin gap at the IQH-CSL boundary, consistent with the topological criticality expected there.
This model therefore produces a phase diagram hosting weakly correlated band topology (IQH), intrinsic topological order (CSL), and conventional symmetry breaking (AFM) side by side.
Upon doping any of these three parent phases, a single chiral topological superconductor with spin Chern number $C_s=-2$ occupies nearly the entire phase diagram.
DMRG, DQMC, and weak-coupling perturbation theory agree on its pairing symmetry. This shows that doping the topological band of our model robustly induces the same paired state throughout the phase diagram regardless of the character of the underlying correlated half-filled state and its magnetic fluctuations, whether an IQH, a spin liquid or a weakly fluctuating AFM.
\end{abstract}
\maketitle

\textbf{Introduction.---}
How superconductivity (SC) arises from the interplay of correlation and topology is a central question in condensed matter physics, dating back to the cuprates, where SC emerges upon doping a topologically trivial antiferromagnetic (AFM) Mott insulator~\cite{Anderson1987RVB,Keimer2015HighTc,LeeNagaosaWen2006}. Chiral SC, whose gap function lacks time-reversal symmetry has likewise long been predicted to emerge upon doping a chiral spin liquid (CSL)~\cite{Laughlin1988AnyonSC,WenWilczekZee1989ChiralSpinSC,LeeFisher1989AnyonSC}, a topologically ordered Mott parent state that supports anyonic excitations and chiral edge states~\cite{Balents2010SpinLiquids,KalmeyerLaughlin1987CSL,WenWilczekZee1989ChiralSpinSC} and has by now been numerically identified in kagom\'e and triangular-lattice magnets~\cite{He2014KagomeCSL,Gong2014KagomeCSL,Bauer2014KagomeCSL,Szasz2020TriangularCSL,Chen2022TriangularCSL,Zhou2022,Wietek2021TriangularMott}. The CSL can also be viewed as a spin analogue of the fractional Chern insulator (FCI), in the sense that they realize the same intrinsic topological order, observed in moir\'e materials~\cite{Cai2023FQAH,Zeng2023FQAH,Park2023FQAH,Xu2023FQAH,Lu2024GrapheneFQAH}, where signatures of SC~\cite{Xu2025MoTe2SC} sparked a flurry of theoretical works~\cite{ShiSenthil2024DopingFQAH,Pichler2025AnyonSC,Kim2025BeyondPairing,Nosov2026AnyonSCPlateau,Guerci2025FlatChernBandSC,ShiSenthil2025NonAbelianTSC}. Because of the FCI-CSL correspondence, SC obtained by doping a CSL would shed light on doped-FCI experiments, in addition to providing an interesting avenue for spin and charge intertwinement.

In realistic models, however, doping a CSL does not generically yield SC. Pairing has so far been reported mainly in the strong Mott regime of $t$-$J$-type models~\cite{Jiang2020DopedCSL,Jiang2021npjDopedQSL,Huang2022ChiralNematicSC,Huang2023DopedTriangularTJ}, whereas in the weak or intermediate Mott regime, precisely where the CSL arises in Hubbard models, doping tends to yield either a chiral metal or charge order~\cite{Song2021DopingCSL,Zhu2022DopedMottTriangular,Peng2021DopedKagome}. Recent work proposes that topological criticality can energetically stabilize such doped SC~\cite{Divic2025AnyonSuperconductivity,Han2026AnyonSuperfluidity,Pichler2025AnyonSC}, with numerical support in triangular-lattice Hofstadter-Hubbard models~\cite{Chen2026TriangularHofstadterSC,Kuhlenkamp2025RobustSuperconductivity,Niu2026ThermodynamicLimit,Gallegos2026IQHCSL}. Finding a realistic model that realizes chiral SC upon doping, and clarifying how the pairing mechanism relates to the intrinsic characters of the parent states remain open and important problems.

Here we study a Hofstadter-Hubbard model on the square lattice~\cite{Shaffer2021HofstadterSC,Shaffer2022HofstadterSC}.
Starting from a $t$-$t'$-$U$ Hubbard model of the kind proposed for twisted square homobilayers~\cite{Eugenio2025SquareHomobilayers}, with a tunable next-nearest-neighbor hopping ratio $t'/t$, we include a $\pi$ flux through every elementary square plaquette. Such an order-one flux is out of reach in ordinary crystals, yet attainable at laboratory fields in moir\'e systems because of the large unit cell~\cite{Hofstadter1976,Dean2013Hofstadter}. At $t'=0$ the undoped model undergoes a well-studied $U$-driven direct transition from a Dirac semimetal to the AFM at $U_c\simeq5.6$, in the chiral Heisenberg Gross-Neveu universality class~\cite{ChangScalettar2012,ParisenToldin2015,OtsukaYunokiSorella2016}. The frustration brought by a finite $t'$, and the behavior under hole doping, have by contrast not been systematically explored.

Combining the density-matrix renormalization group (DMRG) with a parton analysis, we find that tuning $t'$ and $U$ at half-filling brings an integer quantum Hall (IQH) phase, a CSL, and an AFM into close proximity through a two-parameter ground-state phase diagram.
This model realizes weakly correlated band topology, intrinsic topological order, and conventional symmetry breaking side by side.
As a finite-temperature complement to this ground-state picture, determinant quantum Monte Carlo (DQMC) resolves the charge and spin fluctuations, finding (i) signs of a shrinking or vanishing charge gap at the IQH-CSL boundary while (ii) the spin gap stays open.
That boundary appears to realize topological criticality that supports singlet Cooper pairs as cheap charge excitations, since they carry charge $2e$ but no spin. %
Upon doping, a chiral SC with quantized spin Chern number $C_s=-2$ covers nearly the entire phase diagram, descending from the weakly correlated IQH phase, the topologically ordered CSL, as well as the trivial AFM. %
This challenges the common expectation that a paired state naturally inherits its parent's topology and calls for a pairing mechanism that applies across distinct parent phases.

\begin{figure}[t]
\centering
\includegraphics[width=\columnwidth]{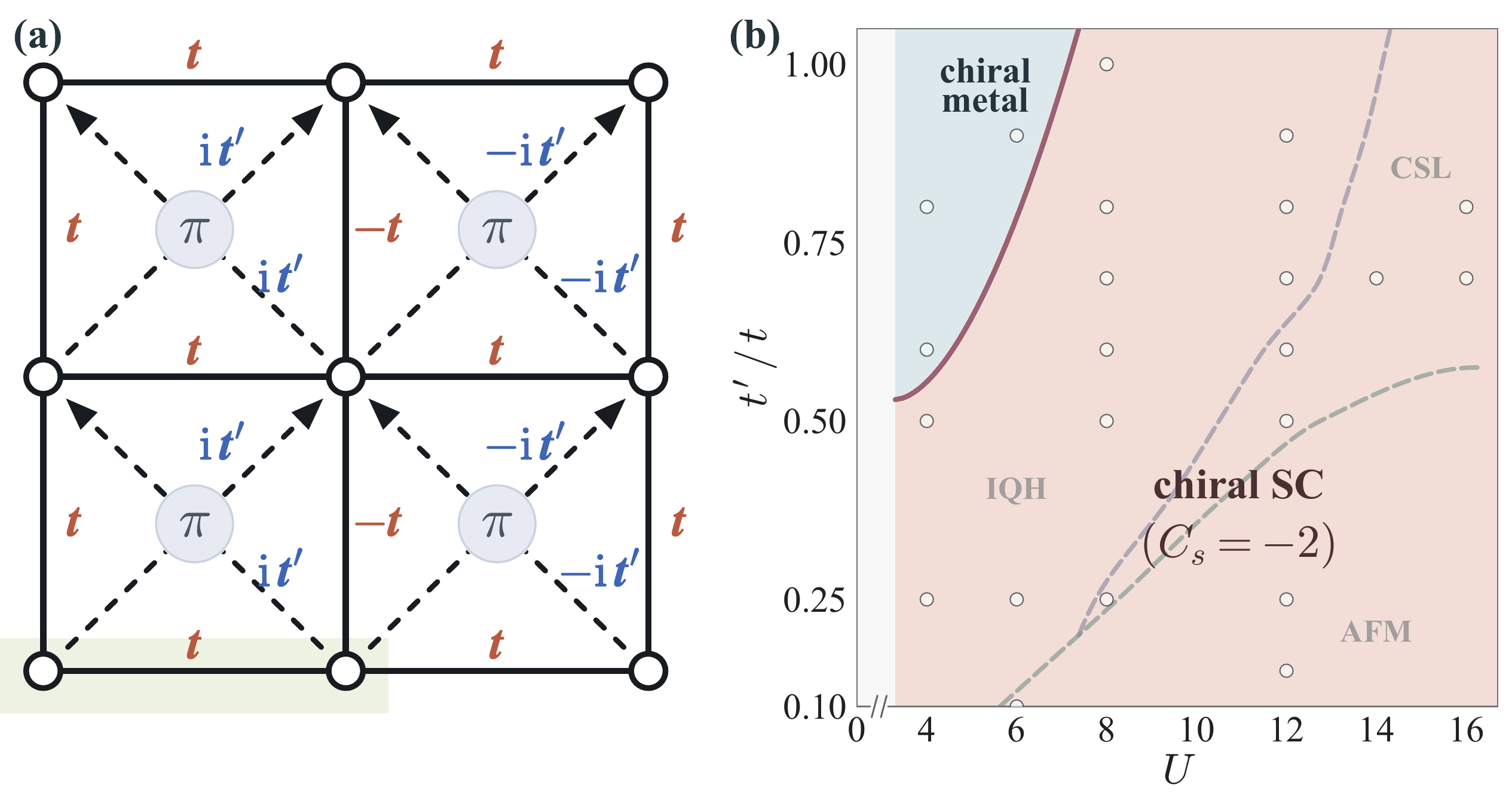}
\caption{(a) Square-lattice Hofstadter-Hubbard model in the Landau gauge. Real NN amplitudes $\pm t$ are staggered along $\hat y$ by $(-1)^{i_x}$; dashed NNN bonds carry purely imaginary $\pm it'$ oriented by the arrows, and every square plaquette encloses flux $\pi$. The shaded region is the magnetic unit cell. (b) DMRG phase diagram at $\delta=1/12$, showing a chiral SC with $C_s=-2$ and a chiral metal; circles mark the simulated points. Dashed lines reproduce the half-filling phase boundaries of \figref{fig:undoped}(a).}
\label{fig:model}
\end{figure}

\textbf{Model and methods.---}
We study the Hamiltonian
\begin{equation}
H=\sum_{i,j,\sigma}\Big(-t_{ij}\,c^{\dagger}_{i\sigma}c_{j\sigma}+\mathrm{H.c.}\Big)
+U\sum_{i}n_{i\up}n_{i\dn},
\label{eq:model}
\end{equation}
with real nearest-neighbor (NN) hopping $\pm t$ and purely imaginary next-nearest-neighbor (NNN) hopping $\pm it'$ arranged so that each square plaquette encloses flux $\pi$. Throughout, we work in the Landau gauge of \figref{fig:model}(a), with $t_{i,i+\hat x}=t$, $t_{i,i+\hat y}=(-1)^{i_x}t$, and $t_{i,i\mp\hat x+\hat y}=\pm i(-1)^{i_x}t'$. At $t'=0$ the band structure is the gapless $\pi$-flux Dirac semimetal~\cite{AffleckMarston1988}; a finite $t'$ gaps the Dirac nodes into Chern bands of total spin-summed Chern number $-2\sgn(t')$. %
Unless stated otherwise, we use $t'>0$; time reversal sends $t'\to-t'$ and reverses all chiralities.

We combine ground-state DMRG with finite-temperature DQMC, setting $t=1$ as the energy unit. For DMRG we use $N=L_x\times L_y$ cylinders with $L_y=6$ legs in the main text ($L_y=3$ and $4$ in \appref{app:dmrg_ly34}), and periodic (open) boundary conditions along $y$ ($x$), enforcing charge-$U(1)$ and spin-$SU(2)$ symmetries~\cite{White1992DMRG,McCulloch2007NonabelianDMRG}. Bond dimensions reach $M=16000$ $SU(2)$ multiplets, equivalent to around $48000$ $U(1)$ states resulting in a DMRG truncation error $\sim 10^{-6}$. Doped simulations are performed at hole doping $\delta=1/12$ on $L_x=24$, and $32$ cylinders.  For DQMC we use the standard determinant algorithm~\cite{Blankenbecler1981DQMC,White1989DQMC} on $L\times L$ tori with $L=6$ and $8$. The complex NNN hopping breaks the usual bipartite sign-free condition at half filling;
the resulting sign problem~\cite{Loh1990Sign} is most severe near the phase boundaries and toward large $t'$ and $U$ [\figref{fig:sign}], restricting the accessible temperatures to $T\gtrsim t/5$.

\begin{figure}[t]
\centering
\includegraphics[width=\columnwidth]{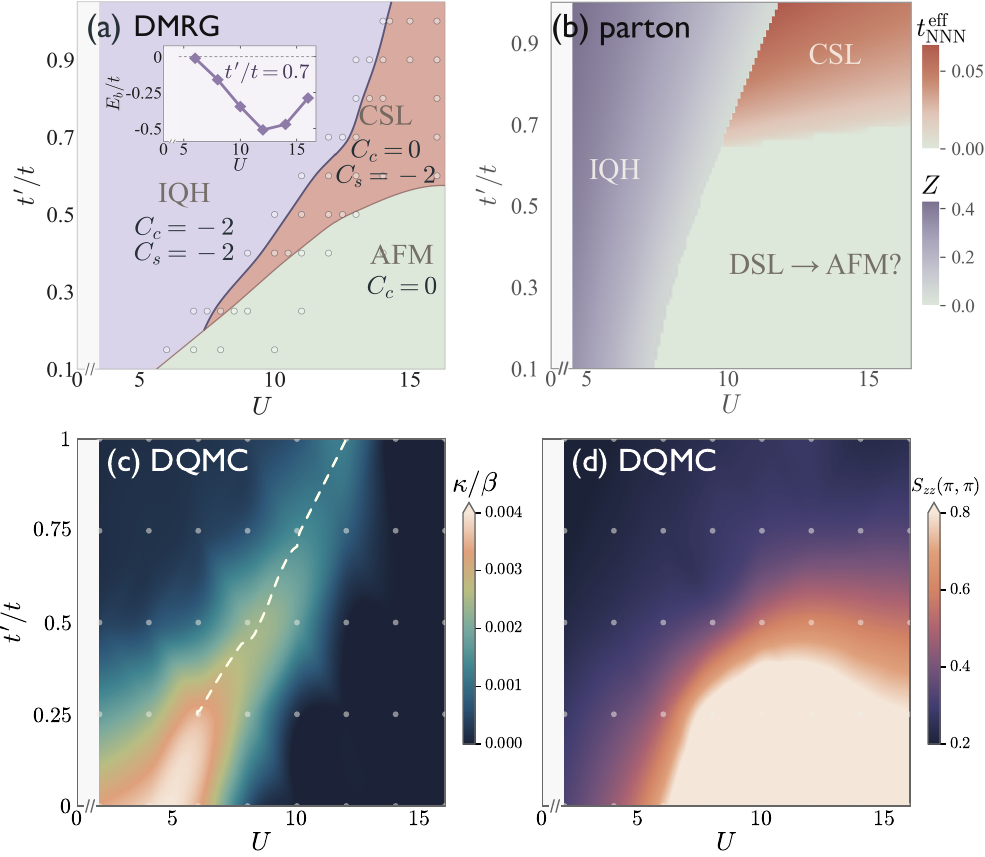}
\caption{Half-filling phase diagram. (a) DMRG phase diagram on $L_y=6$ cylinders from the charge and spin Chern numbers, with IQH ($C_c=C_s=-2$), CSL ($C_c=0$, $C_s=-2$), and AFM ($C_c=0$, $C_s$ undefined); circles mark the simulated points. Inset: two-electron binding energy $E_b$ versus $U$ at $t'/t=0.7$. (b) $U(1)$ slave-rotor mean field. The purple scale shows the rotor condensate $Z$ and the red scale the effective NNN spinon hopping $t^{\rm eff}_{\rm NNN}$ in units of $t'$, which distinguishes the CSL (nonzero) from the Dirac spin liquid (zero) inside the Mott regime, the latter expected to confine into the AFM. (c),(d) DQMC results on a $6\times6$ torus at $T=t/4$,
showing (c) the charge compressibility divided by $\beta$, $\kappa/\beta=(\langle\hat N^2\rangle-\langle\hat N\rangle^2)/N$, where $\hat N=\sum_i(n_{i\up}+n_{i\dn})$, and (d) the staggered spin structure factor $S_{zz}(\pi,\pi)$. Dots mark the simulated points and white dashed lines are guides to the eye. %
}
\label{fig:undoped}
\end{figure}

\textbf{Undoped phase diagram.---}
The half-filling phase diagram from DMRG is shown in \figref{fig:undoped}(a). The primary diagnostics are the charge and spin Chern numbers, $C_c$ and $C_s$, measured by adiabatic insertion of a flux $\theta_F$ through the cylinder~\cite{Laughlin1981Pump,Zaletel2014FluxInsertion,Gong2014KagomeCSL,Chen2026TriangularHofstadterSC}. In $U(1)_{\rm charge}\times U(1)_{\rm spin}$ simulations we impose the twisted boundary condition $\hat c_{(x,L_y+1),\sigma}=e^{iq^{a}_{\sigma}\theta_F}\,\hat c_{(x,1),\sigma}$, with $q^{c}_{\sigma}=1$ and $q^{s}_{\sigma}=\sigma=\pm1$ for $a=c,s$, and read off $C_{a}=\Delta Q_{a}\big|_{\theta_F=0}^{2\pi}$ from the charge $n_\up+n_\dn$ and spin $n_\up-n_\dn$ accumulated near one edge. Three phases are identified. At small $U$ the IQH phase has $C_c=C_s=-2$; at large $U$ the CSL has $C_c=0$ and $C_s=-2$, its spin sector still carrying the quantized topological response, while the AFM has $C_c=0$ and a closed spin gap that leaves $C_s$ undefined. We further check each phase through the spin structure factor $S(\bm q)=N^{-1}\sum_{i,j}e^{i\bm q\cdot(\bm r_i-\bm r_j)}\langle \bm S_i\cdot\bm S_j\rangle$, finding quasi-long-range (short-range) spin correlations in the AFM (IQH and CSL) phase [see Sec.~\ref{app:dmrg_half} of the Supplemental Material (SM)~\cite{SuppMaterial}].

This phase diagram agrees qualitatively with a $U(1)$ slave-rotor mean-field theory~\cite{FlorensGeorges2004SlaveRotor,Song2023SlaveRotorTMD}, shown in \figref{fig:undoped}(b) and detailed in \appref{app:rotor}. The electron fractionalizes into a charged rotor and a neutral fermionic spinon, $c^{\dagger}_{i\sigma}=f^{\dagger}_{i\sigma}e^{i\theta_i}$, the spinons hopping with the rotor-renormalized amplitude $t^{\rm eff}_{ij}=t_{ij}\langle e^{i(\theta_i-\theta_j)}\rangle$. The rotor condensate $Z=|\langle e^{i\theta_i}\rangle|^{2}$, which equals the electron quasiparticle weight, determines the Mott boundary, which separates the electronic IQH phase at $Z\neq0$ from the Mott regime at $Z=0$. Within the Mott regime, nonzero $t^{\rm eff}_{\rm NNN}$ keeps the spinons in gapped Chern bands and thus yields the CSL, whereas vanishing $t^{\rm eff}_{\rm NNN}$ gives a gapless $\pi$-flux Dirac spin liquid. However, the compact $U(1)$ gauge field makes the Dirac spin liquid unstable on the square lattice, where monopole proliferation confines it to AFM order~\cite{Hermele2004U1SpinLiquids,Song2019CompetingOrders,Song2020MonopolesDSL}, consistent with the AFM that DMRG finds in the same region. In the $SU(2)$ extension of this framework, the transition between the IQH phase and the CSL can be continuous, as on the triangular lattice~\cite{Divic2025AnyonSuperconductivity,Gallegos2026IQHCSL}, and is captured by a Higgs-Chern-Simons theory whose critical mode is the charge-sector rotor, so the charge gap closes while the massive spinon Chern bands keep the spin gap open [\appref{app:su2k}]. This hierarchy of gaps, with a closing charge gap but an open spin gap, makes it energetically cheaper, upon doping, to inject charge as Cooper pairs than as single electrons; it follows that near the criticality a paired condensate is favored over a metal.

We test this gap hierarchy through finite-temperature charge and spin fluctuations from DQMC, complementing ground-state DMRG.
We infer the charge and spin gaps from the charge compressibility $\kappa=\beta\big(\langle\hat N^2\rangle-\langle\hat N\rangle^2\big)/N$ and the staggered spin structure factor $S_{zz}(\pi,\pi)=N^{-1}\sum_{i,j}e^{i\bm Q\cdot(\bm r_i-\bm r_j)}\langle S_i^zS_j^z\rangle$, respectively, where $\bm Q=(\pi,\pi)$;
a charge (resp.\ spin) gap exponentially suppresses $\kappa$ (resp.\ $S_{zz}({\bm Q}$) at low temperature.
Even at $T=t/4$ (the sign problem prevents us from reaching lower temperatures), we find that the compressibility [\figref{fig:undoped}(c)] is enhanced along the IQH-CSL phase boundary---on \figref{fig:undoped}(c) the white dashed line traces its peak---while the staggered spin response grows only upon entering the AFM region [\figref{fig:undoped}(d)], with momentum-resolved spin correlations peaked at ${\bm Q}=(\pi,\pi)$ [\figref{fig:spin}(a)].

Finally, we note that ground-state DMRG also supports this gap hierarchy at $t'/t=0.7$: the two-electron binding energy $E_b=E_0(N_e+2)+E_0(N_e)-2E_0(N_e+1)$, with $E_0(N_e)$ the ground-state energy for $N_e$ electrons, varies nonmonotonically with $U$ and is most negative near the IQH-CSL transition [inset of \figref{fig:undoped}(a)], while the spin gap decreases monotonically with $U$ (\appref{app:dmrg_gap}). These trends are consistent with a continuous IQH-CSL transition %
where we expect
the charge gap to close while the spin gap persists.

\begin{figure}[t]
\centering
\includegraphics[width=\columnwidth]{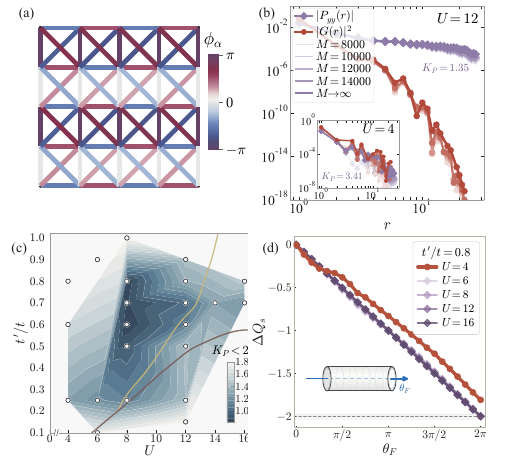}
\caption{DMRG characterization of the doped phases at $\delta=1/12$. (a) Relative pairing phase $\phi_\alpha(\bm r)$ on NN and NNN bonds in the Landau gauge at $U=12$ and $t'/t=0.8$. (b) Pair-field correlation $|P_{yy}(r)|$ (purple diamonds) and squared single-particle Green's function $|G(r)|^{2}$ (red circles) at $t'/t=0.8$ and $U=12$, for finite bond dimensions $M$ (light to dark) and after the $M\to\infty$ extrapolation (thick lines); $K_P$ is the power-law exponent of the extrapolated $|P_{yy}(r)|$. Inset: the same comparison at $U=4$. (c) Map of the pairing exponent $K_P$ over the doped phase diagram, interpolated between the simulated points (open circles) where $K_P<2$, with darker color marking smaller $K_P$; the gold and brown curves reproduce the half-filling phase boundaries of \figref{fig:undoped}(a). (d) Spin $\Delta Q_s$ pumped to the cylinder edge versus inserted spin flux $\theta_F$ at $t'/t=0.8$ for $U=4$--$16$; the inset sketches the flux insertion. The curves reach $\Delta Q_s=-2$ at $\theta_F=2\pi$ for $U\ge6$, but not for $U=4$. %
}
\label{fig:doped}
\end{figure}

\textbf{Chiral superconductivity upon doping.---}
We find that the DMRG phase diagram at $\delta=1/12$ doping [\figref{fig:model}(b)] is dominated by a single chiral SC phase; only at weak coupling and large $t'$ does a chiral metal appear, with no discernible pair binding. Two correlation functions distinguish the two doped phases: the pair-field correlation $P_{\alpha\beta}(r)=\langle\hat\Delta^{\dagger}_{\alpha}(x_0,y_0)\,\hat\Delta_{\beta}(x_0+r,y_0)\rangle$, and the single-particle Green's function $G(r)=\sum_{\sigma}\langle c^{\dagger}_{\sigma}(x_0,y_0)\,c_{\sigma}(x_0+r,y_0)\rangle$, with $\bm\alpha$ and $\bm\beta$ denoting NN or NNN bond vectors, and where the singlet bond pair operator is $\hat\Delta_{\alpha}(\bm r)=\big(c_{\bm r\up}c_{\bm r+\bm\alpha\dn}-c_{\bm r\dn}c_{\bm r+\bm\alpha\up}\big)/\sqrt2$. \figref{fig:doped}(b) compares the $\hat y$-bond pair correlation $|P_{yy}(r)|$ with $|G(r)|^{2}$ at $t'/t=0.8$, each extrapolated to infinite bond dimension, $M\to\infty$ (\appref{app:dmrg_supp}). At $U=12$ the two decay rates are clearly separated: the extrapolated pair correlation follows a slow power law, $|P_{yy}(r)|\sim r^{-K_P}$ with $K_P=1.35$, while $|G(r)|^{2}$ falls off much faster, signaling strong pair binding in the SC. At $U=4$, by contrast, the two curves decay at nearly the same rate and $K_P$ exceeds $2$ [inset of \figref{fig:doped}(b)], with no sign of pair binding, consistent with the chiral metal of \figref{fig:model}(b). We identify the SC by $K_P<2$~\cite{Giamarchi2004,Jiang2021npjDopedQSL} (\appref{app:dmrg_supp}), and \figref{fig:doped}(c) maps $K_P$ over the SC region. The smallest $K_P$ values lie at $t'/t$ and $U$ not far from the half-filling boundary between the IQH phase and the CSL.

The pairing phase of the SC, $\phi_{\alpha}(\bm r)=\arg\langle\hat\Delta^{\dagger}_{b_0}\,\hat\Delta_{\alpha}(\bm r)\rangle$, measured relative to a reference bond $b_0$ in the Landau gauge (\appref{app:model}), is not uniform. The same pattern appears throughout the SC region,
exemplified at $U=12$ and $t'/t=0.8$ in \figref{fig:doped}(a).

We identify the topological character of this SC by its spin Chern number $C_s=-2$, quantized throughout the SC phase [\figref{fig:doped}(d) and \appref{app:dmrg_chern}].
By contrast, in the chiral metal at $U=4$, the pumped spin fails to quantize, consistent with a closed spin gap. This $C_s$ invariant coincides with that of the IQH and CSL parents, and even doping the topologically trivial AFM produces the SC with $C_s=-2$.

\begin{figure}[t]
 \centering
 \includegraphics[width=\columnwidth]{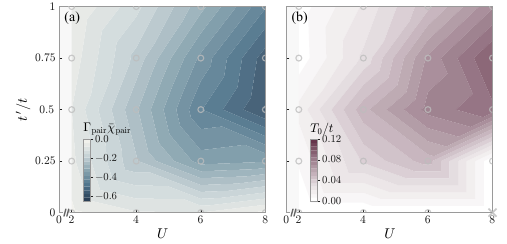}
 \caption{\label{fig:dqmcmain} Finite-temperature pairing tendency across the $U$-$t'$ plane, from DQMC on a $6\times6$ lattice at $\delta=4\%$.
 (a) Pairing-vertex indicator $\Gamma_{\mathrm{pair}}\bar{\chi}_{\mathrm{pair}}$ at $T/t=1/4$.
 (b) Extrapolated pairing scale $T_0/t$, defined by the zero crossing of $1/\chi_{\mathrm{pair}}$ from a linear fit to the two lowest available temperatures at each parameter point.
 Gray circles mark simulated points; colors are linear interpolations, and the gray cross marks a discarded extrapolation with large uncertainty.}
\end{figure}

Finite-temperature DQMC probes singlet pairing fluctuations, yielding results qualitatively consistent with ground-state DMRG.
Doping worsens the sign problem, so the doped DQMC scans reach only $U/t\le8$, which sit below the range covered by DMRG in \figref{fig:model}(b), and we work at hole dopings $\delta=2\%$--$10\%$. The low-temperature results are qualitatively unchanged throughout this doping range.
We measure the unequal-time pairing correlation function $\langle T_\tau\hat\Delta^{\dagger}_{b_0}(\beta/2)\, \hat\Delta_\alpha(\bm r,0)\rangle$ at the midpoint imaginary time $\tau=\beta/2$: at finite temperature it is more sensitive to soft modes in the thermal energy window $k_BT$ than the equal-time ($\tau=0$) and static ($\omega=0$) correlators~\cite{wang2025finite,Chen2026TriangularHofstadterSC,Li2026FluctuatingPDW}.
We extract from it the pairing phase pattern, which grows progressively more structured on cooling and, at the lowest temperatures reached, matches the pattern found by DMRG [\figref{fig:doped}(a)] over most of the accessible regime, except at small $U$ and when $t'\to0$ [\figsref{fig:8x8phase}--\ref{fig:phases3}].

Two measures quantify the low-temperature pairing fluctuations:
(i) a characteristic temperature scale $T_0$, which we extract from $\chi_{\mathrm{pair}}=\beta\langle T_\tau\hat\Delta^{\dagger}_{\mathrm{pair}}(\beta/2)\hat\Delta_{\mathrm{pair}}(0)\rangle$ where $\hat\Delta_{\mathrm{pair}}=N^{-1/2}\sum_{\bm r,\alpha}e^{-i\phi_\alpha(\bm r)}\hat\Delta_\alpha(\bm r)$ is the gap function of the channel fixed by the phase pattern $\phi_\alpha(\bm r)$ illustrated in \figref{fig:doped}(a).
A linear fit, $1/\chi_{\mathrm{pair}}=a(T-T_0)$, defines $T_0$ [\figref{fig:dqmcmain}(b)] and a finite $T_0$ marks a tendency towards a pairing instability at low temperature.
(ii) the dimensionless pairing-vertex indicator $\Gamma_{\mathrm{pair}}\bar{\chi}_{\mathrm{pair}}=\bar{\chi}_{\mathrm{pair}}/\chi_{\mathrm{pair}}-1$ [\figref{fig:dqmcmain}(a)], with $\bar{\chi}_{\mathrm{pair}}$ the no-vertex bubble built from the dressed single-particle propagators and $\Gamma_{\mathrm{pair}}$ the effective pairing vertex: the more negative it is, the stronger the interaction-induced effective attraction~\cite{White1989PairingVertices}. %
This indicator is the finite-temperature analogue of the DMRG comparison between $|P_{yy}(r)|$ and $|G(r)|^{2}$ in \figref{fig:doped}(b).
Over most of the accessible parameter regime, $T_0$ is finite and $\Gamma_{\mathrm{pair}}\bar{\chi}_{\mathrm{pair}}$ is negative,
apart from small $U$ or as $t'\to0$.
Despite the different accessible regimes, the strongest DQMC pairing enhancement and highest $T_0$ broadly coincide with the region of smallest $K_P$ found by DMRG [\figref{fig:doped}(c)].
Results for other temperatures, parameters, and $8\times8$ lattices are in \appref{app:dqmc_pair}.

The two unbiased numerical methods thus agree across a wide region of the doped phase diagram. In DQMC, the pairing spectral weight in the channel with the chiral phase pattern transfers progressively to low energy on cooling, and in the DMRG ground state this weight is frozen into SC quasi-long-range order.

\begin{figure}[t]
\centering
\includegraphics[width=0.9\columnwidth]{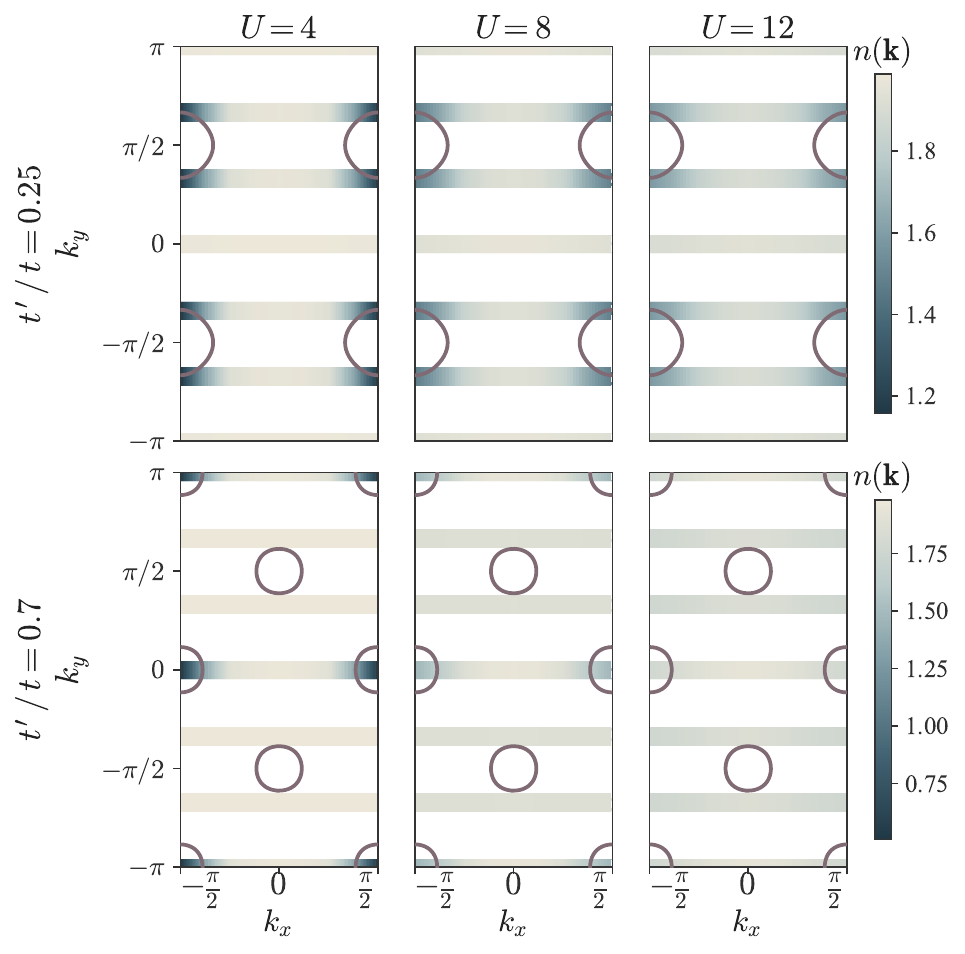}
\caption{Electron momentum distribution $n(\bm k)$ in the magnetic Brillouin zone, measured by DMRG at $\delta=1/12$ for $t'/t=0.25$ (top row) and $t'/t=0.7$ (bottom row) at $U=4$, $8$, and $12$ (columns); each row shares one color scale. The horizontal stripes are the discrete transverse momenta of the $L_y=6$ cylinder, interpolated only along $k_x$. Gray contours show the noninteracting Fermi surface at the same $t'$.}
\label{fig:nk}
\end{figure}

\textbf{Discussion.---}
The $\pi$-flux square-lattice Hofstadter-Hubbard model hosts topologically distinct half-filled phases and, upon doping, chiral topological SC with quantized spin Chern number $C_s=-2$ over a wide parameter range.

The SC obtained over the topologically trivial AFM is adiabatically connected to the one obtained by doping the CSL or IQH,
making neither a topologically nontrivial nor a spin-liquid parent a prior requirement.
This broadens the search space for unconventional SC in experimental systems.

The electron momentum distribution $n(\bm k)$ measured by DMRG [\figref{fig:nk}] gives further insight into the doped states. At weak coupling, $n(\bm k)$ retains a rather sharp Fermi-surface structure, indicating that the electron remains a reasonably well-defined quasiparticle. At low doping we can therefore analyze pairing perturbatively in $U$. Virtual interband fluctuations across the Chern-band gap generate an effective attraction~\cite{CrepelFu2021,Crepel2022InterbandSC,Chen2026TriangularHofstadterSC} and with it a SC instability of the Fermi surface (\appref{app:KL}).
The pairing phase pattern and Bogoliubov-de Gennes spin Chern number $C_s=-2$ agree with DMRG [\figsref{fig:doped}(a), \ref{fig:doped}(d)].
Upon increasing $U$, the Fermi-surface features of $n(\bm k)$ blur  [\figref{fig:nk}], the electron loses its well-defined quasiparticle character and a perturbative description is no longer appropriate.
In the parton framework, a gauge transformation recasts the spinon ansatz as an explicit chiral resonating-valence-bond (RVB) paired state.
Upon doping, this neutral paired state is expected to acquire charge and become a genuine Cooper pair, which would produce the chiral SC with the same inherited $C_s=-2$ (\appref{app:parton_pairing}).

The same $n(\bm k)$ data [\figref{fig:nk}] clarify why a chiral metal, rather than the SC, appears at weak coupling and large $t'$ in the simulations. For $t'<t/2$ the momentum cuts of the six-leg cylinder largely overlap with the two-dimensional Fermi pockets (gray contours in \figref{fig:nk}), whereas for $t'>t/2$ two pockets---those which intersect the momentum line $k_y=\pm\pi/2$---are missed by the $n(\bm k)$ sampling, because the six-leg grid does not contain those lines, unlike four-leg cylinders (\appref{app:sixleg}). A bosonization analysis of the pockets which {\em do} overlap with the DMRG momentum cuts (\appref{app:sixleg}) shows that in this geometry the DMRG results map to two nearly decoupled spinful chains. In turn, at the cylinder lengths accessible to the simulations, the RG flow can be cut off before any spin gap opens, leaving a metal. The chiral metal observed at small $U$ and large $t'$ may therefore be an artifact of the cylinder geometry and need not survive in the two-dimensional thermodynamic limit and SC there is not excluded. Since the ``missing'' pockets matter mainly where the Fermi surface controls the physics, whereas local physics dominates at large $U$, we expect the DMRG results outside the weak-coupling, large-$t'$ corner to carry over to two dimensions.

An open question is that of doping at $t'=0$, where the convergence of the Hubbard model is hard to achieve [\appref{app:dmrg_t0}]. This motivates revisiting the problem in a smaller Hilbert space, such as that of the Hofstadter $t$-$J$ model. What makes this limit worth studying is that time-reversal symmetry is restored there, so a chiral SC surviving doping would have to break that symmetry spontaneously.

\addtocontents{toc}{\string\tocdepth@munge}

\begin{acknowledgments}
{\textbf{Acknowledgements.---}}
J.-X.Z.\ and L.S.\ were funded by the European Research Council (ERC) under the European Union’s Horizon 2020 research and innovation program (Grant Agreement No.\ 853116, acronym TRANSPORT).
W.O.W. acknowledges support from the Simons Collaboration on Ultra-Quantum Matter, which is a grant from the Simons Foundation (Grant No. 651440), and from the Gordon and Betty Moore Foundation through Grant GBMF14251 to the American Chemical Society for the ACS--Moore Postdoctoral Bridge Award in Quantum Materials.
DQMC simulations made use of computational facilities purchased with funds from the NSF (CNS-1725797) and administered by the Center for Scientific Computing (CSC) supported by the California NanoSystems Institute and the Materials Research Science and Engineering Center (MRSEC; NSF DMR 2308708) at UC Santa Barbara. It has also benefited from the Mitchel Postdoctoral Scholar Career Development Fund.
L.B. was funded by the Air Force Office of Scientific Research (AFOSR) (award FA9550-22-1-0432).  D.N.S. is  supported by  the DOE, Office of BES under Grant No. DE-FG02-06ER46305 for large scale DMRG  simulations.  D.N.S. also acknowledges  travel support through  NSF grant PHY-2309135 to KITP.
\end{acknowledgments}

\newpage
\bibliographystyle{apsrev4-2-titles}
\bibliography{main}

\begin{thebibliography}{73}%
\makeatletter
\providecommand \@ifxundefined [1]{%
 \@ifx{#1\undefined}
}%
\providecommand \@ifnum [1]{%
 \ifnum #1\expandafter \@firstoftwo
 \else \expandafter \@secondoftwo
 \fi
}%
\providecommand \@ifx [1]{%
 \ifx #1\expandafter \@firstoftwo
 \else \expandafter \@secondoftwo
 \fi
}%
\providecommand \natexlab [1]{#1}%
\providecommand \enquote  [1]{``#1''}%
\providecommand \bibnamefont  [1]{#1}%
\providecommand \bibfnamefont [1]{#1}%
\providecommand \citenamefont [1]{#1}%
\providecommand \href@noop [0]{\@secondoftwo}%
\providecommand \href [0]{\begingroup \@sanitize@url \@href}%
\providecommand \@href[1]{\@@startlink{#1}\@@href}%
\providecommand \@@href[1]{\endgroup#1\@@endlink}%
\providecommand \@sanitize@url [0]{\catcode `\\12\catcode `\$12\catcode `\&12\catcode `\#12\catcode `\^12\catcode `\_12\catcode `\%12\relax}%
\providecommand \@@startlink[1]{}%
\providecommand \@@endlink[0]{}%
\providecommand \url  [0]{\begingroup\@sanitize@url \@url }%
\providecommand \@url [1]{\endgroup\@href {#1}{\urlprefix }}%
\providecommand \urlprefix  [0]{URL }%
\providecommand \Eprint [0]{\href }%
\providecommand \doibase [0]{https://doi.org/}%
\providecommand \selectlanguage [0]{\@gobble}%
\providecommand \bibinfo  [0]{\@secondoftwo}%
\providecommand \bibfield  [0]{\@secondoftwo}%
\providecommand \translation [1]{[#1]}%
\providecommand \BibitemOpen [0]{}%
\providecommand \bibitemStop [0]{}%
\providecommand \bibitemNoStop [0]{.\EOS\space}%
\providecommand \EOS [0]{\spacefactor3000\relax}%
\providecommand \BibitemShut  [1]{\csname bibitem#1\endcsname}%
\let\auto@bib@innerbib\@empty
\bibitem [{\citenamefont {Anderson}(1987)}]{Anderson1987RVB}%
  \BibitemOpen
  \bibfield  {author} {\bibinfo {author} {\bibfnamefont {P.~W.}\ \bibnamefont {Anderson}},\ }\bibfield  {title} {\bibinfo {title} {The resonating valence bond state in {La$_2$CuO$_4$} and superconductivity},\ }\href {https://doi.org/10.1126/science.235.4793.1196} {\bibfield  {journal} {\bibinfo  {journal} {Science}\ }\textbf {\bibinfo {volume} {235}},\ \bibinfo {pages} {1196} (\bibinfo {year} {1987})}\BibitemShut {NoStop}%
\bibitem [{\citenamefont {Keimer}\ \emph {et~al.}(2015)\citenamefont {Keimer}, \citenamefont {Kivelson}, \citenamefont {Norman}, \citenamefont {Uchida},\ and\ \citenamefont {Zaanen}}]{Keimer2015HighTc}%
  \BibitemOpen
  \bibfield  {author} {\bibinfo {author} {\bibfnamefont {B.}~\bibnamefont {Keimer}}, \bibinfo {author} {\bibfnamefont {S.~A.}\ \bibnamefont {Kivelson}}, \bibinfo {author} {\bibfnamefont {M.~R.}\ \bibnamefont {Norman}}, \bibinfo {author} {\bibfnamefont {S.}~\bibnamefont {Uchida}},\ and\ \bibinfo {author} {\bibfnamefont {J.}~\bibnamefont {Zaanen}},\ }\bibfield  {title} {\bibinfo {title} {From quantum matter to high-temperature superconductivity in copper oxides},\ }\href {https://doi.org/10.1038/nature14165} {\bibfield  {journal} {\bibinfo  {journal} {Nature}\ }\textbf {\bibinfo {volume} {518}},\ \bibinfo {pages} {179} (\bibinfo {year} {2015})}\BibitemShut {NoStop}%
\bibitem [{\citenamefont {Lee}\ \emph {et~al.}(2006)\citenamefont {Lee}, \citenamefont {Nagaosa},\ and\ \citenamefont {Wen}}]{LeeNagaosaWen2006}%
  \BibitemOpen
  \bibfield  {author} {\bibinfo {author} {\bibfnamefont {P.~A.}\ \bibnamefont {Lee}}, \bibinfo {author} {\bibfnamefont {N.}~\bibnamefont {Nagaosa}},\ and\ \bibinfo {author} {\bibfnamefont {X.-G.}\ \bibnamefont {Wen}},\ }\bibfield  {title} {\bibinfo {title} {Doping a {Mott} insulator: Physics of high-temperature superconductivity},\ }\href {https://doi.org/10.1103/RevModPhys.78.17} {\bibfield  {journal} {\bibinfo  {journal} {Reviews of Modern Physics}\ }\textbf {\bibinfo {volume} {78}},\ \bibinfo {pages} {17} (\bibinfo {year} {2006})}\BibitemShut {NoStop}%
\bibitem [{\citenamefont {Laughlin}(1988)}]{Laughlin1988AnyonSC}%
  \BibitemOpen
  \bibfield  {author} {\bibinfo {author} {\bibfnamefont {R.~B.}\ \bibnamefont {Laughlin}},\ }\bibfield  {title} {\bibinfo {title} {Superconducting ground state of noninteracting particles obeying fractional statistics},\ }\href {https://doi.org/10.1103/PhysRevLett.60.2677} {\bibfield  {journal} {\bibinfo  {journal} {Physical Review Letters}\ }\textbf {\bibinfo {volume} {60}},\ \bibinfo {pages} {2677} (\bibinfo {year} {1988})}\BibitemShut {NoStop}%
\bibitem [{\citenamefont {Wen}\ \emph {et~al.}(1989)\citenamefont {Wen}, \citenamefont {Wilczek},\ and\ \citenamefont {Zee}}]{WenWilczekZee1989ChiralSpinSC}%
  \BibitemOpen
  \bibfield  {author} {\bibinfo {author} {\bibfnamefont {X.~G.}\ \bibnamefont {Wen}}, \bibinfo {author} {\bibfnamefont {F.}~\bibnamefont {Wilczek}},\ and\ \bibinfo {author} {\bibfnamefont {A.}~\bibnamefont {Zee}},\ }\bibfield  {title} {\bibinfo {title} {Chiral spin states and superconductivity},\ }\href {https://doi.org/10.1103/PhysRevB.39.11413} {\bibfield  {journal} {\bibinfo  {journal} {Physical Review B}\ }\textbf {\bibinfo {volume} {39}},\ \bibinfo {pages} {11413} (\bibinfo {year} {1989})}\BibitemShut {NoStop}%
\bibitem [{\citenamefont {Lee}\ and\ \citenamefont {Fisher}(1989)}]{LeeFisher1989AnyonSC}%
  \BibitemOpen
  \bibfield  {author} {\bibinfo {author} {\bibfnamefont {D.-H.}\ \bibnamefont {Lee}}\ and\ \bibinfo {author} {\bibfnamefont {M.~P.~A.}\ \bibnamefont {Fisher}},\ }\bibfield  {title} {\bibinfo {title} {Anyon superconductivity and the fractional quantum {Hall} effect},\ }\href {https://doi.org/10.1103/PhysRevLett.63.903} {\bibfield  {journal} {\bibinfo  {journal} {Physical Review Letters}\ }\textbf {\bibinfo {volume} {63}},\ \bibinfo {pages} {903} (\bibinfo {year} {1989})}\BibitemShut {NoStop}%
\bibitem [{\citenamefont {Balents}(2010)}]{Balents2010SpinLiquids}%
  \BibitemOpen
  \bibfield  {author} {\bibinfo {author} {\bibfnamefont {L.}~\bibnamefont {Balents}},\ }\bibfield  {title} {\bibinfo {title} {Spin liquids in frustrated magnets},\ }\href {https://doi.org/10.1038/nature08917} {\bibfield  {journal} {\bibinfo  {journal} {Nature}\ }\textbf {\bibinfo {volume} {464}},\ \bibinfo {pages} {199} (\bibinfo {year} {2010})}\BibitemShut {NoStop}%
\bibitem [{\citenamefont {Kalmeyer}\ and\ \citenamefont {Laughlin}(1987)}]{KalmeyerLaughlin1987CSL}%
  \BibitemOpen
  \bibfield  {author} {\bibinfo {author} {\bibfnamefont {V.}~\bibnamefont {Kalmeyer}}\ and\ \bibinfo {author} {\bibfnamefont {R.~B.}\ \bibnamefont {Laughlin}},\ }\bibfield  {title} {\bibinfo {title} {Equivalence of the resonating-valence-bond and fractional quantum {Hall} states},\ }\href {https://doi.org/10.1103/PhysRevLett.59.2095} {\bibfield  {journal} {\bibinfo  {journal} {Physical Review Letters}\ }\textbf {\bibinfo {volume} {59}},\ \bibinfo {pages} {2095} (\bibinfo {year} {1987})}\BibitemShut {NoStop}%
\bibitem [{\citenamefont {He}\ \emph {et~al.}(2014)\citenamefont {He}, \citenamefont {Sheng},\ and\ \citenamefont {Chen}}]{He2014KagomeCSL}%
  \BibitemOpen
  \bibfield  {author} {\bibinfo {author} {\bibfnamefont {Y.-C.}\ \bibnamefont {He}}, \bibinfo {author} {\bibfnamefont {D.~N.}\ \bibnamefont {Sheng}},\ and\ \bibinfo {author} {\bibfnamefont {Y.}~\bibnamefont {Chen}},\ }\bibfield  {title} {\bibinfo {title} {Chiral spin liquid in a frustrated anisotropic kagome {Heisenberg} model},\ }\href {https://doi.org/10.1103/PhysRevLett.112.137202} {\bibfield  {journal} {\bibinfo  {journal} {Physical Review Letters}\ }\textbf {\bibinfo {volume} {112}},\ \bibinfo {pages} {137202} (\bibinfo {year} {2014})}\BibitemShut {NoStop}%
\bibitem [{\citenamefont {Gong}\ \emph {et~al.}(2014)\citenamefont {Gong}, \citenamefont {Zhu},\ and\ \citenamefont {Sheng}}]{Gong2014KagomeCSL}%
  \BibitemOpen
  \bibfield  {author} {\bibinfo {author} {\bibfnamefont {S.-S.}\ \bibnamefont {Gong}}, \bibinfo {author} {\bibfnamefont {W.}~\bibnamefont {Zhu}},\ and\ \bibinfo {author} {\bibfnamefont {D.~N.}\ \bibnamefont {Sheng}},\ }\bibfield  {title} {\bibinfo {title} {Emergent chiral spin liquid: Fractional quantum {Hall} effect in a kagome {Heisenberg} model},\ }\href {https://doi.org/10.1038/srep06317} {\bibfield  {journal} {\bibinfo  {journal} {Scientific Reports}\ }\textbf {\bibinfo {volume} {4}},\ \bibinfo {pages} {6317} (\bibinfo {year} {2014})}\BibitemShut {NoStop}%
\bibitem [{\citenamefont {Bauer}\ \emph {et~al.}(2014)\citenamefont {Bauer}, \citenamefont {Cincio}, \citenamefont {Keller}, \citenamefont {Dolfi}, \citenamefont {Vidal}, \citenamefont {Trebst},\ and\ \citenamefont {Ludwig}}]{Bauer2014KagomeCSL}%
  \BibitemOpen
  \bibfield  {author} {\bibinfo {author} {\bibfnamefont {B.}~\bibnamefont {Bauer}}, \bibinfo {author} {\bibfnamefont {L.}~\bibnamefont {Cincio}}, \bibinfo {author} {\bibfnamefont {B.~P.}\ \bibnamefont {Keller}}, \bibinfo {author} {\bibfnamefont {M.}~\bibnamefont {Dolfi}}, \bibinfo {author} {\bibfnamefont {G.}~\bibnamefont {Vidal}}, \bibinfo {author} {\bibfnamefont {S.}~\bibnamefont {Trebst}},\ and\ \bibinfo {author} {\bibfnamefont {A.~W.~W.}\ \bibnamefont {Ludwig}},\ }\bibfield  {title} {\bibinfo {title} {Chiral spin liquid and emergent anyons in a kagome lattice {Mott} insulator},\ }\href {https://doi.org/10.1038/ncomms6137} {\bibfield  {journal} {\bibinfo  {journal} {Nature Communications}\ }\textbf {\bibinfo {volume} {5}},\ \bibinfo {pages} {5137} (\bibinfo {year} {2014})}\BibitemShut {NoStop}%
\bibitem [{\citenamefont {Szasz}\ \emph {et~al.}(2020)\citenamefont {Szasz}, \citenamefont {Motruk}, \citenamefont {Zaletel},\ and\ \citenamefont {Moore}}]{Szasz2020TriangularCSL}%
  \BibitemOpen
  \bibfield  {author} {\bibinfo {author} {\bibfnamefont {A.}~\bibnamefont {Szasz}}, \bibinfo {author} {\bibfnamefont {J.}~\bibnamefont {Motruk}}, \bibinfo {author} {\bibfnamefont {M.~P.}\ \bibnamefont {Zaletel}},\ and\ \bibinfo {author} {\bibfnamefont {J.~E.}\ \bibnamefont {Moore}},\ }\bibfield  {title} {\bibinfo {title} {Chiral spin liquid phase of the triangular lattice {Hubbard} model: A density matrix renormalization group study},\ }\href {https://doi.org/10.1103/PhysRevX.10.021042} {\bibfield  {journal} {\bibinfo  {journal} {Physical Review X}\ }\textbf {\bibinfo {volume} {10}},\ \bibinfo {pages} {021042} (\bibinfo {year} {2020})}\BibitemShut {NoStop}%
\bibitem [{\citenamefont {Chen}\ \emph {et~al.}(2022)\citenamefont {Chen}, \citenamefont {Chen}, \citenamefont {Gong}, \citenamefont {Sheng}, \citenamefont {Li},\ and\ \citenamefont {Weichselbaum}}]{Chen2022TriangularCSL}%
  \BibitemOpen
  \bibfield  {author} {\bibinfo {author} {\bibfnamefont {B.-B.}\ \bibnamefont {Chen}}, \bibinfo {author} {\bibfnamefont {Z.}~\bibnamefont {Chen}}, \bibinfo {author} {\bibfnamefont {S.-S.}\ \bibnamefont {Gong}}, \bibinfo {author} {\bibfnamefont {D.~N.}\ \bibnamefont {Sheng}}, \bibinfo {author} {\bibfnamefont {W.}~\bibnamefont {Li}},\ and\ \bibinfo {author} {\bibfnamefont {A.}~\bibnamefont {Weichselbaum}},\ }\bibfield  {title} {\bibinfo {title} {Quantum spin liquid with emergent chiral order in the triangular-lattice {Hubbard} model},\ }\href {https://doi.org/10.1103/PhysRevB.106.094420} {\bibfield  {journal} {\bibinfo  {journal} {Physical Review B}\ }\textbf {\bibinfo {volume} {106}},\ \bibinfo {pages} {094420} (\bibinfo {year} {2022})}\BibitemShut {NoStop}%
\bibitem [{\citenamefont {Zhou}\ \emph {et~al.}(2022)\citenamefont {Zhou}, \citenamefont {Sheng},\ and\ \citenamefont {Kim}}]{Zhou2022}%
  \BibitemOpen
  \bibfield  {author} {\bibinfo {author} {\bibfnamefont {Y.}~\bibnamefont {Zhou}}, \bibinfo {author} {\bibfnamefont {D.~N.}\ \bibnamefont {Sheng}},\ and\ \bibinfo {author} {\bibfnamefont {E.-A.}\ \bibnamefont {Kim}},\ }\bibfield  {title} {\bibinfo {title} {Quantum phases of transition metal dichalcogenide moir\'e systems},\ }\href {https://doi.org/10.1103/PhysRevLett.128.157602} {\bibfield  {journal} {\bibinfo  {journal} {Phys. Rev. Lett.}\ }\textbf {\bibinfo {volume} {128}},\ \bibinfo {pages} {157602} (\bibinfo {year} {2022})}\BibitemShut {NoStop}%
\bibitem [{\citenamefont {Wietek}\ \emph {et~al.}(2021)\citenamefont {Wietek}, \citenamefont {Rossi}, \citenamefont {{\v S}imkovic}, \citenamefont {Klett}, \citenamefont {Hansmann}, \citenamefont {Ferrero}, \citenamefont {Stoudenmire}, \citenamefont {Sch{\"a}fer},\ and\ \citenamefont {Georges}}]{Wietek2021TriangularMott}%
  \BibitemOpen
  \bibfield  {author} {\bibinfo {author} {\bibfnamefont {A.}~\bibnamefont {Wietek}}, \bibinfo {author} {\bibfnamefont {R.}~\bibnamefont {Rossi}}, \bibinfo {author} {\bibfnamefont {F.}~\bibnamefont {{\v S}imkovic}}, \bibinfo {author} {\bibfnamefont {M.}~\bibnamefont {Klett}}, \bibinfo {author} {\bibfnamefont {P.}~\bibnamefont {Hansmann}}, \bibinfo {author} {\bibfnamefont {M.}~\bibnamefont {Ferrero}}, \bibinfo {author} {\bibfnamefont {E.~M.}\ \bibnamefont {Stoudenmire}}, \bibinfo {author} {\bibfnamefont {T.}~\bibnamefont {Sch{\"a}fer}},\ and\ \bibinfo {author} {\bibfnamefont {A.}~\bibnamefont {Georges}},\ }\bibfield  {title} {\bibinfo {title} {Mott insulating states with competing orders in the triangular lattice {Hubbard} model},\ }\href {https://doi.org/10.1103/PhysRevX.11.041013} {\bibfield  {journal} {\bibinfo  {journal} {Physical Review X}\ }\textbf {\bibinfo {volume} {11}},\ \bibinfo {pages} {041013} (\bibinfo {year} {2021})}\BibitemShut {NoStop}%
\bibitem [{\citenamefont {Cai}\ \emph {et~al.}(2023)\citenamefont {Cai}, \citenamefont {Anderson}, \citenamefont {Wang} \emph {et~al.}}]{Cai2023FQAH}%
  \BibitemOpen
  \bibfield  {author} {\bibinfo {author} {\bibfnamefont {J.}~\bibnamefont {Cai}}, \bibinfo {author} {\bibfnamefont {E.}~\bibnamefont {Anderson}}, \bibinfo {author} {\bibfnamefont {C.}~\bibnamefont {Wang}}, \emph {et~al.},\ }\bibfield  {title} {\bibinfo {title} {Signatures of fractional quantum anomalous {Hall} states in twisted {MoTe}$_2$},\ }\href {https://doi.org/10.1038/s41586-023-06289-w} {\bibfield  {journal} {\bibinfo  {journal} {Nature}\ }\textbf {\bibinfo {volume} {622}},\ \bibinfo {pages} {63} (\bibinfo {year} {2023})}\BibitemShut {NoStop}%
\bibitem [{\citenamefont {Zeng}\ \emph {et~al.}(2023)\citenamefont {Zeng}, \citenamefont {Xia}, \citenamefont {Kang} \emph {et~al.}}]{Zeng2023FQAH}%
  \BibitemOpen
  \bibfield  {author} {\bibinfo {author} {\bibfnamefont {Y.}~\bibnamefont {Zeng}}, \bibinfo {author} {\bibfnamefont {Z.}~\bibnamefont {Xia}}, \bibinfo {author} {\bibfnamefont {K.}~\bibnamefont {Kang}}, \emph {et~al.},\ }\bibfield  {title} {\bibinfo {title} {Thermodynamic evidence of fractional {Chern} insulator in moir\'e {MoTe}$_2$},\ }\href {https://doi.org/10.1038/s41586-023-06452-3} {\bibfield  {journal} {\bibinfo  {journal} {Nature}\ }\textbf {\bibinfo {volume} {622}},\ \bibinfo {pages} {69} (\bibinfo {year} {2023})}\BibitemShut {NoStop}%
\bibitem [{\citenamefont {Park}\ \emph {et~al.}(2023)\citenamefont {Park}, \citenamefont {Cai}, \citenamefont {Anderson} \emph {et~al.}}]{Park2023FQAH}%
  \BibitemOpen
  \bibfield  {author} {\bibinfo {author} {\bibfnamefont {H.}~\bibnamefont {Park}}, \bibinfo {author} {\bibfnamefont {J.}~\bibnamefont {Cai}}, \bibinfo {author} {\bibfnamefont {E.}~\bibnamefont {Anderson}}, \emph {et~al.},\ }\bibfield  {title} {\bibinfo {title} {Observation of fractionally quantized anomalous {Hall} effect},\ }\href {https://doi.org/10.1038/s41586-023-06536-0} {\bibfield  {journal} {\bibinfo  {journal} {Nature}\ }\textbf {\bibinfo {volume} {622}},\ \bibinfo {pages} {74} (\bibinfo {year} {2023})}\BibitemShut {NoStop}%
\bibitem [{\citenamefont {Xu}\ \emph {et~al.}(2023)\citenamefont {Xu}, \citenamefont {Sun}, \citenamefont {Jia} \emph {et~al.}}]{Xu2023FQAH}%
  \BibitemOpen
  \bibfield  {author} {\bibinfo {author} {\bibfnamefont {F.}~\bibnamefont {Xu}}, \bibinfo {author} {\bibfnamefont {Z.}~\bibnamefont {Sun}}, \bibinfo {author} {\bibfnamefont {T.}~\bibnamefont {Jia}}, \emph {et~al.},\ }\bibfield  {title} {\bibinfo {title} {Observation of integer and fractional quantum anomalous {Hall} effects in twisted bilayer {MoTe}$_2$},\ }\href {https://doi.org/10.1103/PhysRevX.13.031037} {\bibfield  {journal} {\bibinfo  {journal} {Physical Review X}\ }\textbf {\bibinfo {volume} {13}},\ \bibinfo {pages} {031037} (\bibinfo {year} {2023})}\BibitemShut {NoStop}%
\bibitem [{\citenamefont {Lu}\ \emph {et~al.}(2024)\citenamefont {Lu}, \citenamefont {Han}, \citenamefont {Yao}, \citenamefont {Reddy}, \citenamefont {Yang}, \citenamefont {Seo}, \citenamefont {Watanabe}, \citenamefont {Taniguchi}, \citenamefont {Fu},\ and\ \citenamefont {Ju}}]{Lu2024GrapheneFQAH}%
  \BibitemOpen
  \bibfield  {author} {\bibinfo {author} {\bibfnamefont {Z.}~\bibnamefont {Lu}}, \bibinfo {author} {\bibfnamefont {T.}~\bibnamefont {Han}}, \bibinfo {author} {\bibfnamefont {Y.}~\bibnamefont {Yao}}, \bibinfo {author} {\bibfnamefont {A.~P.}\ \bibnamefont {Reddy}}, \bibinfo {author} {\bibfnamefont {J.}~\bibnamefont {Yang}}, \bibinfo {author} {\bibfnamefont {J.}~\bibnamefont {Seo}}, \bibinfo {author} {\bibfnamefont {K.}~\bibnamefont {Watanabe}}, \bibinfo {author} {\bibfnamefont {T.}~\bibnamefont {Taniguchi}}, \bibinfo {author} {\bibfnamefont {L.}~\bibnamefont {Fu}},\ and\ \bibinfo {author} {\bibfnamefont {L.}~\bibnamefont {Ju}},\ }\bibfield  {title} {\bibinfo {title} {Fractional quantum anomalous {Hall} effect in multilayer graphene},\ }\href@noop {} {\bibfield  {journal} {\bibinfo  {journal} {Nature}\ }\textbf {\bibinfo {volume} {626}},\ \bibinfo {pages} {759} (\bibinfo {year} {2024})}\BibitemShut {NoStop}%
\bibitem [{\citenamefont {Xu}\ \emph {et~al.}(2025)\citenamefont {Xu}, \citenamefont {Sun}, \citenamefont {Li}, \citenamefont {Zheng}, \citenamefont {Xu}, \citenamefont {Gao}, \citenamefont {Jia}, \citenamefont {Su}, \citenamefont {Watanabe}, \citenamefont {Taniguchi}, \citenamefont {Tong}, \citenamefont {Lu}, \citenamefont {Jia}, \citenamefont {Shi}, \citenamefont {Jiang}, \citenamefont {Lin}, \citenamefont {Zhang}, \citenamefont {Zhang}, \citenamefont {Lei}, \citenamefont {Liu},\ and\ \citenamefont {Li}}]{Xu2025MoTe2SC}%
  \BibitemOpen
  \bibfield  {author} {\bibinfo {author} {\bibfnamefont {F.}~\bibnamefont {Xu}}, \bibinfo {author} {\bibfnamefont {Z.}~\bibnamefont {Sun}}, \bibinfo {author} {\bibfnamefont {J.}~\bibnamefont {Li}}, \bibinfo {author} {\bibfnamefont {C.}~\bibnamefont {Zheng}}, \bibinfo {author} {\bibfnamefont {C.}~\bibnamefont {Xu}}, \bibinfo {author} {\bibfnamefont {J.}~\bibnamefont {Gao}}, \bibinfo {author} {\bibfnamefont {T.}~\bibnamefont {Jia}}, \bibinfo {author} {\bibfnamefont {Y.}~\bibnamefont {Su}}, \bibinfo {author} {\bibfnamefont {K.}~\bibnamefont {Watanabe}}, \bibinfo {author} {\bibfnamefont {T.}~\bibnamefont {Taniguchi}}, \bibinfo {author} {\bibfnamefont {B.}~\bibnamefont {Tong}}, \bibinfo {author} {\bibfnamefont {L.}~\bibnamefont {Lu}}, \bibinfo {author} {\bibfnamefont {J.}~\bibnamefont {Jia}}, \bibinfo {author} {\bibfnamefont {Z.}~\bibnamefont {Shi}}, \bibinfo {author} {\bibfnamefont {S.}~\bibnamefont {Jiang}}, \bibinfo {author} {\bibfnamefont {J.}~\bibnamefont {Lin}}, \bibinfo {author} {\bibfnamefont {Y.}~\bibnamefont {Zhang}}, \bibinfo {author} {\bibfnamefont {Y.}~\bibnamefont {Zhang}}, \bibinfo {author} {\bibfnamefont {S.}~\bibnamefont {Lei}}, \bibinfo {author} {\bibfnamefont {X.}~\bibnamefont {Liu}},\ and\ \bibinfo {author} {\bibfnamefont {T.}~\bibnamefont {Li}},\ }\href {https://arxiv.org/abs/2504.06972} {\bibinfo {title} {Signatures of unconventional superconductivity near reentrant and fractional quantum anomalous {Hall} insulators}} (\bibinfo {year} {2025}),\ \Eprint {https://arxiv.org/abs/2504.06972} {arXiv:2504.06972 [cond-mat.mes-hall]} \BibitemShut {NoStop}%
\bibitem [{\citenamefont {Shi}\ and\ \citenamefont {Senthil}(2025{\natexlab{a}})}]{ShiSenthil2024DopingFQAH}%
  \BibitemOpen
  \bibfield  {author} {\bibinfo {author} {\bibfnamefont {Z.~D.}\ \bibnamefont {Shi}}\ and\ \bibinfo {author} {\bibfnamefont {T.}~\bibnamefont {Senthil}},\ }\bibfield  {title} {\bibinfo {title} {Doping a fractional quantum anomalous {Hall} insulator},\ }\href@noop {} {\bibfield  {journal} {\bibinfo  {journal} {Physical Review X}\ }\textbf {\bibinfo {volume} {15}},\ \bibinfo {pages} {031069} (\bibinfo {year} {2025}{\natexlab{a}})},\ \Eprint {https://arxiv.org/abs/2409.20567} {arXiv:2409.20567} \BibitemShut {NoStop}%
\bibitem [{\citenamefont {Pichler}\ \emph {et~al.}(2026)\citenamefont {Pichler}, \citenamefont {Kuhlenkamp}, \citenamefont {Knap},\ and\ \citenamefont {Vishwanath}}]{Pichler2025AnyonSC}%
  \BibitemOpen
  \bibfield  {author} {\bibinfo {author} {\bibfnamefont {F.}~\bibnamefont {Pichler}}, \bibinfo {author} {\bibfnamefont {C.}~\bibnamefont {Kuhlenkamp}}, \bibinfo {author} {\bibfnamefont {M.}~\bibnamefont {Knap}},\ and\ \bibinfo {author} {\bibfnamefont {A.}~\bibnamefont {Vishwanath}},\ }\bibfield  {title} {\bibinfo {title} {Microscopic mechanism of anyon superconductivity emerging from fractional {Chern} insulators},\ }\href {https://doi.org/10.1016/j.newton.2025.100340} {\bibfield  {journal} {\bibinfo  {journal} {Newton}\ }\textbf {\bibinfo {volume} {2}},\ \bibinfo {pages} {100340} (\bibinfo {year} {2026})},\ \Eprint {https://arxiv.org/abs/2506.08000} {arXiv:2506.08000} \BibitemShut {NoStop}%
\bibitem [{\citenamefont {Kim}\ \emph {et~al.}(2025)\citenamefont {Kim}, \citenamefont {Timmel}, \citenamefont {Ju},\ and\ \citenamefont {Wen}}]{Kim2025BeyondPairing}%
  \BibitemOpen
  \bibfield  {author} {\bibinfo {author} {\bibfnamefont {M.}~\bibnamefont {Kim}}, \bibinfo {author} {\bibfnamefont {A.}~\bibnamefont {Timmel}}, \bibinfo {author} {\bibfnamefont {L.}~\bibnamefont {Ju}},\ and\ \bibinfo {author} {\bibfnamefont {X.-G.}\ \bibnamefont {Wen}},\ }\bibfield  {title} {\bibinfo {title} {Topological chiral superconductivity beyond pairing in a {Fermi} liquid},\ }\href {https://doi.org/10.1103/PhysRevB.111.014508} {\bibfield  {journal} {\bibinfo  {journal} {Physical Review B}\ }\textbf {\bibinfo {volume} {111}},\ \bibinfo {pages} {014508} (\bibinfo {year} {2025})}\BibitemShut {NoStop}%
\bibitem [{\citenamefont {Nosov}\ \emph {et~al.}(2026)\citenamefont {Nosov}, \citenamefont {Han},\ and\ \citenamefont {Khalaf}}]{Nosov2026AnyonSCPlateau}%
  \BibitemOpen
  \bibfield  {author} {\bibinfo {author} {\bibfnamefont {P.~A.}\ \bibnamefont {Nosov}}, \bibinfo {author} {\bibfnamefont {Z.}~\bibnamefont {Han}},\ and\ \bibinfo {author} {\bibfnamefont {E.}~\bibnamefont {Khalaf}},\ }\bibfield  {title} {\bibinfo {title} {Anyon superconductivity and plateau transitions in doped fractional quantum anomalous {Hall} insulators},\ }\href {https://doi.org/10.1103/6bgj-bfdn} {\bibfield  {journal} {\bibinfo  {journal} {Physical Review Letters}\ }\textbf {\bibinfo {volume} {136}},\ \bibinfo {pages} {106501} (\bibinfo {year} {2026})},\ \Eprint {https://arxiv.org/abs/2506.02108} {arXiv:2506.02108} \BibitemShut {NoStop}%
\bibitem [{\citenamefont {Guerci}\ \emph {et~al.}(2025)\citenamefont {Guerci}, \citenamefont {Abouelkomsan},\ and\ \citenamefont {Fu}}]{Guerci2025FlatChernBandSC}%
  \BibitemOpen
  \bibfield  {author} {\bibinfo {author} {\bibfnamefont {D.}~\bibnamefont {Guerci}}, \bibinfo {author} {\bibfnamefont {A.}~\bibnamefont {Abouelkomsan}},\ and\ \bibinfo {author} {\bibfnamefont {L.}~\bibnamefont {Fu}},\ }\bibfield  {title} {\bibinfo {title} {From fractionalization to chiral topological superconductivity in a flat {Chern} band},\ }\href@noop {} {\bibfield  {journal} {\bibinfo  {journal} {Physical Review Letters}\ }\textbf {\bibinfo {volume} {135}},\ \bibinfo {pages} {186601} (\bibinfo {year} {2025})},\ \Eprint {https://arxiv.org/abs/2506.10938} {arXiv:2506.10938} \BibitemShut {NoStop}%
\bibitem [{\citenamefont {Shi}\ and\ \citenamefont {Senthil}(2025{\natexlab{b}})}]{ShiSenthil2025NonAbelianTSC}%
  \BibitemOpen
  \bibfield  {author} {\bibinfo {author} {\bibfnamefont {Z.~D.}\ \bibnamefont {Shi}}\ and\ \bibinfo {author} {\bibfnamefont {T.}~\bibnamefont {Senthil}},\ }\href {https://arxiv.org/abs/2512.17996} {\bibinfo {title} {Non-{Abelian} topological superconductivity from melting {Abelian} fractional {Chern} insulators}} (\bibinfo {year} {2025}{\natexlab{b}}),\ \Eprint {https://arxiv.org/abs/2512.17996} {arXiv:2512.17996 [cond-mat.str-el]} \BibitemShut {NoStop}%
\bibitem [{\citenamefont {Jiang}\ and\ \citenamefont {Jiang}(2020)}]{Jiang2020DopedCSL}%
  \BibitemOpen
  \bibfield  {author} {\bibinfo {author} {\bibfnamefont {Y.-F.}\ \bibnamefont {Jiang}}\ and\ \bibinfo {author} {\bibfnamefont {H.-C.}\ \bibnamefont {Jiang}},\ }\bibfield  {title} {\bibinfo {title} {Topological superconductivity in the doped chiral spin liquid on the triangular lattice},\ }\href {https://doi.org/10.1103/PhysRevLett.125.157002} {\bibfield  {journal} {\bibinfo  {journal} {Physical Review Letters}\ }\textbf {\bibinfo {volume} {125}},\ \bibinfo {pages} {157002} (\bibinfo {year} {2020})}\BibitemShut {NoStop}%
\bibitem [{\citenamefont {Jiang}(2021)}]{Jiang2021npjDopedQSL}%
  \BibitemOpen
  \bibfield  {author} {\bibinfo {author} {\bibfnamefont {H.-C.}\ \bibnamefont {Jiang}},\ }\bibfield  {title} {\bibinfo {title} {Superconductivity in the doped quantum spin liquid on the triangular lattice},\ }\href {https://doi.org/10.1038/s41535-021-00375-w} {\bibfield  {journal} {\bibinfo  {journal} {npj Quantum Materials}\ }\textbf {\bibinfo {volume} {6}},\ \bibinfo {pages} {71} (\bibinfo {year} {2021})}\BibitemShut {NoStop}%
\bibitem [{\citenamefont {Huang}\ and\ \citenamefont {Sheng}(2022)}]{Huang2022ChiralNematicSC}%
  \BibitemOpen
  \bibfield  {author} {\bibinfo {author} {\bibfnamefont {Y.}~\bibnamefont {Huang}}\ and\ \bibinfo {author} {\bibfnamefont {D.~N.}\ \bibnamefont {Sheng}},\ }\bibfield  {title} {\bibinfo {title} {Topological chiral and nematic superconductivity by doping {Mott} insulators on triangular lattice},\ }\href {https://doi.org/10.1103/PhysRevX.12.031009} {\bibfield  {journal} {\bibinfo  {journal} {Physical Review X}\ }\textbf {\bibinfo {volume} {12}},\ \bibinfo {pages} {031009} (\bibinfo {year} {2022})}\BibitemShut {NoStop}%
\bibitem [{\citenamefont {Huang}\ \emph {et~al.}(2023)\citenamefont {Huang}, \citenamefont {Gong},\ and\ \citenamefont {Sheng}}]{Huang2023DopedTriangularTJ}%
  \BibitemOpen
  \bibfield  {author} {\bibinfo {author} {\bibfnamefont {Y.}~\bibnamefont {Huang}}, \bibinfo {author} {\bibfnamefont {S.-S.}\ \bibnamefont {Gong}},\ and\ \bibinfo {author} {\bibfnamefont {D.~N.}\ \bibnamefont {Sheng}},\ }\bibfield  {title} {\bibinfo {title} {Quantum phase diagram and spontaneously emergent topological chiral superconductivity in doped triangular-lattice {Mott} insulators},\ }\href {https://doi.org/10.1103/PhysRevLett.130.136003} {\bibfield  {journal} {\bibinfo  {journal} {Physical Review Letters}\ }\textbf {\bibinfo {volume} {130}},\ \bibinfo {pages} {136003} (\bibinfo {year} {2023})}\BibitemShut {NoStop}%
\bibitem [{\citenamefont {Song}\ \emph {et~al.}(2021)\citenamefont {Song}, \citenamefont {Vishwanath},\ and\ \citenamefont {Zhang}}]{Song2021DopingCSL}%
  \BibitemOpen
  \bibfield  {author} {\bibinfo {author} {\bibfnamefont {X.-Y.}\ \bibnamefont {Song}}, \bibinfo {author} {\bibfnamefont {A.}~\bibnamefont {Vishwanath}},\ and\ \bibinfo {author} {\bibfnamefont {Y.-H.}\ \bibnamefont {Zhang}},\ }\bibfield  {title} {\bibinfo {title} {Doping the chiral spin liquid: Topological superconductor or chiral metal},\ }\href {https://doi.org/10.1103/PhysRevB.103.165138} {\bibfield  {journal} {\bibinfo  {journal} {Physical Review B}\ }\textbf {\bibinfo {volume} {103}},\ \bibinfo {pages} {165138} (\bibinfo {year} {2021})}\BibitemShut {NoStop}%
\bibitem [{\citenamefont {Zhu}\ \emph {et~al.}(2022)\citenamefont {Zhu}, \citenamefont {Sheng},\ and\ \citenamefont {Vishwanath}}]{Zhu2022DopedMottTriangular}%
  \BibitemOpen
  \bibfield  {author} {\bibinfo {author} {\bibfnamefont {Z.}~\bibnamefont {Zhu}}, \bibinfo {author} {\bibfnamefont {D.~N.}\ \bibnamefont {Sheng}},\ and\ \bibinfo {author} {\bibfnamefont {A.}~\bibnamefont {Vishwanath}},\ }\bibfield  {title} {\bibinfo {title} {Doped {Mott} insulators in the triangular-lattice {Hubbard} model},\ }\href {https://doi.org/10.1103/PhysRevB.105.205110} {\bibfield  {journal} {\bibinfo  {journal} {Physical Review B}\ }\textbf {\bibinfo {volume} {105}},\ \bibinfo {pages} {205110} (\bibinfo {year} {2022})}\BibitemShut {NoStop}%
\bibitem [{\citenamefont {Peng}\ \emph {et~al.}(2021)\citenamefont {Peng}, \citenamefont {Jiang}, \citenamefont {Sheng},\ and\ \citenamefont {Jiang}}]{Peng2021DopedKagome}%
  \BibitemOpen
  \bibfield  {author} {\bibinfo {author} {\bibfnamefont {C.}~\bibnamefont {Peng}}, \bibinfo {author} {\bibfnamefont {Y.-F.}\ \bibnamefont {Jiang}}, \bibinfo {author} {\bibfnamefont {D.-N.}\ \bibnamefont {Sheng}},\ and\ \bibinfo {author} {\bibfnamefont {H.-C.}\ \bibnamefont {Jiang}},\ }\bibfield  {title} {\bibinfo {title} {Doping quantum spin liquids on the kagome lattice},\ }\href {https://doi.org/10.1002/qute.202000126} {\bibfield  {journal} {\bibinfo  {journal} {Advanced Quantum Technologies}\ }\textbf {\bibinfo {volume} {4}},\ \bibinfo {pages} {2000126} (\bibinfo {year} {2021})}\BibitemShut {NoStop}%
\bibitem [{\citenamefont {Divic}\ \emph {et~al.}(2025)\citenamefont {Divic}, \citenamefont {Cr{\'e}pel}, \citenamefont {Soejima}, \citenamefont {Song}, \citenamefont {Millis}, \citenamefont {Zaletel},\ and\ \citenamefont {Vishwanath}}]{Divic2025AnyonSuperconductivity}%
  \BibitemOpen
  \bibfield  {author} {\bibinfo {author} {\bibfnamefont {S.}~\bibnamefont {Divic}}, \bibinfo {author} {\bibfnamefont {V.}~\bibnamefont {Cr{\'e}pel}}, \bibinfo {author} {\bibfnamefont {T.}~\bibnamefont {Soejima}}, \bibinfo {author} {\bibfnamefont {X.-Y.}\ \bibnamefont {Song}}, \bibinfo {author} {\bibfnamefont {A.~J.}\ \bibnamefont {Millis}}, \bibinfo {author} {\bibfnamefont {M.~P.}\ \bibnamefont {Zaletel}},\ and\ \bibinfo {author} {\bibfnamefont {A.}~\bibnamefont {Vishwanath}},\ }\bibfield  {title} {\bibinfo {title} {Anyon superconductivity from topological criticality in a {Hofstadter-Hubbard} model},\ }\href {https://doi.org/10.1073/pnas.2426680122} {\bibfield  {journal} {\bibinfo  {journal} {Proceedings of the National Academy of Sciences}\ }\textbf {\bibinfo {volume} {122}},\ \bibinfo {pages} {e2426680122} (\bibinfo {year} {2025})}\BibitemShut {NoStop}%
\bibitem [{\citenamefont {Han}\ \emph {et~al.}(2026)\citenamefont {Han}, \citenamefont {Wang}, \citenamefont {Dong}, \citenamefont {Zaletel},\ and\ \citenamefont {Vishwanath}}]{Han2026AnyonSuperfluidity}%
  \BibitemOpen
  \bibfield  {author} {\bibinfo {author} {\bibfnamefont {Z.}~\bibnamefont {Han}}, \bibinfo {author} {\bibfnamefont {T.}~\bibnamefont {Wang}}, \bibinfo {author} {\bibfnamefont {Z.}~\bibnamefont {Dong}}, \bibinfo {author} {\bibfnamefont {M.~P.}\ \bibnamefont {Zaletel}},\ and\ \bibinfo {author} {\bibfnamefont {A.}~\bibnamefont {Vishwanath}},\ }\bibfield  {title} {\bibinfo {title} {Anyon superfluidity of excitons in quantum {Hall} bilayers},\ }\href {https://doi.org/10.1103/xs7g-dmn8} {\bibfield  {journal} {\bibinfo  {journal} {Physical Review X}\ }\textbf {\bibinfo {volume} {16}},\ \bibinfo {pages} {021044} (\bibinfo {year} {2026})}\BibitemShut {NoStop}%
\bibitem [{\citenamefont {Chen}\ \emph {et~al.}(2026)\citenamefont {Chen}, \citenamefont {Wang}, \citenamefont {Zhang}, \citenamefont {Balents},\ and\ \citenamefont {Sheng}}]{Chen2026TriangularHofstadterSC}%
  \BibitemOpen
  \bibfield  {author} {\bibinfo {author} {\bibfnamefont {F.}~\bibnamefont {Chen}}, \bibinfo {author} {\bibfnamefont {W.~O.}\ \bibnamefont {Wang}}, \bibinfo {author} {\bibfnamefont {J.-X.}\ \bibnamefont {Zhang}}, \bibinfo {author} {\bibfnamefont {L.}~\bibnamefont {Balents}},\ and\ \bibinfo {author} {\bibfnamefont {D.~N.}\ \bibnamefont {Sheng}},\ }\bibfield  {title} {\bibinfo {title} {Topological chiral superconductivity in the triangular-lattice {Hofstadter-Hubbard} model},\ }\href {https://doi.org/10.1103/tm9q-w5y7} {\bibfield  {journal} {\bibinfo  {journal} {Physical Review Letters}\ }\textbf {\bibinfo {volume} {136}},\ \bibinfo {pages} {086503} (\bibinfo {year} {2026})}\BibitemShut {NoStop}%
\bibitem [{\citenamefont {Kuhlenkamp}\ \emph {et~al.}(2025)\citenamefont {Kuhlenkamp}, \citenamefont {Divic}, \citenamefont {Zaletel}, \citenamefont {Soejima},\ and\ \citenamefont {Vishwanath}}]{Kuhlenkamp2025RobustSuperconductivity}%
  \BibitemOpen
  \bibfield  {author} {\bibinfo {author} {\bibfnamefont {C.}~\bibnamefont {Kuhlenkamp}}, \bibinfo {author} {\bibfnamefont {S.}~\bibnamefont {Divic}}, \bibinfo {author} {\bibfnamefont {M.~P.}\ \bibnamefont {Zaletel}}, \bibinfo {author} {\bibfnamefont {T.}~\bibnamefont {Soejima}},\ and\ \bibinfo {author} {\bibfnamefont {A.}~\bibnamefont {Vishwanath}},\ }\href {https://arxiv.org/abs/2509.02675} {\bibinfo {title} {Robust superconductivity upon doping chiral spin liquid and {Chern} insulators in a {Hubbard-Hofstadter} model}} (\bibinfo {year} {2025}),\ \Eprint {https://arxiv.org/abs/2509.02675} {arXiv:2509.02675 [cond-mat.str-el]} \BibitemShut {NoStop}%
\bibitem [{\citenamefont {Niu}\ \emph {et~al.}(2026)\citenamefont {Niu}, \citenamefont {Sheng},\ and\ \citenamefont {Peng}}]{Niu2026ThermodynamicLimit}%
  \BibitemOpen
  \bibfield  {author} {\bibinfo {author} {\bibfnamefont {S.}~\bibnamefont {Niu}}, \bibinfo {author} {\bibfnamefont {D.~N.}\ \bibnamefont {Sheng}},\ and\ \bibinfo {author} {\bibfnamefont {Y.}~\bibnamefont {Peng}},\ }\href {https://arxiv.org/abs/2512.21503} {\bibinfo {title} {Thermodynamic-limit evidence for chiral superconductivity induced by doping chiral topological phases}} (\bibinfo {year} {2026}),\ \Eprint {https://arxiv.org/abs/2512.21503} {arXiv:2512.21503 [cond-mat.str-el]} \BibitemShut {NoStop}%
\bibitem [{\citenamefont {Gallegos}\ \emph {et~al.}(2026)\citenamefont {Gallegos}, \citenamefont {Magaldi}, \citenamefont {Millis},\ and\ \citenamefont {White}}]{Gallegos2026IQHCSL}%
  \BibitemOpen
  \bibfield  {author} {\bibinfo {author} {\bibfnamefont {C.~A.}\ \bibnamefont {Gallegos}}, \bibinfo {author} {\bibfnamefont {R.~M.}\ \bibnamefont {Magaldi}}, \bibinfo {author} {\bibfnamefont {A.~J.}\ \bibnamefont {Millis}},\ and\ \bibinfo {author} {\bibfnamefont {S.~R.}\ \bibnamefont {White}},\ }\bibfield  {title} {\bibinfo {title} {Quantum {Hall} to chiral spin liquid transition in a triangular lattice {Hofstadter-Hubbard} model},\ }\href {https://doi.org/10.1103/xv65-m7m8} {\bibfield  {journal} {\bibinfo  {journal} {Physical Review B}\ }\textbf {\bibinfo {volume} {113}},\ \bibinfo {pages} {064412} (\bibinfo {year} {2026})},\ \Eprint {https://arxiv.org/abs/2510.19907} {arXiv:2510.19907 [cond-mat.str-el]} \BibitemShut {NoStop}%
\bibitem [{\citenamefont {Shaffer}\ \emph {et~al.}(2021)\citenamefont {Shaffer}, \citenamefont {Wang},\ and\ \citenamefont {Santos}}]{Shaffer2021HofstadterSC}%
  \BibitemOpen
  \bibfield  {author} {\bibinfo {author} {\bibfnamefont {D.}~\bibnamefont {Shaffer}}, \bibinfo {author} {\bibfnamefont {J.}~\bibnamefont {Wang}},\ and\ \bibinfo {author} {\bibfnamefont {L.~H.}\ \bibnamefont {Santos}},\ }\bibfield  {title} {\bibinfo {title} {Theory of {Hofstadter} superconductors},\ }\href {https://doi.org/10.1103/PhysRevB.104.184501} {\bibfield  {journal} {\bibinfo  {journal} {Physical Review B}\ }\textbf {\bibinfo {volume} {104}},\ \bibinfo {pages} {184501} (\bibinfo {year} {2021})}\BibitemShut {NoStop}%
\bibitem [{\citenamefont {Shaffer}\ \emph {et~al.}(2022)\citenamefont {Shaffer}, \citenamefont {Wang},\ and\ \citenamefont {Santos}}]{Shaffer2022HofstadterSC}%
  \BibitemOpen
  \bibfield  {author} {\bibinfo {author} {\bibfnamefont {D.}~\bibnamefont {Shaffer}}, \bibinfo {author} {\bibfnamefont {J.}~\bibnamefont {Wang}},\ and\ \bibinfo {author} {\bibfnamefont {L.~H.}\ \bibnamefont {Santos}},\ }\bibfield  {title} {\bibinfo {title} {Unconventional self-similar {Hofstadter} superconductivity from repulsive interactions},\ }\href {https://doi.org/10.1038/s41467-022-35316-z} {\bibfield  {journal} {\bibinfo  {journal} {Nature Communications}\ }\textbf {\bibinfo {volume} {13}},\ \bibinfo {pages} {7785} (\bibinfo {year} {2022})}\BibitemShut {NoStop}%
\bibitem [{\citenamefont {Eugenio}\ \emph {et~al.}(2025)\citenamefont {Eugenio}, \citenamefont {Luo}, \citenamefont {Vishwanath},\ and\ \citenamefont {Volkov}}]{Eugenio2025SquareHomobilayers}%
  \BibitemOpen
  \bibfield  {author} {\bibinfo {author} {\bibfnamefont {P.~M.}\ \bibnamefont {Eugenio}}, \bibinfo {author} {\bibfnamefont {Z.-X.}\ \bibnamefont {Luo}}, \bibinfo {author} {\bibfnamefont {A.}~\bibnamefont {Vishwanath}},\ and\ \bibinfo {author} {\bibfnamefont {P.~A.}\ \bibnamefont {Volkov}},\ }\bibfield  {title} {\bibinfo {title} {Tunable {$t\text{-}t^{\prime}\text{-}U$} {Hubbard} models in twisted square homobilayers},\ }\href {https://doi.org/10.1103/PhysRevLett.134.236503} {\bibfield  {journal} {\bibinfo  {journal} {Physical Review Letters}\ }\textbf {\bibinfo {volume} {134}},\ \bibinfo {pages} {236503} (\bibinfo {year} {2025})}\BibitemShut {NoStop}%
\bibitem [{\citenamefont {Hofstadter}(1976)}]{Hofstadter1976}%
  \BibitemOpen
  \bibfield  {author} {\bibinfo {author} {\bibfnamefont {D.~R.}\ \bibnamefont {Hofstadter}},\ }\bibfield  {title} {\bibinfo {title} {Energy levels and wave functions of {Bloch} electrons in rational and irrational magnetic fields},\ }\href {https://doi.org/10.1103/PhysRevB.14.2239} {\bibfield  {journal} {\bibinfo  {journal} {Physical Review B}\ }\textbf {\bibinfo {volume} {14}},\ \bibinfo {pages} {2239} (\bibinfo {year} {1976})}\BibitemShut {NoStop}%
\bibitem [{\citenamefont {Dean}\ \emph {et~al.}(2013)\citenamefont {Dean}, \citenamefont {Wang}, \citenamefont {Maher}, \citenamefont {Forsythe}, \citenamefont {Ghahari}, \citenamefont {Gao}, \citenamefont {Katoch}, \citenamefont {Ishigami}, \citenamefont {Moon}, \citenamefont {Koshino}, \citenamefont {Taniguchi}, \citenamefont {Watanabe}, \citenamefont {Shepard}, \citenamefont {Hone},\ and\ \citenamefont {Kim}}]{Dean2013Hofstadter}%
  \BibitemOpen
  \bibfield  {author} {\bibinfo {author} {\bibfnamefont {C.~R.}\ \bibnamefont {Dean}}, \bibinfo {author} {\bibfnamefont {L.}~\bibnamefont {Wang}}, \bibinfo {author} {\bibfnamefont {P.}~\bibnamefont {Maher}}, \bibinfo {author} {\bibfnamefont {C.}~\bibnamefont {Forsythe}}, \bibinfo {author} {\bibfnamefont {F.}~\bibnamefont {Ghahari}}, \bibinfo {author} {\bibfnamefont {Y.}~\bibnamefont {Gao}}, \bibinfo {author} {\bibfnamefont {J.}~\bibnamefont {Katoch}}, \bibinfo {author} {\bibfnamefont {M.}~\bibnamefont {Ishigami}}, \bibinfo {author} {\bibfnamefont {P.}~\bibnamefont {Moon}}, \bibinfo {author} {\bibfnamefont {M.}~\bibnamefont {Koshino}}, \bibinfo {author} {\bibfnamefont {T.}~\bibnamefont {Taniguchi}}, \bibinfo {author} {\bibfnamefont {K.}~\bibnamefont {Watanabe}}, \bibinfo {author} {\bibfnamefont {K.~L.}\ \bibnamefont {Shepard}}, \bibinfo {author} {\bibfnamefont {J.}~\bibnamefont {Hone}},\ and\ \bibinfo {author} {\bibfnamefont {P.}~\bibnamefont {Kim}},\ }\bibfield  {title} {\bibinfo {title} {Hofstadter's butterfly and the fractal quantum {Hall} effect in moir{\'e} superlattices},\ }\href {https://doi.org/10.1038/nature12186} {\bibfield  {journal} {\bibinfo  {journal} {Nature}\ }\textbf {\bibinfo {volume} {497}},\ \bibinfo {pages} {598} (\bibinfo {year} {2013})}\BibitemShut {NoStop}%
\bibitem [{\citenamefont {Chang}\ and\ \citenamefont {Scalettar}(2012)}]{ChangScalettar2012}%
  \BibitemOpen
  \bibfield  {author} {\bibinfo {author} {\bibfnamefont {C.-C.}\ \bibnamefont {Chang}}\ and\ \bibinfo {author} {\bibfnamefont {R.~T.}\ \bibnamefont {Scalettar}},\ }\bibfield  {title} {\bibinfo {title} {Quantum disordered phase near the {Mott} transition in the staggered-flux {Hubbard} model on a square lattice},\ }\href {https://doi.org/10.1103/PhysRevLett.109.026404} {\bibfield  {journal} {\bibinfo  {journal} {Physical Review Letters}\ }\textbf {\bibinfo {volume} {109}},\ \bibinfo {pages} {026404} (\bibinfo {year} {2012})}\BibitemShut {NoStop}%
\bibitem [{\citenamefont {Parisen~Toldin}\ \emph {et~al.}(2015)\citenamefont {Parisen~Toldin}, \citenamefont {Hohenadler}, \citenamefont {Assaad},\ and\ \citenamefont {Herbut}}]{ParisenToldin2015}%
  \BibitemOpen
  \bibfield  {author} {\bibinfo {author} {\bibfnamefont {F.}~\bibnamefont {Parisen~Toldin}}, \bibinfo {author} {\bibfnamefont {M.}~\bibnamefont {Hohenadler}}, \bibinfo {author} {\bibfnamefont {F.~F.}\ \bibnamefont {Assaad}},\ and\ \bibinfo {author} {\bibfnamefont {I.~F.}\ \bibnamefont {Herbut}},\ }\bibfield  {title} {\bibinfo {title} {Fermionic quantum criticality in honeycomb and $\pi$-flux {Hubbard} models: Finite-size scaling of renormalization-group-invariant observables from quantum {Monte} {Carlo}},\ }\href {https://doi.org/10.1103/PhysRevB.91.165108} {\bibfield  {journal} {\bibinfo  {journal} {Physical Review B}\ }\textbf {\bibinfo {volume} {91}},\ \bibinfo {pages} {165108} (\bibinfo {year} {2015})}\BibitemShut {NoStop}%
\bibitem [{\citenamefont {Otsuka}\ \emph {et~al.}(2016)\citenamefont {Otsuka}, \citenamefont {Yunoki},\ and\ \citenamefont {Sorella}}]{OtsukaYunokiSorella2016}%
  \BibitemOpen
  \bibfield  {author} {\bibinfo {author} {\bibfnamefont {Y.}~\bibnamefont {Otsuka}}, \bibinfo {author} {\bibfnamefont {S.}~\bibnamefont {Yunoki}},\ and\ \bibinfo {author} {\bibfnamefont {S.}~\bibnamefont {Sorella}},\ }\bibfield  {title} {\bibinfo {title} {Universal quantum criticality in the metal-insulator transition of two-dimensional interacting {Dirac} electrons},\ }\href {https://doi.org/10.1103/PhysRevX.6.011029} {\bibfield  {journal} {\bibinfo  {journal} {Physical Review X}\ }\textbf {\bibinfo {volume} {6}},\ \bibinfo {pages} {011029} (\bibinfo {year} {2016})}\BibitemShut {NoStop}%
\bibitem [{\citenamefont {Affleck}\ and\ \citenamefont {Marston}(1988)}]{AffleckMarston1988}%
  \BibitemOpen
  \bibfield  {author} {\bibinfo {author} {\bibfnamefont {I.}~\bibnamefont {Affleck}}\ and\ \bibinfo {author} {\bibfnamefont {J.~B.}\ \bibnamefont {Marston}},\ }\bibfield  {title} {\bibinfo {title} {Large-$n$ limit of the {Heisenberg-Hubbard} model: Implications for high-{$T_c$} superconductors},\ }\href {https://doi.org/10.1103/PhysRevB.37.3774} {\bibfield  {journal} {\bibinfo  {journal} {Physical Review B}\ }\textbf {\bibinfo {volume} {37}},\ \bibinfo {pages} {3774} (\bibinfo {year} {1988})}\BibitemShut {NoStop}%
\bibitem [{\citenamefont {White}(1992)}]{White1992DMRG}%
  \BibitemOpen
  \bibfield  {author} {\bibinfo {author} {\bibfnamefont {S.~R.}\ \bibnamefont {White}},\ }\bibfield  {title} {\bibinfo {title} {Density matrix formulation for quantum renormalization groups},\ }\href {https://doi.org/10.1103/PhysRevLett.69.2863} {\bibfield  {journal} {\bibinfo  {journal} {Physical Review Letters}\ }\textbf {\bibinfo {volume} {69}},\ \bibinfo {pages} {2863} (\bibinfo {year} {1992})}\BibitemShut {NoStop}%
\bibitem [{\citenamefont {McCulloch}(2007)}]{McCulloch2007NonabelianDMRG}%
  \BibitemOpen
  \bibfield  {author} {\bibinfo {author} {\bibfnamefont {I.~P.}\ \bibnamefont {McCulloch}},\ }\bibfield  {title} {\bibinfo {title} {From density-matrix renormalization group to matrix product states},\ }\href {https://doi.org/10.1088/1742-5468/2007/10/P10014} {\bibfield  {journal} {\bibinfo  {journal} {Journal of Statistical Mechanics: Theory and Experiment}\ }\textbf {\bibinfo {volume} {2007}},\ \bibinfo {pages} {P10014} (\bibinfo {year} {2007})}\BibitemShut {NoStop}%
\bibitem [{\citenamefont {Blankenbecler}\ \emph {et~al.}(1981)\citenamefont {Blankenbecler}, \citenamefont {Scalapino},\ and\ \citenamefont {Sugar}}]{Blankenbecler1981DQMC}%
  \BibitemOpen
  \bibfield  {author} {\bibinfo {author} {\bibfnamefont {R.}~\bibnamefont {Blankenbecler}}, \bibinfo {author} {\bibfnamefont {D.~J.}\ \bibnamefont {Scalapino}},\ and\ \bibinfo {author} {\bibfnamefont {R.~L.}\ \bibnamefont {Sugar}},\ }\bibfield  {title} {\bibinfo {title} {Monte {Carlo} calculations of coupled boson-fermion systems. {I}},\ }\href {https://doi.org/10.1103/PhysRevD.24.2278} {\bibfield  {journal} {\bibinfo  {journal} {Physical Review D}\ }\textbf {\bibinfo {volume} {24}},\ \bibinfo {pages} {2278} (\bibinfo {year} {1981})}\BibitemShut {NoStop}%
\bibitem [{\citenamefont {White}\ \emph {et~al.}(1989{\natexlab{a}})\citenamefont {White}, \citenamefont {Scalapino}, \citenamefont {Sugar}, \citenamefont {Loh}, \citenamefont {Gubernatis},\ and\ \citenamefont {Scalettar}}]{White1989DQMC}%
  \BibitemOpen
  \bibfield  {author} {\bibinfo {author} {\bibfnamefont {S.~R.}\ \bibnamefont {White}}, \bibinfo {author} {\bibfnamefont {D.~J.}\ \bibnamefont {Scalapino}}, \bibinfo {author} {\bibfnamefont {R.~L.}\ \bibnamefont {Sugar}}, \bibinfo {author} {\bibfnamefont {E.~Y.}\ \bibnamefont {Loh}}, \bibinfo {author} {\bibfnamefont {J.~E.}\ \bibnamefont {Gubernatis}},\ and\ \bibinfo {author} {\bibfnamefont {R.~T.}\ \bibnamefont {Scalettar}},\ }\bibfield  {title} {\bibinfo {title} {Numerical study of the two-dimensional {Hubbard} model},\ }\href {https://doi.org/10.1103/PhysRevB.40.506} {\bibfield  {journal} {\bibinfo  {journal} {Physical Review B}\ }\textbf {\bibinfo {volume} {40}},\ \bibinfo {pages} {506} (\bibinfo {year} {1989}{\natexlab{a}})}\BibitemShut {NoStop}%
\bibitem [{\citenamefont {Loh}\ \emph {et~al.}(1990)\citenamefont {Loh}, \citenamefont {Gubernatis}, \citenamefont {Scalettar}, \citenamefont {White}, \citenamefont {Scalapino},\ and\ \citenamefont {Sugar}}]{Loh1990Sign}%
  \BibitemOpen
  \bibfield  {author} {\bibinfo {author} {\bibfnamefont {E.~Y.}\ \bibnamefont {Loh}, \bibfnamefont {Jr.}}, \bibinfo {author} {\bibfnamefont {J.~E.}\ \bibnamefont {Gubernatis}}, \bibinfo {author} {\bibfnamefont {R.~T.}\ \bibnamefont {Scalettar}}, \bibinfo {author} {\bibfnamefont {S.~R.}\ \bibnamefont {White}}, \bibinfo {author} {\bibfnamefont {D.~J.}\ \bibnamefont {Scalapino}},\ and\ \bibinfo {author} {\bibfnamefont {R.~L.}\ \bibnamefont {Sugar}},\ }\bibfield  {title} {\bibinfo {title} {Sign problem in the numerical simulation of many-electron systems},\ }\href {https://doi.org/10.1103/PhysRevB.41.9301} {\bibfield  {journal} {\bibinfo  {journal} {Physical Review B}\ }\textbf {\bibinfo {volume} {41}},\ \bibinfo {pages} {9301} (\bibinfo {year} {1990})}\BibitemShut {NoStop}%
\bibitem [{\citenamefont {Laughlin}(1981)}]{Laughlin1981Pump}%
  \BibitemOpen
  \bibfield  {author} {\bibinfo {author} {\bibfnamefont {R.~B.}\ \bibnamefont {Laughlin}},\ }\bibfield  {title} {\bibinfo {title} {Quantized {Hall} conductivity in two dimensions},\ }\href {https://doi.org/10.1103/PhysRevB.23.5632} {\bibfield  {journal} {\bibinfo  {journal} {Physical Review B}\ }\textbf {\bibinfo {volume} {23}},\ \bibinfo {pages} {5632} (\bibinfo {year} {1981})}\BibitemShut {NoStop}%
\bibitem [{\citenamefont {Zaletel}\ \emph {et~al.}(2014)\citenamefont {Zaletel}, \citenamefont {Mong},\ and\ \citenamefont {Pollmann}}]{Zaletel2014FluxInsertion}%
  \BibitemOpen
  \bibfield  {author} {\bibinfo {author} {\bibfnamefont {M.~P.}\ \bibnamefont {Zaletel}}, \bibinfo {author} {\bibfnamefont {R.~S.~K.}\ \bibnamefont {Mong}},\ and\ \bibinfo {author} {\bibfnamefont {F.}~\bibnamefont {Pollmann}},\ }\bibfield  {title} {\bibinfo {title} {Flux insertion, entanglement, and quantized responses},\ }\href {https://doi.org/10.1088/1742-5468/2014/10/P10007} {\bibfield  {journal} {\bibinfo  {journal} {Journal of Statistical Mechanics: Theory and Experiment}\ }\textbf {\bibinfo {volume} {2014}},\ \bibinfo {pages} {P10007} (\bibinfo {year} {2014})}\BibitemShut {NoStop}%
\bibitem [{Sup()}]{SuppMaterial}%
  \BibitemOpen
  \href@noop {} {}\bibinfo {note} {See Supplemental Material for more technical details and additional supporting data.}\BibitemShut {Stop}%
\bibitem [{\citenamefont {Florens}\ and\ \citenamefont {Georges}(2004)}]{FlorensGeorges2004SlaveRotor}%
  \BibitemOpen
  \bibfield  {author} {\bibinfo {author} {\bibfnamefont {S.}~\bibnamefont {Florens}}\ and\ \bibinfo {author} {\bibfnamefont {A.}~\bibnamefont {Georges}},\ }\bibfield  {title} {\bibinfo {title} {Slave-rotor mean-field theories of strongly correlated systems and the {Mott} transition in finite dimensions},\ }\href {https://doi.org/10.1103/PhysRevB.70.035114} {\bibfield  {journal} {\bibinfo  {journal} {Physical Review B}\ }\textbf {\bibinfo {volume} {70}},\ \bibinfo {pages} {035114} (\bibinfo {year} {2004})}\BibitemShut {NoStop}%
\bibitem [{\citenamefont {Song}\ \emph {et~al.}(2023)\citenamefont {Song}, \citenamefont {Seifert}, \citenamefont {Luo},\ and\ \citenamefont {Balents}}]{Song2023SlaveRotorTMD}%
  \BibitemOpen
  \bibfield  {author} {\bibinfo {author} {\bibfnamefont {Z.}~\bibnamefont {Song}}, \bibinfo {author} {\bibfnamefont {U.~F.~P.}\ \bibnamefont {Seifert}}, \bibinfo {author} {\bibfnamefont {Z.-X.}\ \bibnamefont {Luo}},\ and\ \bibinfo {author} {\bibfnamefont {L.}~\bibnamefont {Balents}},\ }\bibfield  {title} {\bibinfo {title} {Mott insulators in moir{\'e} transition metal dichalcogenides at fractional fillings: Slave-rotor mean-field theory},\ }\href {https://doi.org/10.1103/PhysRevB.108.155109} {\bibfield  {journal} {\bibinfo  {journal} {Physical Review B}\ }\textbf {\bibinfo {volume} {108}},\ \bibinfo {pages} {155109} (\bibinfo {year} {2023})}\BibitemShut {NoStop}%
\bibitem [{\citenamefont {Hermele}\ \emph {et~al.}(2004)\citenamefont {Hermele}, \citenamefont {Senthil}, \citenamefont {Fisher}, \citenamefont {Lee}, \citenamefont {Nagaosa},\ and\ \citenamefont {Wen}}]{Hermele2004U1SpinLiquids}%
  \BibitemOpen
  \bibfield  {author} {\bibinfo {author} {\bibfnamefont {M.}~\bibnamefont {Hermele}}, \bibinfo {author} {\bibfnamefont {T.}~\bibnamefont {Senthil}}, \bibinfo {author} {\bibfnamefont {M.~P.~A.}\ \bibnamefont {Fisher}}, \bibinfo {author} {\bibfnamefont {P.~A.}\ \bibnamefont {Lee}}, \bibinfo {author} {\bibfnamefont {N.}~\bibnamefont {Nagaosa}},\ and\ \bibinfo {author} {\bibfnamefont {X.-G.}\ \bibnamefont {Wen}},\ }\bibfield  {title} {\bibinfo {title} {Stability of {$U(1)$} spin liquids in two dimensions},\ }\href {https://doi.org/10.1103/PhysRevB.70.214437} {\bibfield  {journal} {\bibinfo  {journal} {Physical Review B}\ }\textbf {\bibinfo {volume} {70}},\ \bibinfo {pages} {214437} (\bibinfo {year} {2004})}\BibitemShut {NoStop}%
\bibitem [{\citenamefont {Song}\ \emph {et~al.}(2019)\citenamefont {Song}, \citenamefont {Wang}, \citenamefont {Vishwanath},\ and\ \citenamefont {He}}]{Song2019CompetingOrders}%
  \BibitemOpen
  \bibfield  {author} {\bibinfo {author} {\bibfnamefont {X.-Y.}\ \bibnamefont {Song}}, \bibinfo {author} {\bibfnamefont {C.}~\bibnamefont {Wang}}, \bibinfo {author} {\bibfnamefont {A.}~\bibnamefont {Vishwanath}},\ and\ \bibinfo {author} {\bibfnamefont {Y.-C.}\ \bibnamefont {He}},\ }\bibfield  {title} {\bibinfo {title} {Unifying description of competing orders in two-dimensional quantum magnets},\ }\href {https://doi.org/10.1038/s41467-019-11727-3} {\bibfield  {journal} {\bibinfo  {journal} {Nature Communications}\ }\textbf {\bibinfo {volume} {10}},\ \bibinfo {pages} {4254} (\bibinfo {year} {2019})}\BibitemShut {NoStop}%
\bibitem [{\citenamefont {Song}\ \emph {et~al.}(2020)\citenamefont {Song}, \citenamefont {He}, \citenamefont {Vishwanath},\ and\ \citenamefont {Wang}}]{Song2020MonopolesDSL}%
  \BibitemOpen
  \bibfield  {author} {\bibinfo {author} {\bibfnamefont {X.-Y.}\ \bibnamefont {Song}}, \bibinfo {author} {\bibfnamefont {Y.-C.}\ \bibnamefont {He}}, \bibinfo {author} {\bibfnamefont {A.}~\bibnamefont {Vishwanath}},\ and\ \bibinfo {author} {\bibfnamefont {C.}~\bibnamefont {Wang}},\ }\bibfield  {title} {\bibinfo {title} {From spinon band topology to the symmetry quantum numbers of monopoles in {Dirac} spin liquids},\ }\href {https://doi.org/10.1103/PhysRevX.10.011033} {\bibfield  {journal} {\bibinfo  {journal} {Physical Review X}\ }\textbf {\bibinfo {volume} {10}},\ \bibinfo {pages} {011033} (\bibinfo {year} {2020})}\BibitemShut {NoStop}%
\bibitem [{\citenamefont {Giamarchi}(2004)}]{Giamarchi2004}%
  \BibitemOpen
  \bibfield  {author} {\bibinfo {author} {\bibfnamefont {T.}~\bibnamefont {Giamarchi}},\ }\href@noop {} {\emph {\bibinfo {title} {Quantum Physics in One Dimension}}}\ (\bibinfo  {publisher} {Oxford University Press},\ \bibinfo {address} {Oxford},\ \bibinfo {year} {2004})\BibitemShut {NoStop}%
\bibitem [{\citenamefont {Wang}\ and\ \citenamefont {Devereaux}(2025)}]{wang2025finite}%
  \BibitemOpen
  \bibfield  {author} {\bibinfo {author} {\bibfnamefont {W.~O.}\ \bibnamefont {Wang}}\ and\ \bibinfo {author} {\bibfnamefont {T.~P.}\ \bibnamefont {Devereaux}},\ }\href {https://arxiv.org/abs/2510.16616} {\bibinfo {title} {Finite-temperature signatures of underlying superconductivity in the electron-doped hubbard model}} (\bibinfo {year} {2025}),\ \Eprint {https://arxiv.org/abs/2510.16616} {arXiv:2510.16616 [cond-mat.supr-con]} \BibitemShut {NoStop}%
\bibitem [{\citenamefont {Li}\ \emph {et~al.}(2026)\citenamefont {Li}, \citenamefont {Qi},\ and\ \citenamefont {Li}}]{Li2026FluctuatingPDW}%
  \BibitemOpen
  \bibfield  {author} {\bibinfo {author} {\bibfnamefont {Q.}~\bibnamefont {Li}}, \bibinfo {author} {\bibfnamefont {Y.}~\bibnamefont {Qi}},\ and\ \bibinfo {author} {\bibfnamefont {W.}~\bibnamefont {Li}},\ }\href {https://arxiv.org/abs/2604.16293} {\bibinfo {title} {Fluctuating pair density wave in finite-temperature phase diagram of the $t$-$t'$ {H}ubbard model}} (\bibinfo {year} {2026}),\ \Eprint {https://arxiv.org/abs/2604.16293} {arXiv:2604.16293 [cond-mat.str-el]} \BibitemShut {NoStop}%
\bibitem [{\citenamefont {White}\ \emph {et~al.}(1989{\natexlab{b}})\citenamefont {White}, \citenamefont {Scalapino}, \citenamefont {Sugar}, \citenamefont {Bickers},\ and\ \citenamefont {Scalettar}}]{White1989PairingVertices}%
  \BibitemOpen
  \bibfield  {author} {\bibinfo {author} {\bibfnamefont {S.~R.}\ \bibnamefont {White}}, \bibinfo {author} {\bibfnamefont {D.~J.}\ \bibnamefont {Scalapino}}, \bibinfo {author} {\bibfnamefont {R.~L.}\ \bibnamefont {Sugar}}, \bibinfo {author} {\bibfnamefont {N.~E.}\ \bibnamefont {Bickers}},\ and\ \bibinfo {author} {\bibfnamefont {R.~T.}\ \bibnamefont {Scalettar}},\ }\bibfield  {title} {\bibinfo {title} {Attractive and repulsive pairing interaction vertices for the two-dimensional hubbard model},\ }\href {https://doi.org/10.1103/PhysRevB.39.839} {\bibfield  {journal} {\bibinfo  {journal} {Physical Review B}\ }\textbf {\bibinfo {volume} {39}},\ \bibinfo {pages} {839} (\bibinfo {year} {1989}{\natexlab{b}})}\BibitemShut {NoStop}%
\bibitem [{\citenamefont {Cr{\'e}pel}\ and\ \citenamefont {Fu}(2021)}]{CrepelFu2021}%
  \BibitemOpen
  \bibfield  {author} {\bibinfo {author} {\bibfnamefont {V.}~\bibnamefont {Cr{\'e}pel}}\ and\ \bibinfo {author} {\bibfnamefont {L.}~\bibnamefont {Fu}},\ }\bibfield  {title} {\bibinfo {title} {New mechanism and exact theory of superconductivity from strong repulsive interaction},\ }\href {https://doi.org/10.1126/sciadv.abh2233} {\bibfield  {journal} {\bibinfo  {journal} {Science Advances}\ }\textbf {\bibinfo {volume} {7}},\ \bibinfo {pages} {eabh2233} (\bibinfo {year} {2021})}\BibitemShut {NoStop}%
\bibitem [{\citenamefont {Cr{\'e}pel}\ \emph {et~al.}(2022)\citenamefont {Cr{\'e}pel}, \citenamefont {Cea}, \citenamefont {Fu},\ and\ \citenamefont {Guinea}}]{Crepel2022InterbandSC}%
  \BibitemOpen
  \bibfield  {author} {\bibinfo {author} {\bibfnamefont {V.}~\bibnamefont {Cr{\'e}pel}}, \bibinfo {author} {\bibfnamefont {T.}~\bibnamefont {Cea}}, \bibinfo {author} {\bibfnamefont {L.}~\bibnamefont {Fu}},\ and\ \bibinfo {author} {\bibfnamefont {F.}~\bibnamefont {Guinea}},\ }\bibfield  {title} {\bibinfo {title} {Unconventional superconductivity due to interband polarization},\ }\href {https://doi.org/10.1103/PhysRevB.105.094506} {\bibfield  {journal} {\bibinfo  {journal} {Physical Review B}\ }\textbf {\bibinfo {volume} {105}},\ \bibinfo {pages} {094506} (\bibinfo {year} {2022})}\BibitemShut {NoStop}%
\bibitem [{\citenamefont {Kohn}\ and\ \citenamefont {Luttinger}(1965)}]{KohnLuttinger1965}%
  \BibitemOpen
  \bibfield  {author} {\bibinfo {author} {\bibfnamefont {W.}~\bibnamefont {Kohn}}\ and\ \bibinfo {author} {\bibfnamefont {J.~M.}\ \bibnamefont {Luttinger}},\ }\bibfield  {title} {\bibinfo {title} {New mechanism for superconductivity},\ }\href {https://doi.org/10.1103/PhysRevLett.15.524} {\bibfield  {journal} {\bibinfo  {journal} {Physical Review Letters}\ }\textbf {\bibinfo {volume} {15}},\ \bibinfo {pages} {524} (\bibinfo {year} {1965})}\BibitemShut {NoStop}%
\bibitem [{\citenamefont {Fukui}\ \emph {et~al.}(2005)\citenamefont {Fukui}, \citenamefont {Hatsugai},\ and\ \citenamefont {Suzuki}}]{FukuiHatsugaiSuzuki2005}%
  \BibitemOpen
  \bibfield  {author} {\bibinfo {author} {\bibfnamefont {T.}~\bibnamefont {Fukui}}, \bibinfo {author} {\bibfnamefont {Y.}~\bibnamefont {Hatsugai}},\ and\ \bibinfo {author} {\bibfnamefont {H.}~\bibnamefont {Suzuki}},\ }\bibfield  {title} {\bibinfo {title} {Chern numbers in discretized {Brillouin} zone: Efficient method of computing (spin) {Hall} conductances},\ }\href {https://doi.org/10.1143/JPSJ.74.1674} {\bibfield  {journal} {\bibinfo  {journal} {Journal of the Physical Society of Japan}\ }\textbf {\bibinfo {volume} {74}},\ \bibinfo {pages} {1674} (\bibinfo {year} {2005})}\BibitemShut {NoStop}%
\bibitem [{\citenamefont {Balents}\ and\ \citenamefont {Fisher}(1996)}]{BalentsFisher1996}%
  \BibitemOpen
  \bibfield  {author} {\bibinfo {author} {\bibfnamefont {L.}~\bibnamefont {Balents}}\ and\ \bibinfo {author} {\bibfnamefont {M.~P.~A.}\ \bibnamefont {Fisher}},\ }\bibfield  {title} {\bibinfo {title} {Weak-coupling phase diagram of the two-chain {H}ubbard model},\ }\href {https://doi.org/10.1103/PhysRevB.53.12133} {\bibfield  {journal} {\bibinfo  {journal} {Phys. Rev. B}\ }\textbf {\bibinfo {volume} {53}},\ \bibinfo {pages} {12133} (\bibinfo {year} {1996})}\BibitemShut {NoStop}%
\bibitem [{\citenamefont {Lin}\ \emph {et~al.}(1997)\citenamefont {Lin}, \citenamefont {Balents},\ and\ \citenamefont {Fisher}}]{LinBalentsFisher1997}%
  \BibitemOpen
  \bibfield  {author} {\bibinfo {author} {\bibfnamefont {H.-H.}\ \bibnamefont {Lin}}, \bibinfo {author} {\bibfnamefont {L.}~\bibnamefont {Balents}},\ and\ \bibinfo {author} {\bibfnamefont {M.~P.~A.}\ \bibnamefont {Fisher}},\ }\bibfield  {title} {\bibinfo {title} {$n$-chain {H}ubbard model in weak coupling},\ }\href {https://doi.org/10.1103/PhysRevB.56.6569} {\bibfield  {journal} {\bibinfo  {journal} {Phys. Rev. B}\ }\textbf {\bibinfo {volume} {56}},\ \bibinfo {pages} {6569} (\bibinfo {year} {1997})}\BibitemShut {NoStop}%
\bibitem [{\citenamefont {Song}\ \emph {et~al.}(2025)\citenamefont {Song}, \citenamefont {Seifert}, \citenamefont {Balents},\ and\ \citenamefont {Jiang}}]{Song2025moire}%
  \BibitemOpen
  \bibfield  {author} {\bibinfo {author} {\bibfnamefont {Z.}~\bibnamefont {Song}}, \bibinfo {author} {\bibfnamefont {U.~F.~P.}\ \bibnamefont {Seifert}}, \bibinfo {author} {\bibfnamefont {L.}~\bibnamefont {Balents}},\ and\ \bibinfo {author} {\bibfnamefont {H.-C.}\ \bibnamefont {Jiang}},\ }\href@noop {} {\bibinfo {title} {Emergent magnetism and spin liquids in an extended {H}ubbard description of moir\'e bilayers}} (\bibinfo {year} {2025}),\ \Eprint {https://arxiv.org/abs/2505.06339} {arXiv:2505.06339 [cond-mat.str-el]} \BibitemShut {NoStop}%
\end{thebibliography}%

\clearpage
\widetext

\begin{center}
\textbf{\large Supplemental Material for ``Chiral Superconductivity Indifferent to Parent Topology in the Square-Lattice Hofstadter-Hubbard Model''}
\end{center}
\vspace{1mm}
\setcounter{secnumdepth}{2}
\setcounter{section}{0}
\renewcommand{\thesection}{\Alph{section}}
\renewcommand{\thesubsection}{\arabic{subsection}}
\renewcommand{\thefigure}{S\arabic{figure}}
\renewcommand{\theequation}{S\arabic{equation}}
\renewcommand{\thetable}{S\arabic{table}}

\renewcommand{\theHfigure}{S\arabic{figure}}
\renewcommand{\theHequation}{S\arabic{equation}}
\renewcommand{\theHtable}{S\arabic{table}}
\setcounter{figure}{0}
\setcounter{equation}{0}
\setcounter{table}{0}
\addtocontents{toc}{\string\tocdepth@restore}

In this Supplemental Material we collect the model conventions, the analytic derivations, and additional DQMC and DMRG results supporting the main text. All analytic calculations are
performed directly in the Landau gauge used by the DMRG and DQMC simulations,
so that no gauge conversion intervenes anywhere between the analytic and the numerical results. Sec.~\ref{app:model} defines the Landau-gauge conventions and the band structure; Sec.~\ref{app:rotor} presents the slave-rotor self-consistent calculation;
Sec.~\ref{app:su2k} derives the $SU(2)_k$ Higgs-Chern-Simons critical theory of
the IQH-CSL transition; Sec.~\ref{app:KL} contains the Kohn-Luttinger analysis
and the Bogoliubov-de Gennes topology of the weak-coupling superconductor;
Sec.~\ref{app:sixleg} uses bosonization to analyze the weak-coupling behavior of the six-leg cylinder at large $t'$;
Sec.~\ref{app:parton} presents a strong-coupling analysis of the doped chiral spin liquid; Sec.~\ref{app:dqmc_supp} presents additional DQMC results and the pairing formalism used in the main text; and Sec.~\ref{app:dmrg_supp} collects additional DMRG results at half filling and at $\delta=1/12$, including the spin structure factors, the spin gap at half filling, the density profiles and scalar spin chirality of the doped phases, the electron and spin correlation lengths across the doped phase diagram, the pair binding on three- and four-leg cylinders, and the correlation functions at representative points of the chiral metal and of the time-reversal-symmetric line $t'=0$.

\setcounter{tocdepth}{2}
\tableofcontents

\section{Model and gauge conventions}
\label{app:model}

All analytic calculations in this Supplemental Material are performed
directly in the Landau gauge of \figref{fig:model}(a), which is also the
gauge used by the DMRG and DQMC simulations. The hopping amplitudes are
\begin{equation}
\begin{aligned}
t_{i,i+\hat x}&=t, &
t_{i,i+\hat y}&=(-1)^{i_x}\,t,\\
t_{i,i-\hat x+\hat y}&=i(-1)^{i_x}\,t', &
t_{i,i+\hat x+\hat y}&=-i(-1)^{i_x}\,t',
\end{aligned}
\label{eq:landau}
\end{equation}
with $i=(i_x,i_y)$ the site coordinates. The real NN amplitudes thread
flux $\pi$ through every square plaquette, and the purely imaginary NNN
amplitudes $\pm it'$ add flux $\pm\pi/2$ to the elementary triangles.

Bond pairing amplitudes are not gauge invariant (under a local phase
redefinition $c_{i\sigma}\to c_{i\sigma}e^{i\vartheta_i}$ the singlet bond
amplitude $\langle c_{i\up}c_{j\dn}-c_{i\dn}c_{j\up}\rangle$ acquires
the phase $e^{i(\vartheta_i+\vartheta_j)}$), so all bond-phase patterns quoted below,
in particular the pairing pattern compared with \figref{fig:doped}(a) in
\secref{app:KL}, are tied to the gauge choice of \eqnref{eq:landau}.

\begin{figure}[t]
\centering
\includegraphics[width=0.75\textwidth]{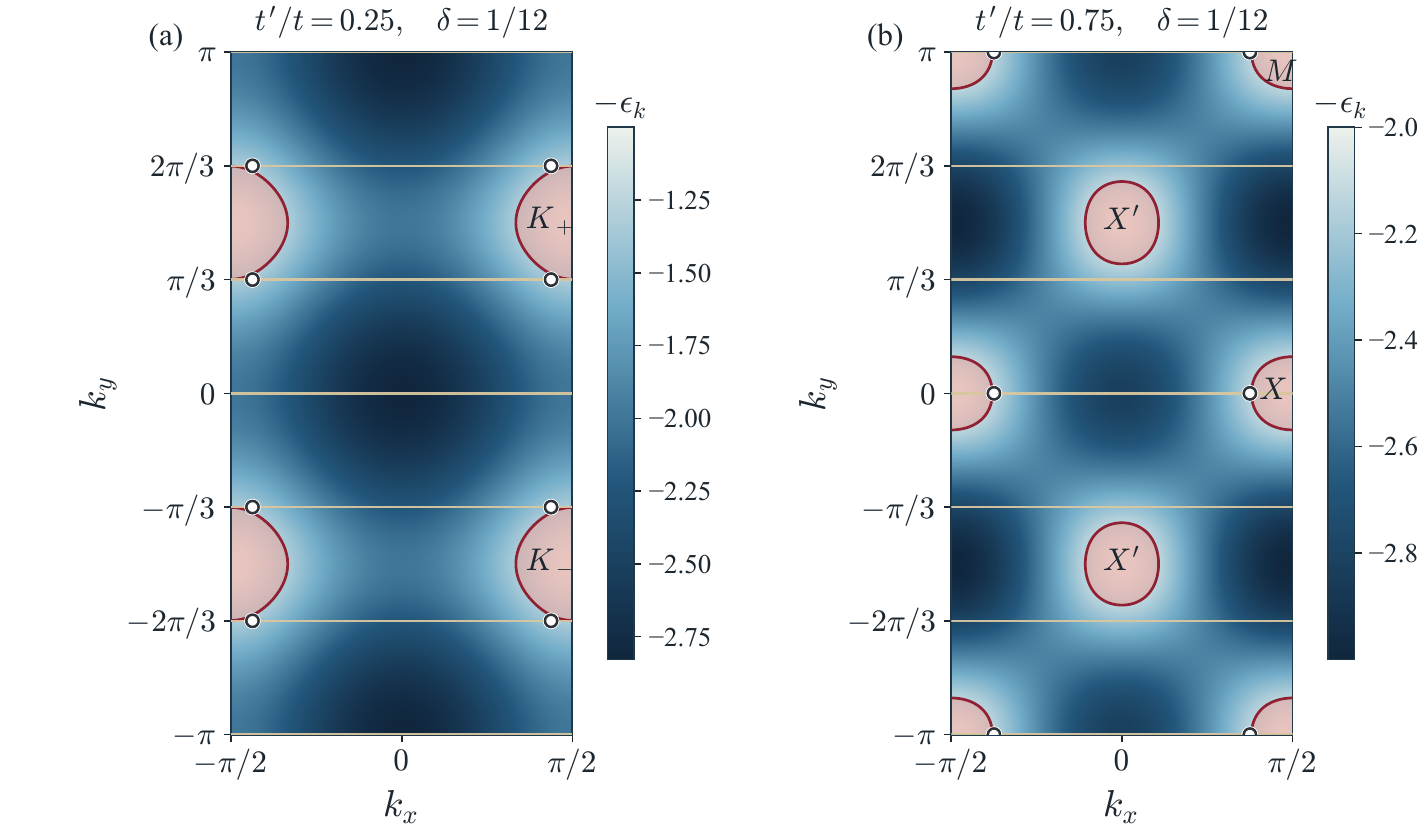}
\caption{Fermi surfaces at hole doping $\delta=1/12$ in the Landau gauge,
overlaid on the lower band $-\epsilon_{\bm k}$. (a)
$t'=0.25$: two near-circular hole pockets per spin (dark red
contours, shaded interior), centered on the massive Dirac points
$K_\pm=(\pi/2,\pm\pi/2)$. (b) $t'=0.75$: four elliptical
pockets per spin, centered at $X=(\pi/2,0)$, $M=(\pi/2,\pi)$, and
$X'_\pm=(0,\pm\pi/2)$. Yellow lines mark the six momentum cuts
$k_y=0,\pm\pi/3,\pm2\pi/3,\pi$ allowed on the $L_y=6$ cylinder
[\eqnref{eq:kycuts}], and open circles mark the quasi-one-dimensional
Fermi points where the cuts cross the pockets. In (b) the cuts $k_y=0,\pi$ pass exactly through the
$X$ and $M$ pockets, while the two $X'$ pockets are absent on the cylinder
[\secref{app:sixleg}].
}
\label{fig:landaufs}
\end{figure}

Under this gauge, the natural magnetic unit cell is doubled along $\hat x$. Sublattice
$A$ occupies the even columns and sublattice $B$ the odd columns, with
offsets $\bm d_A=(0,0)$ and $\bm d_B=(1,0)$ and primitive vectors $\bm a_1=(2,0)$ and
$\bm a_2=(0,1)$. The reciprocal vectors are $\bm G_1=(\pi,0)$ and $\bm G_2=(0,2\pi)$, and
the magnetic Brillouin zone (MBZ) is the rectangle $k_x\in[-\pi/2,\pi/2]$,
$k_y\in[-\pi,\pi]$ of area $2\pi^2$. For momentum-space calculations we use
the Fourier convention
\begin{equation}
c_{\bm R,\gamma,\sigma}
=
\frac{1}{\sqrt{N_c}}\sum_{\bm k}c_{\bm k,\gamma,\sigma}\,e^{i\bm k\cdot(\bm R+\bm d_\gamma)},
\qquad \gamma=A,B,
\label{eq:fourier}
\end{equation}
with $N_c$ the number of magnetic unit cells, $\bm R$ the unit-cell position, and $\bm k$ in the magnetic Brillouin zone. In
the basis $\psi_{\bm k\sigma}=(c_{\bm k,A,\sigma},c_{\bm k,B,\sigma})^T$ the kinetic term
reads $H_t=\sum_{\bm k\sigma}\psi^\dagger_{\bm k\sigma}h_{\bm k}\psi_{\bm k\sigma}$ with
\begin{equation}
h_{\bm k}=-2t\cos k_x\,\tau_x+4t'\sin k_x\sin k_y\,\tau_y-2t\cos k_y\,\tau_z,
\label{eq:bloch}
\end{equation}
where $\tau_{x,y,z}$ act in sublattice space and we set the chemical potential
to zero at half filling. Because the offsets
$\bm d_\gamma$ enter the Fourier phases, the embedded Bloch basis is sewn across
the MBZ boundary by $c_{\bm k+\bm G,\gamma}=D_\gamma(\bm G)\,c_{\bm k,\gamma}$ with
$D_\gamma(\bm G)=e^{-i\bm G\cdot\bm d_\gamma}$; in particular
$D(\bm G_1)=\mathrm{diag}(1,-1)=\tau_z$ and $h_{\bm k+\bm G_1}=\tau_z h_{\bm k}\tau_z$. The two bands
$\pm\epsilon_{\bm k}$ follow from
\begin{equation}
\epsilon_{\bm k}^2
=4t^2\big(\cos^2k_x+\cos^2k_y\big)+16t'^2\sin^2k_x\,\sin^2k_y.
\label{eq:dispersion}
\end{equation}
Diagonalization
defines the band basis $c_{\bm k,\gamma,\sigma}=\sum_{m=\pm}w_{\gamma
m}(\bm k)\,c_{\bm k,m,\sigma}$ used in \secref{app:KL}, and we use the
high-symmetry points $\Gamma=(0,0)$, $X=(\pi/2,0)$, $M=(\pi/2,\pi)$,
$Y=(0,\pi)$, $X'_\pm=(0,\pm\pi/2)$, and $K_{\pm}=(\pi/2,\pm\pi/2)$
[\figref{fig:landaufs}].
At $t'=0$ the spectrum has two inequivalent
Dirac nodes at $K_\pm$; a finite $t'$ gaps them into massive Dirac cones. For $|t'|<t/2$,
\begin{equation}
\epsilon(K_\pm+\bm q)\simeq\sqrt{m_D^2+v_F^2\,\bm q^2},
\qquad
m_D=4|t'|,
\quad
v_F=2\sqrt{t^2-4t'^2},
\label{eq:diracmass}
\end{equation}
and the occupied band acquires Chern number $-\sgn(t')$ per spin, i.e.,
$-2\sgn(t')$ including spin degeneracy.

Upon hole doping the carriers enter pockets at the top of the lower
band, and the position of that band top changes across $|t'|=t/2$. For
$|t'|<t/2$ it sits at the two massive Dirac points $K_\pm$, while for
$|t'|>t/2$ it jumps to the four inequivalent points $X$, $M$, and
$X'_\pm=(0,\pm\pi/2)$, so the number and location of the low-energy pockets
differ qualitatively between the two regimes. The resulting Fermi surfaces at
$\delta=1/12$ are shown in \figref{fig:landaufs}, with two pockets per spin
centered at $K_\pm$ for $t'=0.25$, and four pockets per spin centered at $X$, $M$, and $X'_\pm$ for $t'=0.75$. This
change of Fermi-surface topology controls the two pairing regimes of
\secref{app:KL}.

On the $L_y=6$ cylinders used in DMRG the transverse momentum is
quantized,
\begin{equation}
k_y=\frac{2\pi n}{L_y}\in\Big\{0,\ \pm\frac{\pi}{3},\ \pm\frac{2\pi}{3},\
\pi\Big\},
\label{eq:kycuts}
\end{equation}
and the MBZ collapses onto the six cuts overlaid in
\figref{fig:landaufs}. The pocket counting of \figref{fig:landaufs} survives only partially on the cylinder.
For $t'=0.25$ the two $K_\pm$ pockets are sampled by the adjacent cuts
$k_y=\pm\pi/3,\pm2\pi/3$, while for $t'=0.75$ the cuts hit the $X$ and $M$
pockets exactly; the two $X'$ pockets are missed altogether, a six-leg
size effect whose consequences for the weak-coupling physics at $t'=0.75$
are analyzed in \secref{app:sixleg}.

\section{Slave-rotor theory of the half-filled phase diagram}
\label{app:halffilling}

\subsection{$U(1)$ slave-rotor mean-field theory}
\label{app:rotor}

This subsection supplies the construction and the self-consistency details of
the slave-rotor mean field behind \figref{fig:undoped}(b). It is
convenient to absorb a chemical-potential
shift and work with the particle-hole symmetric form of the interaction,
\begin{equation}
\frac U2\sum_i(n_i-1)^2
=U\sum_i n_{i\up}n_{i\dn}+\frac U2\sum_i(1-n_i),
\label{eq:HV}
\end{equation}
which differs from the Hubbard term of \eqnref{eq:model} only by
a constant and a shift of the chemical potential. In the rotor representation
$c^{\dagger}_{i\sigma}=f^{\dagger}_{i\sigma}e^{i\theta_i}$ the three local
charge states map onto rotor angular momenta as
$\ket{n_i=2}\leftrightarrow L_i=+1$, $\ket{n_i=1}\leftrightarrow L_i=0$, and
$\ket{n_i=0}\leftrightarrow L_i=-1$, so that $n_i=L_i+1$ and
\eqnref{eq:HV} becomes $\frac U2\sum_iL_i^2$. Equivalently, the rotor angular
momentum $L_i$, conjugate to the rotor phase $\theta_i$, measures the charge
relative to half filling,
\begin{equation}
L_i=\sum_{\sigma}f^{\dagger}_{i\sigma}f_{i\sigma}-1,
\qquad
\big[e^{i\theta_i},L_j\big]=-\delta_{ij}\,e^{i\theta_i}.
\label{eq:constraint}
\end{equation}%
Attaching the rotor phases to
the Landau-gauge amplitudes of \eqnref{eq:landau}, the kinetic term reads
\begin{equation}
H_t=-\sum_{i,j,\sigma}\Big[t_{ij}\,e^{i(\theta_i-\theta_j)}
f^{\dagger}_{i\sigma}f_{j\sigma}+\mathrm{H.c.}\Big],
\label{eq:rotorkinetic}
\end{equation}
and the mean-field decoupling replaces the two factors on every bond by
their averages,
\begin{equation}
\begin{aligned}
K_{ij}&=\sum_\sigma t_{ij}\big\langle f^{\dagger}_{i\sigma}f_{j\sigma}
\big\rangle,\\
t^{\rm eff}_{ij}&=t_{ij}\big\langle e^{i(\theta_i-\theta_j)}\big\rangle,
\end{aligned}
\label{eq:decouple}
\end{equation}
which produces the two mean-field Hamiltonians
\begin{equation}
\begin{aligned}
H_f&=-\sum_{i,j}\Big[\sum_{\sigma}t^{\rm eff}_{ij}f^{\dagger}_{i\sigma}f_{j\sigma}+\mathrm{H.c.}\Big],\\
H_\theta&=-\sum_{i,j}\Big[K_{ij}\,e^{i(\theta_i-\theta_j)}+\mathrm{H.c.}\Big]+\frac U2\sum_iL_i^2,
\end{aligned}
\label{eq:mf}
\end{equation}%
together with the constant
$\sum_{ij}K_{ij}\,t^{\rm eff}_{ij}/t_{ij}+\mathrm{H.c.}$ that corrects the
double counting of the bond energy below. At generic filling the constraint
of \eqnref{eq:constraint} is enforced on average by a Lagrange multiplier
$\lambda$ coupled to $L_i$, i.e., by adding $\lambda\sum_iL_i$ to the rotor sector
and $-\lambda\sum_{i\sigma}f^{\dagger}_{i\sigma}f_{i\sigma}$ to the spinon sector,
with $\lambda$ fixed by the number equation
\begin{equation}
\langle L_i\rangle+1
=\sum_\sigma\big\langle f^{\dagger}_{i\sigma}f_{i\sigma}\big\rangle
=\frac{2}{N}\sum_{\bm k,m}^{\rm occ}1;
\label{eq:constraintB}
\end{equation}
throughout this subsection $N$ denotes the number of sites, as in the main text, and all
energies are per site.
At half filling particle-hole symmetry fixes $\lambda=0$, which we use throughout. On the uniform saddle the bond fields take one real value per
bond type, $K^{(1)}$ (NN) and $K^{(2)}$ (NNN), with coordination factors
$z_1=z_2=2$ counting the bonds of each type per site.

On the weak-coupling side we write $\langle
e^{i\theta_i}\rangle=\sqrt{Z_i}\,e^{i\varphi_i}$, so that
$t^{\rm eff}_{ij}=t_{ij}\sqrt{Z_iZ_j}\,e^{i(\varphi_i-\varphi_j)}$, and
specialize to the uniform saddle $Z_i=Z$, $\varphi_i=0$, where $Z$ is
the rotor condensate fraction, equal to the electron quasiparticle residue.
The rotor sector of \eqnref{eq:mf} then reduces to independent
single-site problems,
\begin{equation}
\begin{aligned}
H^{(i)}_\theta&=-2\mathcal K\sqrt Z\big(e^{i\theta_i}+e^{-i\theta_i}\big)+\frac U2 L_i^2+2\mathcal KZ,\\
\mathcal K&\equiv\sum_a z_aK^{(a)}.
\end{aligned}
\label{eq:rotorsite}
\end{equation}
Truncating the local charge basis to $L_i\in\{0,\pm1\}$, which is
quantitatively accurate at the couplings of interest, \eqnref{eq:rotorsite}
becomes the $3\times3$ matrix in the basis $(L_i=-1,0,+1)$
\begin{equation}
H^{(i)}_\theta=
\begin{pmatrix}
\frac U2+2\mathcal KZ & -2\mathcal K\sqrt Z & 0\\
-2\mathcal K\sqrt Z & 2\mathcal KZ & -2\mathcal K\sqrt Z\\
0 & -2\mathcal K\sqrt Z & \frac U2+2\mathcal KZ
\end{pmatrix},
\label{eq:rotormatrix}
\end{equation}
whose lowest eigenvalue defines the rotor energy $E_\theta(\mathcal K,Z)$ and
whose ground state yields $\langle L_i\rangle$. At half filling the
antisymmetric combination $(\ket{-1}-\ket{+1})/\sqrt2$ decouples, and
diagonalizing the remaining $2\times2$ block in the basis
$\{(\ket{-1}+\ket{+1})/\sqrt2,\ket{0}\}$ gives the closed form
\begin{equation}
E_\theta(\mathcal K,Z)
=2\mathcal KZ+\frac U4
-\sqrt{\frac{U^2}{16}+8\mathcal K^2Z}\,,
\label{eq:rotorenergy}
\end{equation}
whose ground state has $\langle L_i\rangle=0$, consistent with $\lambda=0$ at
half filling. The spinon sector is the
Landau-gauge band problem of \eqnref{eq:bloch} with $t\to tZ$ and
$t'\to t'Z$, giving bands $\epsilon_{\bm k,m}=mZ\epsilon_{\bm k}$ ($m=\pm$) and the
energy density $E_f=\frac{2}{N}\sum_{\bm k,m}^{\rm occ}\epsilon_{\bm k,m}
+2z_1K^{(1)}Z+2z_2K^{(2)}Z$, where the prefactor $2$ is the spin
degeneracy and the last two terms compensate the double counting introduced by
the decoupling. The bond fields follow from the Hellmann-Feynman theorem,
$\partial E_f/\partial t^{\mathrm{eff}}_a=-(2z_a/t_a)K^{(a)}$, with $t^{\mathrm{eff}}_1=t^{\mathrm{eff}}_{\mathrm{NN}}=tZ$ and $t^{\mathrm{eff}}_2=t^{\mathrm{eff}}_{\mathrm{NNN}}=t'Z$. In the Landau gauge the
dispersion \eqnref{eq:dispersion} separates as
\begin{equation}
\begin{aligned}
\epsilon_{\bm k}^2&=2(t^{\mathrm{eff}}_{\mathrm{NN}})^{2}A_{\bm k}+4(t^{\mathrm{eff}}_{\mathrm{NNN}})^{2}B_{\bm k},\\
A_{\bm k}&\equiv2+\cos2k_x+\cos2k_y,\\
B_{\bm k}&\equiv(1-\cos2k_x)(1-\cos2k_y),
\end{aligned}
\label{eq:ABkernels}
\end{equation}
so that
$\partial\epsilon_{\bm k}/\partial t^{\mathrm{eff}}_{\mathrm{NN}}=2t^{\mathrm{eff}}_{\mathrm{NN}} A_{\bm k}/\epsilon_{\bm k}$ and
$\partial\epsilon_{\bm k}/\partial t^{\mathrm{eff}}_{\mathrm{NNN}}=4t^{\mathrm{eff}}_{\mathrm{NNN}}B_{\bm k}/\epsilon_{\bm k}$, and the
Hellmann-Feynman relation evaluates to
\begin{equation}
K^{(1)}=-\frac{t^{2}Z}{z_1}\,\frac{2}{N}\sum_{\bm k,m}^{\rm occ}
\frac{A_{\bm k}}{\epsilon_{\bm k,m}},
\qquad
K^{(2)}=-\frac{t'^{2}Z}{z_2}\,\frac{4}{N}\sum_{\bm k,m}^{\rm occ}
\frac{B_{\bm k}}{\epsilon_{\bm k,m}},
\label{eq:Kfields}
\end{equation}
where the sums run over occupied states, $\epsilon_{\bm k,m}<0$, with $\bm k$
in the MBZ of \secref{app:model}. For given $Z$ these
equations close, and $Z$ is then fixed by minimizing $E_\theta+E_f$;
since $E_f$ is linear in $Z$, the minimization balances the linear band-energy
gain against the convex rotor cost \eqnref{eq:rotorenergy}. The
condensate obtained in this way is the purple scale of \figref{fig:undoped}(b), and the Mott boundary is the locus $Z\to0$.

The $Z=0$ saddle is not obtained by naively setting
$Z=0$ above. While the condensate vanishes, the spinon hopping is controlled
by the short-range coherence of the rotor field on a bond, not by rotor
off-diagonal long-range order. The rotor Hamiltonian therefore keeps the bond
fields explicit,
\begin{equation}
H_\theta=-\sum_{\langle ij\rangle}\Big[K^{(1)}e^{i(\theta_i-\theta_j)}+\mathrm{H.c.}\Big]
-\sum_{\langle\!\langle ij\rangle\!\rangle}\Big[K^{(2)}e^{i(\theta_i-\theta_j)}+\mathrm{H.c.}\Big]
+\frac U2\sum_iL_i^2,
\label{eq:Hmott}
\end{equation}
and we treat the bond terms perturbatively in $K^{(a)}/U$ about the product
ground state $\prod_i\ket{L_i=0}$. Each bond operator
$e^{\pm i(\theta_i-\theta_j)}$ creates a virtual pair of rotor charges
$L_i=\pm1$, $L_j=\mp1$ with excitation energy $\frac U2+\frac U2=U$, and the
two orientations of every bond contribute independent intermediate states, so
with $z_a$ bonds of type $a$ per site the energy per site is
\begin{equation}
E_\theta
=-\sum_{a}\frac{2z_a\big(K^{(a)}\big)^{2}}{U}
+O\big(K^3/U^2\big).
\label{eq:mottenergy}
\end{equation}
The
bond coherences then follow from the rotor-sector Hellmann-Feynman relation,
\begin{equation}
\frac{\partial E_\theta}{\partial K^{(a)}}
=\frac{1}{N}\Big\langle\frac{\partial H_\theta}{\partial K^{(a)}}\Big\rangle
=-2z_a\Theta^{(a)}
\;\;\Longrightarrow\;\;
\Theta^{(a)}=\frac{2K^{(a)}}{U},
\label{eq:HFrotor}
\end{equation}
and set the spinon hoppings $t^{\mathrm{eff}}_{\mathrm{NN}}=t\,\Theta^{(1)}$ and
$t^{\mathrm{eff}}_{\mathrm{NNN}}=t'\,\Theta^{(2)}$. Applying the spinon-sector relations of
\eqnref{eq:Kfields} with $Z\to\Theta^{(a)}$ [the spinon energy density
is now
$E_f=\frac{2}{N}\sum_{\bm k,m}^{\rm occ}\epsilon_{\bm k,m}
+2z_1K^{(1)}\Theta^{(1)}+2z_2K^{(2)}\Theta^{(2)}$, with the bands and
the kernels of \eqnref{eq:ABkernels} built from $t^{\mathrm{eff}}_{\mathrm{NN}},t^{\mathrm{eff}}_{\mathrm{NNN}}$] gives
the homogeneous self-consistency conditions
\begin{equation}
\begin{aligned}
z_1\Theta^{(1)}U&=-t^{2}\Theta^{(1)}\,\frac{4}{N}\sum_{\bm k,m}^{\rm occ}
\frac{A_{\bm k}}{\epsilon_{\bm k,m}},\\
z_2\Theta^{(2)}U&=-t'^{2}\Theta^{(2)}\,\frac{8}{N}\sum_{\bm k,m}^{\rm occ}
\frac{B_{\bm k}}{\epsilon_{\bm k,m}},
\end{aligned}
\label{eq:thetasc}
\end{equation}
where now $\epsilon_{\bm k,m}$ is built from $t^{\mathrm{eff}}_{\mathrm{NN}},t^{\mathrm{eff}}_{\mathrm{NNN}}$. Since
$\Theta^{(1)}=\Theta^{(2)}=0$ always solves \eqnref{eq:thetasc}, the physical
saddle is selected by comparing the total energies $E_\theta+E_f$ of the
candidate solutions.

\begin{table}[t]
\caption{Slave-rotor mean-field criteria for the three half-filling phases,
together with their charge and spin Chern numbers (values quoted for $t'>0$).
The rotor condensate $Z$ locates the Mott transition, and the effective NNN spinon hopping $t^{\mathrm{eff}}_{\mathrm{NNN}}$ distinguishes the two Mott insulators. The dash
indicates that $t^{\mathrm{eff}}_{\mathrm{NNN}}$ is trivially finite in the condensed phase and
is not a distinguishing diagnostic there.}
\label{tab:criteria}
\begin{ruledtabular}
\begin{tabular}{lccc}
 & IQH & CSL & DSL$\to$AFM? \\
 & $C_c=C_s=-2$ & $C_c=0$, $C_s=-2$ & $C_c=0$ \\
\colrule
$Z=\big|\langle e^{i\theta_i}\rangle\big|^{2}$ & $\neq0$ & $0$ & $0$ \\
$t^{\mathrm{eff}}_{\mathrm{NNN}}=t'\langle e^{i(\theta_i-\theta_j)}\rangle_{\langle\!\langle ij\rangle\!\rangle}$ & -- & $\neq0$ & $0$ \\
\end{tabular}
\end{ruledtabular}
\end{table}

The identification of the three phases proceeds in
two steps, summarized in Table~\ref{tab:criteria}. The condensed calculation
determines the Mott boundary. A nonzero $Z$ endows the electron with a
quasiparticle weight and, for $t'\neq0$, preserves the Hofstadter Chern bands,
hence the IQH phase. Inside the Mott regime ($Z=0$) the effective NNN spinon hopping $t^{\mathrm{eff}}_{\mathrm{NNN}}=t'\Theta^{(2)}$ distinguishes the two Mott insulators. For $t^{\mathrm{eff}}_{\mathrm{NNN}}\neq0$ the spinons remain in gapped Chern bands, so the spinon topology retains the quantized spin response, and the induced
Chern-Simons term gaps the emergent gauge field, stabilizing the deconfined
CSL. For $t^{\mathrm{eff}}_{\mathrm{NNN}}=0$ the
spinons revert to gapless $\pi$-flux Dirac fermions coupled to a compact
$U(1)$ gauge field, a Dirac spin liquid that is not itself magnetically
ordered; on the square lattice, however, monopole proliferation confines this
saddle and generates N\'eel order~\cite{Hermele2004U1SpinLiquids,Song2019CompetingOrders,Song2020MonopolesDSL},
so we identify it with the topologically trivial AFM observed in DMRG. The effective NNN hopping $t^{\mathrm{eff}}_{\mathrm{NNN}}$
obtained in this way is the red scale of \figref{fig:undoped}(b), and
the resulting mean-field phase diagram is qualitatively consistent with the DMRG result of \figref{fig:undoped}(a).

\subsection{$SU(2)$ Higgs-Chern-Simons critical theory}
\label{app:su2k}

Here we derive the $SU(2)_k$ Higgs-Chern-Simons critical theory invoked in
the main text. At half filling the Landau-gauge hoppings of
\eqnref{eq:landau} satisfy the particle-hole condition
\begin{equation}
t_{ij}=-s_is_j\,t^{*}_{ij},
\qquad
s_i=(-1)^{i_x+i_y},
\label{eq:ph}
\end{equation}
since the NN amplitudes are real ($s_is_j=-1$ on NN bonds) and the NNN
amplitudes are purely imaginary ($s_is_j=+1$ on NNN bonds); this promotes the
charge sector to a pseudospin $SU(2)_c$ symmetry generated by
$\eta^{z}_{i}=\frac12(n_{i}-1)$ and
$\eta^{+}_{i}=s_{i}\,c^{\dagger}_{i\up}c^{\dagger}_{i\dn}$, and permits the
$SU(2)$ slave-rotor factorization
\begin{equation}
\Psi_{i}=
\begin{pmatrix}
c_{i\up} & s_ic^{\dagger}_{i\dn}\\
c_{i\dn} & -s_ic^{\dagger}_{i\up}
\end{pmatrix}
=F_{i}\mathcal Z_i,
\label{eq:su2}
\end{equation}
where $F_{i}$ has the same form with $c\to f$ and $\mathcal Z_i\in SU(2)$ is the charge
rotor; the decomposition carries a local $SU(2)$ gauge redundancy,
$F_{i}\to F_{i}g_{i}^{\dagger}$, $\mathcal Z_i\to g_{i}\mathcal Z_i$. Using \eqnref{eq:ph}, the hopping
term can be written exactly as
$H_t=-\sum_{ij}t_{ij}\tr(\Psi_i^\dagger\Psi_j)$,
and the corresponding action
reorganizes into rotor and spinon bilinears,
\begin{equation}
\begin{aligned}
S_t&=-\int\!d\tau\sum_{ij}|t_{ij}|\tr\big(A_{ij}B_{ij}\big),\\
A_{ij}&=\mathcal Z_j\mathcal Z_i^{\dagger},\\
B_{ij}&=e^{i\phi_{ij}}F_i^{\dagger}F_j,
\end{aligned}
\label{eq:StAB}
\end{equation}
where $\phi_{ij}$ is the Landau-gauge Peierls phase. A complex,
unconstrained $2\times2$ link field $Q_{ij}$ decouples the product,
\begin{equation}
S[Q]=\int\!d\tau\sum_{ij}|t_{ij}|\Big[\tr\big(Q^{\dagger}_{ij}Q_{ij}\big)
-\tr\big(\mathcal Z_i^{\dagger}Q_{ij}\mathcal Z_j\big)
-e^{i\phi_{ij}}\tr\big(F_jQ^{\dagger}_{ij}F_i^{\dagger}\big)\Big],
\label{eq:Qdecouple}
\end{equation}
with gauge covariance $Q_{ij}\to g_iQ_{ij}g_j^{\dagger}$. At the saddle point
connected to the $U=0$ limit, $Q_{ij}=e^{i\phi_{ij}}\langle F_i^\dagger
F_j\rangle=\bar Q_{ij}\openone$ with real $\bar Q_{ij}=\bar Q_1$ ($\bar Q_2$) on NN
(NNN) bonds. The Peierls phase has been stripped
from the rotor sector, so the rotor quadratic form carries zero flux and its
minimum sits at $\bm q=0$. With link scales $J_1=|t|\bar Q_1$ and
$J_2=|t'|\bar Q_2$, integrating out the gapped rotor modes gives the inverse
propagator
\begin{equation}
\mathcal G^{-1}_{\mathcal Z}(\omega,\bm q)=\frac{3}{4U}\omega^2+r_{\mathcal Z,0}
-2J_1(\cos q_x+\cos q_y)-4J_2\cos q_x\cos q_y,
\label{eq:rotorprop}
\end{equation}
which upon expansion about $\bm q=0$ yields the relativistic form
$\frac{3}{4U}\omega^2+r_{\mathcal Z}+(J_1+2J_2)\bm q^2$ with $r_{\mathcal Z}=r_{\mathcal Z,0}-4J_1-4J_2$. The soft
mode is therefore the $2\times2$ matrix field $\mathcal Z(x)\sim \mathcal Z_i$,
transforming under the gauge $SU(2)$ by left multiplication and under the
pseudospin $SU(2)_c$ by right multiplication; a term linear in $\partial_\tau$
is forbidden by the half-filling pseudospin structure. Fluctuations of the
link field supply the gauge field, $Q_{ij}=\bar Q_{ij}\exp(ia^b_{ij}\rho^b/2)$,
which in the continuum produces the covariant derivative
$D_\mu\mathcal Z=(\partial_\mu+\frac{i}{2}a^b_\mu\rho^b)\mathcal Z$ appearing in
\eqnref{eq:LSM} below.

The Chern-Simons level is fixed by the spinon band topology. The spinon mean
field is the two-band problem of \eqnref{eq:bloch} with the effective hoppings
$t^{\mathrm{eff}}_{\mathrm{NN}},t^{\mathrm{eff}}_{\mathrm{NNN}}$; expanding about the two Dirac points
$K_\pm=(\pi/2,\pm\pi/2)$ gives
\begin{equation}
h_f(K_\pm+\bm q)\simeq2t^{\mathrm{eff}}_{\mathrm{NN}}\,q_x\tau_x\pm4t^{\mathrm{eff}}_{\mathrm{NNN}}\,\tau_y\pm2t^{\mathrm{eff}}_{\mathrm{NN}}\,q_y\tau_z,
\label{eq:diracexp}
\end{equation}
so the valley mass (the $\tau_y$ term) and the $q_y$ velocity (the
$\tau_z$ term) reverse sign together between the two valleys, and the two
valleys carry equal Berry-curvature contributions, and the occupied spinon
band has Chern number $C_f=-\sgn(t^{\mathrm{eff}}_{\mathrm{NNN}})$ per gauge doublet.
Coupling the spinons to $a_\mu=a^b_\mu\rho^b/2$ and integrating out the filled
band, the two components of the gauge doublet carry gauge charge $\pm1/2$, so
in a Cartan background $a_\mu^3\rho^3/2$ one obtains $S_{\rm
eff}=\frac{iC_f}{8\pi}\int a^3\wedge da^3$, which matches the non-Abelian
Chern-Simons action restricted to the same background provided
$k=C_f=-\sgn(t^{\mathrm{eff}}_{\mathrm{NNN}})$. Combining the Higgs sector with the induced
Chern-Simons term yields the $SU(2)_k$ Higgs-Chern-Simons theory,
\begin{equation}
\begin{aligned}
\mathcal L={}&\frac{ik}{4\pi}\,\epsilon^{\mu\nu\rho}\tr\Big(a_\mu\partial_\nu a_\rho+\frac{2}{3}a_\mu a_\nu a_\rho\Big)
+\frac{1}{2}\tr\big[(D_\mu\mathcal Z)^{\dagger}D_\mu\mathcal Z\big]\\
&+\frac{r_{\mathcal Z}}{2}\tr\big(\mathcal Z^{\dagger}\mathcal Z\big)
+\frac{\lambda_{\mathcal Z}}{4}\big[\tr\big(\mathcal Z^{\dagger}\mathcal Z\big)\big]^{2}+\cdots,
\end{aligned}
\label{eq:LSM}
\end{equation}
with the Chern-Simons level $k=-\sgn(t^{\mathrm{eff}}_{\mathrm{NNN}})$.

The two phases follow from the sign of the rotor mass. For $r_{\mathcal Z}<0$ the Higgs
field condenses; choosing the gauge $\langle\mathcal Z\rangle=v\,\openone$, the
$SU(2)$ gauge field is completely Higgsed and the spinons recombine with the
rotors into physical electrons. Since the spinon mean field is an integer
quantum Hall Chern band, this side of the transition is the electronic IQH
phase. For $r_{\mathcal Z}>0$ the rotor is gapped, and integrating it out leaves the pure
$SU(2)_k$ Chern-Simons theory, the topological field theory of the
Kalmeyer-Laughlin CSL. Equation~\eqref{eq:LSM} therefore describes a direct
continuous transition between the IQH phase and the $SU(2)_k$ CSL.

The critical theory also identifies which field becomes soft at the
transition. The right $SU(2)_c$ action of $\mathcal Z$ is the continuum
counterpart of the microscopic pseudospin algebra introduced above, where
$\eta^z_{i}=\frac12(n_{i}-1)$ is the charge density relative to half
filling, whereas $\eta^{\pm}_{i}$ create and annihilate on-site Cooper pairs.
The pseudospin components of the critical rotor therefore represent
charge-density fluctuations and charge-$2e$ pair fields, so the critical mode
is a spinless charge-sector rotor. The spin gap can remain open across the
transition because the spinons stay in massive Chern bands, while the charge
(pseudospin) gap closes as $r_{\mathcal Z}\to0$. This provides the field-theoretic reason
that Cooper-pair-like excitations become energetically cheaper than
single-electron excitations near the IQH-CSL transition; it underlies the
gap hierarchy probed by the DQMC data in \figref{fig:undoped}(c),(d)
and the energetic mechanism invoked in the main text for the stability of the
doped superconductor against the chiral metal.

\section{Kohn-Luttinger analysis and BdG topology}
\label{app:KL}

We now dope the weak-coupling side and derive
the pairing quoted in the main text. In the sublattice basis the on-site
interaction $U\sum_{\bm R\gamma}n_{\bm R\gamma\up}n_{\bm R\gamma\dn}$ is repulsive at
first order, so the attraction must arise at second order in
$U$~\cite{KohnLuttinger1965,Shaffer2021HofstadterSC,Shaffer2022HofstadterSC}. The only second-order diagram surviving in the
Cooper channel, given the local bare vertex, is the particle-hole bubble,
which yields the effective pair-scattering vertex between $(\bm k\up,-\bm k\dn)$ and
$(\bm k'\up,-\bm k'\dn)$ with sublattice structure $\gamma\gamma'$,
\begin{equation}
\Gamma^{(2)}_{\gamma\gamma'}(\bm k,\bm k')
=\;\diag{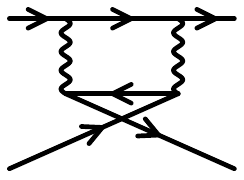}{42pt}\;
=-U^2\sum_{\bm p}\sum_{m,n}
w_{\gamma m}(\bm P)\,w^{*}_{\gamma'm}(\bm P)\,
w_{\gamma'n}(\bm p)\,w^{*}_{\gamma n}(\bm p)\,
\frac{n_F[\xi_m(\bm P)]-n_F[\xi_n(\bm p)]}{\xi_m(\bm P)-\xi_n(\bm p)},
\label{eq:KLvertex}
\end{equation}
where $\bm P\equiv \bm k+\bm k'+\bm p$, $m,n=\pm$ are band indices, $\xi_m$ are the band
energies measured from the chemical potential, $n_F$ is the Fermi function,
and $w_{\gamma m}$ are the Landau-gauge Bloch vectors of \eqnref{eq:bloch}.
In the diagram the straight lines are the external Cooper-pair legs, the wavy
lines are the two bare $U$ vertices, and the internal loop is the
particle-hole bubble carrying momentum $\bm k+\bm k'$. In the lightly doped limit the
Fermi factors force the bubble to connect one occupied lower-band state and
one empty upper-band state, so the attraction is mediated by virtual interband
(exciton) fluctuations across the Chern-band
gap~\cite{CrepelFu2021,Crepel2022InterbandSC}. Explicitly, with
$\epsilon(\bm k)>0$ the band energy of \eqnref{eq:dispersion} and the interband
weight
$\Lambda_{\gamma\gamma'}(\bm q,\bm p)\equiv
w_{\gamma-}(\bm q)w^{*}_{\gamma'-}(\bm q)\,w_{\gamma'+}(\bm p)w^{*}_{\gamma+}(\bm p)$,
\begin{equation}
\Pi_{\gamma\gamma'}(\bm Q)=U^2\sum_{\bm p}
\frac{\Lambda_{\gamma\gamma'}(\bm Q{+}\bm p,\bm p)+\big({-}\leftrightarrow{+}\big)}
{\epsilon(\bm Q{+}\bm p)+\epsilon(\bm p)},
\label{eq:Xbubble}
\end{equation}
and the vertex projected onto the lower band follows by dressing
$\Pi_{\gamma\gamma'}(\bm k+\bm k')$ with the Bloch factors $w_{\gamma-}$ of the four
external legs.

\subsection{Two-valley regime, $t'<t/2$}
\begin{figure}[t]
\centering
\includegraphics[width=0.34\textwidth]{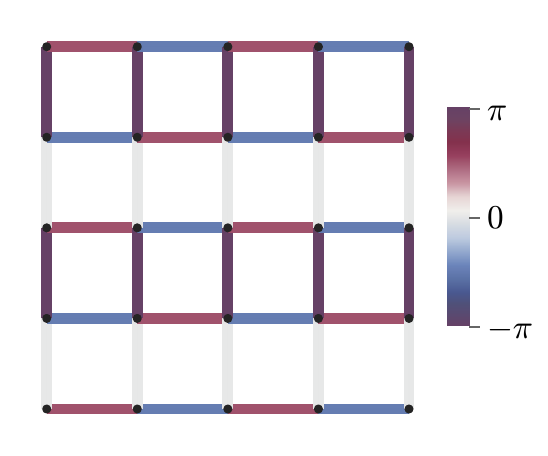}
\caption{Pairing pattern of the Kohn-Luttinger channel \eqnref{eq:KLop} in the Landau gauge. Bond color gives $\arg[\tilde F_{b_0}\tilde F^{*}_{\gamma\gamma'}(\bm d)]$, built from the form factor \eqnref{eq:formfactor} with the same reference bond $b_0$ as in \figref{fig:doped}(a).}
\label{fig:klpattern}
\end{figure}

The doped holes occupy the two
valleys $K_\pm=(\pi/2,\pm\pi/2)$ of \figref{fig:landaufs}(a), with
$-K_+=K_--\bm G_1$ modulo reciprocal vectors, and the particle-hole momenta entering \eqnref{eq:Xbubble} are $0$ and $(0,\pi)$. The low-energy valley pair space is four dimensional. In the Landau gauge it
splits into a pair-momentum-$0$ sector, spanned by $c_{-K_+,\dn}c_{K_+,\up}$
and $c_{-K_-,\dn}c_{K_-,\up}$, and a pair-momentum-$(0,\pi)$ sector, spanned
by $c_{-K_-,\dn}c_{K_+,\up}$ and $c_{-K_+,\dn}c_{K_-,\up}$, with all
operators in the lower band. Diagonalizing the projected vertex over this
space at $t'=0.25$ gives the eigenvalues
$\lambda_{\rm KL}=(-0.11,\,+0.11,\,+0.11,\,+0.11)\,U^2/t$, i.e., a unique
attractive channel with order parameter
\begin{equation}
\hat\Delta
=\frac{1}{\sqrt2}\Big(c_{-K_-,-,\dn}c_{K_+,-,\up}-c_{-K_+,-,\dn}c_{K_-,-,\up}\Big)
=\sum_{(\gamma,\gamma',\bm d)}\tilde F_{\gamma\gamma'}(\bm d)\,\hat\Delta_{\gamma\gamma'}(\bm d),
\label{eq:KLop}
\end{equation}
which carries pair momentum $(0,\pi)$ in this gauge (the Landau-gauge
counterpart of the zero-momentum intervalley singlet of the checkerboard
convention), and is annihilated by the first-order on-site repulsion
[\eqnref{eq:firstorder2} below]. Because $(0,\pi)$ is not a reciprocal
vector of the two-site magnetic cell, the real-space expansion is performed on
the magnetic cell doubled along $\hat y$, with primitive vectors $(2,0)$ and
$(0,2)$ and four sublattices $\gamma\in\{A,B,A',B'\}$, where $A'$ and $B'$
denote the odd-row copies of $A$ and $B$, at offsets $\bm d_{A'}=(0,1)$ and
$\bm d_{B'}=(1,1)$ [$\bm d_A$ and $\bm d_B$ as in \secref{app:model}]. On the
doubled lattice $(0,\pi)$ is a reciprocal vector, so the channel carries zero
momentum of the doubled cell and expands in cell-independent bond operators.
Here
\begin{equation}
\hat\Delta_{\gamma\gamma'}(\bm d)
=\frac{4}{N}\sum_{\bm R}
\big(c_{\bm R,\gamma,\up}c_{\bm R+\bm d,\gamma',\dn}-c_{\bm R,\gamma,\dn}c_{\bm R+\bm d,\gamma',\up}\big)
\label{eq:bondop}
\end{equation}
annihilates a spin singlet on the bond connecting sublattice $\gamma$ in
doubled cell $\bm R$ to sublattice $\gamma'$ in the
cell displaced by $\bm d$, carries no cell dependence
(the $N/4$ doubled cells are summed over uniformly), and the sum
$(\gamma,\gamma',\bm d)$ avoids double counting. Using the
folding identities $c_{-K_\mp,-,\sigma}=-c_{K_\pm,-,\sigma}$, which follow
from the sewing matrices of \secref{app:model}, the channel is equivalently
$\hat\Delta=\frac{1}{\sqrt2}\big(c_{K_-,-,\dn}c_{K_-,-,\up}
-c_{K_+,-,\dn}c_{K_+,-,\up}\big)$; since $e^{-2iK_s\cdot\bm R}=1$ for
every $\bm R$ of the doubled lattice, the cell sum factors out of the
expansion, and the symmetrized form factor takes the compact form
\begin{equation}
\tilde F_{\gamma\gamma'}(\bm d)
=\frac{1}{2\sqrt2}\sum_{s=\pm}s\;
w^{*}_{\gamma-}(K_s)\,w^{*}_{\gamma'-}(K_s)\,
e^{-iK_s\cdot(\bm d+\bm d_\gamma+\bm d_{\gamma'})},
\label{eq:formfactor}
\end{equation}
with $w_-(K_s)=(s\,i,1)^{T}/\sqrt2$ the valley-center Bloch vectors [the
$\bm q\to0$ limit of \eqnref{eq:valleyspinor}], the primed sublattices
carrying the orbital weight of their columns, $w_{A'-}=w_{A-}$ and
$w_{B'-}=w_{B-}$, and $\bm d$ measured as a real-space displacement. As in the checkerboard convention, the bond operator carries no
symmetry information and the entire pairing symmetry resides in
$\tilde F_{\gamma\gamma'}(\bm d)$, with the row alternation of the pattern
entering through the offsets $\bm d_{A'}$ and $\bm d_{B'}$. The resulting
pattern, shown in \figref{fig:klpattern}, is the one measured by DMRG in \figref{fig:doped}(a).

We have shown that the second-order vertex supports the channel
\eqnref{eq:KLop}; next, following the same logic as in the triangular-lattice
analysis of Ref.~\cite{Chen2026TriangularHofstadterSC}, we examine the effect
of the first-order (bare) interaction on this channel. Projecting the on-site
repulsion onto the lower-band pair space gives the positive semidefinite
matrix
\begin{equation}
\begin{aligned}
\Gamma^{(1)}_{jj'}
&=U\,\delta_{\bm Q_j,\bm Q_{j'}}\sum_{\gamma=A,B}\Xi^{*}_{j}(\gamma)\Xi_{j'}(\gamma),\\
\Xi_{j}(\gamma)&\equiv w_{\gamma-}(\bm k^{A}_{j})\,w_{\gamma-}(\bm k^{B}_{j}),
\end{aligned}
\label{eq:firstorder}
\end{equation}
where $(\bm k^{A}_{j},\bm k^{B}_{j})$ are the $(\up,\dn)$ momenta of pair
record $j$, $\bm Q_j=\bm k^{A}_{j}+\bm k^{B}_{j}$ is its pair momentum, conserved modulo
reciprocal vectors, and $\Xi_j(\gamma)$ is the on-site orbital weight of
the pair on sublattice $\gamma$; positive semidefiniteness implies that a
weak-coupling instability can develop only in the kernel of $\Gamma^{(1)}$.
In the four-dimensional two-valley basis ordered as above, the valley-center
spinors give $\Xi(\gamma)=\frac12(-1,1)$ for the first two records and $\Xi(\gamma)=\frac12(1,1)$ for the last
two records, so
\begin{equation}
\Gamma^{(1)}=\frac U2\begin{pmatrix}
1&1&0&0\\
1&1&0&0\\
0&0&1&1\\
0&0&1&1
\end{pmatrix}.
\label{eq:firstorder2}
\end{equation}

Acting on the channel of \eqnref{eq:KLop}, whose components in this
ordered basis are $(0,0,1,-1)/\sqrt2$, this matrix gives zero, so the bare
vertex vanishes identically on the channel and the effective interaction in
this channel is entirely given by the second-order vertex, which is negative,
corresponding to an effective attraction.

\begin{figure}[t]
\centering
\begin{minipage}[c]{0.27\textwidth}
\centering
(a)\\[1pt]
\includegraphics[width=\linewidth]{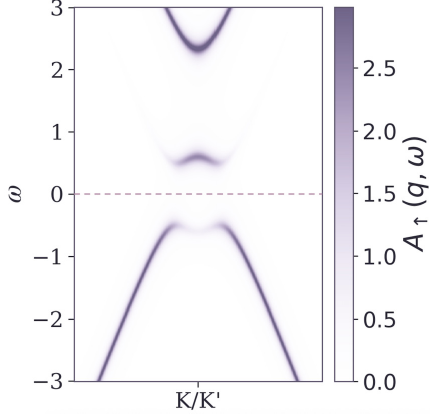}
\end{minipage}\hspace{0.05\textwidth}
\begin{minipage}[c]{0.44\textwidth}
\centering
(b)\\[1pt]
\includegraphics[width=\linewidth]{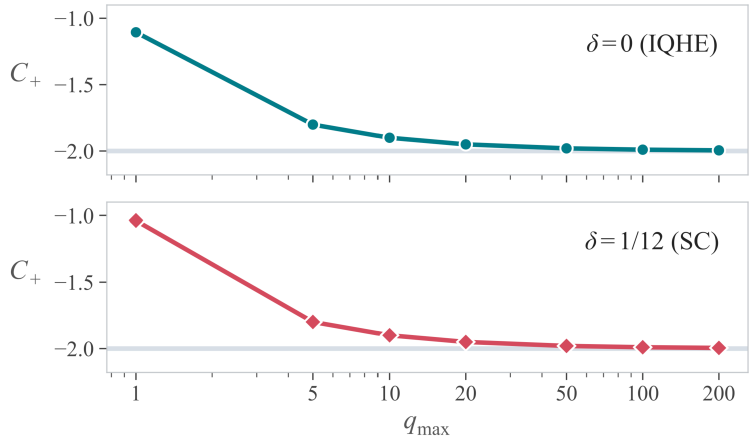}
\end{minipage}
\caption{BdG spectrum and topology of the two-valley superconductor in
the Landau gauge. (a) Electron-sector spectral function of \eqnref{eq:BdG} at
$\Delta_0=0.5$ and $\delta=1/12$, showing a full pairing gap at the hole pockets
centered on the valleys $K_\pm=(\pi/2,\pm\pi/2)$. (b) Wilson-loop Chern
number $C_+$ versus the compactification radius $q_{\max}$ for the undoped
IQH state ($\delta=0$) and the doped superconductor ($\delta=1/12$); both
converge to $C_s=-2$.
}
\label{fig:BdGtopology}
\end{figure}

The spin Chern number of the paired state follows from
the continuum BdG theory of the two valleys. Expanding \eqnref{eq:bloch}
about $K_\pm$ as in \eqnref{eq:diracexp} and adding the chemical potential
gives the valley Hamiltonians
\begin{equation}
h^{K_\pm}_{\bm q}\simeq2t\,q_x\tau_x\pm4t'\tau_y\pm2t\,q_y\tau_z+\mu\tau_0,
\label{eq:valleyham}
\end{equation}
whose lower-band spinors admit the closed form
\begin{equation}
\begin{aligned}
w_{-}(K_\pm+\bm q)&=\frac{1}{\sqrt{2E_{\bm q}(E_{\bm q}\pm2tq_y)}}
\begin{pmatrix}
-2tq_x\pm4it'\\[2pt]
E_{\bm q}\pm2tq_y
\end{pmatrix},\\
E_{\bm q}&=\sqrt{16t'^2+4t^2(q_x^2+q_y^2)}.
\end{aligned}
\label{eq:valleyspinor}
\end{equation}
The pairing of \eqnref{eq:KLop} then enters through the
rank-one orbital matrix
\begin{equation}
W_{\gamma\gamma'}(\bm q)=w_{\gamma-}(K_++\bm q)\,w_{\gamma'-}(K_-+\bm q),
\qquad
W(0)=\frac{1}{2}\big(\tau_0-\tau_y\big),
\label{eq:Wmatrix}
\end{equation}
where the first (second) index refers to the $\up$ ($\dn$) member of the
pair. In the eight-component Nambu basis
$\big(c_{K_++\bm q,\up},\,c_{K_-+\bm q,\up},\,c^{\dagger}_{-K_+-\bm q,\dn},\,
c^{\dagger}_{-K_--\bm q,\dn}\big)$, each entry carrying both sublattices, and
using $h_{-\bm k}=h_{\bm k}$ [\eqnref{eq:bloch}], the BdG Hamiltonian takes the block
form
\begin{equation}
\mathcal H_{\bm q}=
\begin{pmatrix}
h^{K_+}_{\bm q} & 0 & 0 & \Delta_0W(\bm q)\\
0 & h^{K_-}_{\bm q} & -\Delta_0W^{T}(\bm q) & 0\\
0 & -\Delta_0W^{*}(\bm q) & -h^{K_+\,T}_{\bm q} & 0\\
\Delta_0W^{\dagger}(\bm q) & 0 & 0 & -h^{K_-\,T}_{\bm q}
\end{pmatrix},
\label{eq:BdG}
\end{equation}
where the relative minus sign between the two pairing blocks implements
the antisymmetric intervalley channel of \eqnref{eq:KLop}. The
electron-sector spectral function at $\Delta_0=0.5$ and $\delta=1/12$ is fully
gapped at the hole pockets [\figref{fig:BdGtopology}(a)], with a minimum
quasiparticle energy $\simeq\Delta_0$ over the whole $\bm q$ plane.
For an $S_z$-conserving singlet superconductor the two Nambu branches
$\psi_+=(c_\up,c^{\dagger}_\dn)^T$ and $\psi_-=(c_\dn,-c^{\dagger}_\up)^T$
define BdG Chern numbers $C_\pm$, and the spin Chern number is
$C_s=(C_++C_-)/2$, which spin $SU(2)$ symmetry reduces to $C_+$; the chiral
central charge obeys $c_-=C_s$. We evaluate $C_+$ with the non-Abelian
Wilson-loop (Fukui-Hatsugai-Suzuki)
discretization~\cite{FukuiHatsugaiSuzuki2005}: at fixed radius $q=|\bm q|$,
\begin{equation}
\Phi(q)=\operatorname{Arg}\prod_{j=0}^{N_\phi-1}
\frac{\det\big[\mathcal V^{\dagger}(q,\phi_j)\,\mathcal V(q,\phi_{j+1})\big]}
{\big|\det\big[\mathcal V^{\dagger}(q,\phi_j)\,\mathcal V(q,\phi_{j+1})\big]\big|},
\label{eq:wilson}
\end{equation}
with $\mathcal V(q,\phi)$ the matrix of occupied BdG states and $\phi_j=2\pi
j/N_\phi$, and extract $C_+=-[\Phi(q_{\max})-\Phi(0)]/2\pi$ upon compactifying
the momentum plane at radius $q_{\max}$. As shown in
\figref{fig:BdGtopology}(b), both the undoped IQH state
$(\delta,t',\Delta_0)=(0,0.25,0)$ and the doped superconductor $(1/12,0.25,0.5)$
converge to
\begin{equation}
C_s=-2,
\label{eq:Cs}
\end{equation}
so the weak-coupling superconductor inherits the spin Chern number of the
parent Chern band, in agreement with the DMRG measurement in the main text.

\subsection{Four-valley regime, $t'>t/2$}
For $t'>t/2$ the lower band
instead has four hole valleys at $X=(\pi/2,0)$, $M=(\pi/2,\pi)$, and
$X'_\pm=(0,\pm\pi/2)$ [\figref{fig:landaufs}(b)]. Because momenta such as
$-X$ and $-M$ must be folded back to the magnetic Brillouin zone, the
sublattice embedding phases $D_\gamma(\bm G)=e^{-i\bm G\cdot\bm d_\gamma}$ enter,
reducing to $D_A=1$ and $D_B=-1$ for the required foldings, while
$X'_\pm$ are exchanged by inversion without folding. The low-energy valley
pair space is now eight dimensional, again splitting into a pair-momentum-$0$ sector,
spanned by $c_{-X,\dn}c_{X,\up}$, $c_{-M,\dn}c_{M,\up}$, and
$c_{X'_\mp,\dn}c_{X'_\pm,\up}$, and a pair-momentum-$(0,\pi)$ sector, spanned
by $c_{-M,\dn}c_{X,\up}$, $c_{-X,\dn}c_{M,\up}$, and
$c_{X'_\pm,\dn}c_{X'_\pm,\up}$. Diagonalizing the projected vertex over this
space at $t'=0.75$ gives a doubly degenerate leading attractive channel,
$\lambda_{\rm KL}=-0.066\,U^2/t$, lying in the $(0,\pi)$ sector:
\begin{equation}
\begin{aligned}
\hat\Delta_{XM}&=\frac{1}{\sqrt2}\Big(c_{-M,\dn}c_{X,\up}-c_{-X,\dn}c_{M,\up}\Big),\\
\hat\Delta_{X'}&=\frac{1}{\sqrt2}\Big(c_{X'_+,\dn}c_{X'_+,\up}-c_{X'_-,\dn}c_{X'_-,\up}\Big),
\end{aligned}
\label{eq:fourvalleyops}
\end{equation}
with all operators in the lower band; the remaining channels are
repulsive, apart from a weakly attractive pair-momentum-$0$ combination at
$-0.035\,U^2/t$. The two leading channels mix the $\{X,M\}$ and the
$\{X'_+,X'_-\}$ valleys, respectively, and are exactly degenerate at this
order, forming a projective-$C_4$ partner doublet. At quadratic
Kohn-Luttinger order neither the relative amplitude nor the relative phase of
the two channels is fixed; resolving this degeneracy requires higher-order
Ginzburg-Landau terms or additional microscopic input.
As in the two-valley regime, the first-order projection
\eqnref{eq:firstorder} determines which channels survive the bare repulsion.
In the pair-momentum-$0$ basis
$\{c_{-X,\dn}c_{X,\up},\,c_{-M,\dn}c_{M,\up},\,
c_{X'_-,\dn}c_{X'_+,\up},\,c_{X'_+,\dn}c_{X'_-,\up}\}$ and the
$(0,\pi)$ basis
$\{c_{-M,\dn}c_{X,\up},\,c_{-X,\dn}c_{M,\up},\,
c_{X'_+,\dn}c_{X'_+,\up},\,c_{X'_-,\dn}c_{X'_-,\up}\}$, it evaluates to
\begin{equation}
\Gamma^{(1)}_{\bm Q=0}
=\frac U2\begin{pmatrix}
2&0&1&1\\
0&2&1&1\\
1&1&1&1\\
1&1&1&1
\end{pmatrix},
\qquad
\Gamma^{(1)}_{(0,\pi)}
=\frac U2\begin{pmatrix}
0&0&0&0\\
0&0&0&0\\
0&0&1&1\\
0&0&1&1
\end{pmatrix},
\label{eq:firstorder4}
\end{equation}
where the vanishing $X{\leftrightarrow}M$ block follows from the
complete sublattice polarization of the band-edge Bloch states,
$w_-(X)=(1,0)^{T}$ and $w_-(M)=(0,1)^{T}$, which makes the on-site weight
$\Xi(\gamma)$ of an $X$-$M$ pair vanish identically.
In these bases the two leading channels of \eqnref{eq:fourvalleyops}
read $\hat\Delta_{XM}=(1,-1,0,0)/\sqrt2$ and
$\hat\Delta_{X'}=(0,0,1,-1)/\sqrt2$, and both are annihilated by
$\Gamma^{(1)}_{(0,\pi)}$, the former by its vanishing rows and the latter by
antisymmetry; the subleading pair-momentum-$0$ combination at
$-0.035\,U^2/t$, with components $\tfrac12(-1,-1,1,1)$, is likewise
annihilated by $\Gamma^{(1)}_{\bm Q=0}$. As in the
two-valley case, the effective interaction in the surviving channels is
entirely given by the attractive second-order vertex.

\section{Chiral metal on the six-leg cylinder at $t'=0.75$: nearly
decoupled spinful chains}
\label{app:sixleg}

The Kohn-Luttinger vertex of \secref{app:KL} is attractive also in the
four-valley regime, yet the DMRG phase diagram at $\delta=1/12$ finds, at
$t'\simeq0.75$ and weak coupling, a gapless chiral metal with no pair
binding rather than a superconductor [\figref{fig:model}(b)]. This section explains why, by bosonizing the quasi-one-dimensional Fermi-point structure of \secref{app:model}. The explanation rests on two geometric facts: the two sublattice-balanced $X'$ valleys are absent from the six-leg momentum grid, and the surviving $X$ and $M$ pockets are almost completely
sublattice polarized, so that the on-site repulsion couples them only
through a form factor that vanishes in the dilute-doping limit. The
cylinder therefore realizes two nearly decoupled one-dimensional spinful
chains, a C2S2 Luttinger metal on every numerically accessible scale,
where C$n$S$m$ denotes $n$ gapless charge and $m$ gapless spin
modes~\cite{BalentsFisher1996,LinBalentsFisher1997}. We emphasize that the
pocket structure of \figref{fig:landaufs} is that of the two-dimensional
thermodynamic limit, whereas the missing $X'$ pockets observed numerically
are a geometry-dependent feature of the $L_y=6$ cylinder; the metallic
corner of the doped phase diagram therefore cannot be taken as evidence
that a metal survives in the two-dimensional thermodynamic limit.

\subsection{Fermi points and projected interaction}
Of the four hole
pockets of \figref{fig:landaufs}(b), only $X=(\pi/2,0)$ and $M=(\pi/2,\pi)$
lie on the allowed cuts \eqnref{eq:kycuts}, since the two $X'$ pockets sit at
$k_y=\pm\pi/2$, which belongs to the transverse momentum grid only when
$L_y\equiv0\bmod4$. Four- and eight-leg cylinders thus see all four
pockets, and $L_y=6$ is the special geometry that projects out the $X'$ pair. The loss is not innocuous, because at $X'$ the lower-band Bloch vector is sublattice balanced, $|w_{A-}(X')|=|w_{B-}(X')|=1/\sqrt2$
[$h_{X'}=-2t\,\tau_x$ from \eqnref{eq:bloch}], whereas at $X$ and $M$ it is
completely polarized for any $t'$, $w_-(X)=(1,0)^{T}$ and
$w_-(M)=(0,1)^{T}$, the same polarization that annihilated the
$X\leftrightarrow M$ block of the first-order projection
\eqnref{eq:firstorder4}. Of the two degenerate Kohn-Luttinger channels of
\eqnref{eq:fourvalleyops}, $\hat\Delta_{X'}$ is built entirely from the
absent valleys and simply has no support on the cylinder, while
$\hat\Delta_{XM}$ must connect the two polarized pockets, and we show below
that the corresponding bare projected vertices are strongly suppressed at
weak coupling. Since the small-$U$ regime is precisely the one dominated by
this low-energy, long-wavelength pocket physics, the six-leg metal at
$t'=0.75$ is a statement about the $L_y=6$ momentum grid rather than about
the two-dimensional limit.

At $\delta=1/12$ only the two cuts $k_y=0,\pi$
are populated (\secref{app:model}), and the holes occupy $|k_x|\in(k_F,\pi/2]$
on each cut, and each pocket contributes one right- and one left-moving
Fermi point at $k_x=\pm k_F$, four Fermi points per spin in total. The two cuts share a single $k_F$ and a single Fermi velocity $v_F$, so without interactions the cylinder is a four-channel conductor.

In the Fourier convention
\eqnref{eq:fourier} the Hubbard term reads
\begin{equation}
H_U=U\sum_{\bm R,\gamma}n_{\bm R\gamma\up}n_{\bm R\gamma\dn}
=\frac{U}{N_c}\sum_{\bm k_1\bm k_2\bm k_3\bm k_4}\sum_{\bm G}\sum_{\gamma}
c^{\dagger}_{\bm k_1\gamma\up}c^{\dagger}_{\bm k_2\gamma\dn}
c_{\bm k_3\gamma\dn}c_{\bm k_4\gamma\up}\,
\delta_{\bm k_1+\bm k_2,\,\bm k_3+\bm k_4+\bm G}\;e^{-i\bm G\cdot\bm d_\gamma},
\label{eq:HUmom}
\end{equation}
where the cell sum enforces momentum conservation modulo reciprocal vectors
$\bm G$ and the leftover offset phase $e^{-i\bm G\cdot\bm d_\gamma}=D_\gamma(\bm G)$ is
the sewing phase of \secref{app:model}. In the strict weak-coupling regime
the direct interband gap $4t$ justifies projecting out the upper band,
\begin{equation}
c_{\bm k,\gamma,\sigma}\;\to\;u_\gamma(\bm k)\,c_{\bm k\sigma},
\qquad
u_\gamma(\bm k)\equiv w_{\gamma-}(\bm k),
\qquad
c_{\bm k\sigma}\equiv c_{\bm k,-,\sigma},
\label{eq:project}
\end{equation}
after which every vertex carries a sublattice-coherence factor,
\begin{equation}
\begin{aligned}
H_U&\;\to\;\frac{U}{N_c}\sum_{\{\bm k_\ell\}}\sum_{\bm G}
\Lambda(1,2;3,4)\,
c^{\dagger}_{\bm k_1\up}c^{\dagger}_{\bm k_2\dn}c_{\bm k_3\dn}c_{\bm k_4\up}\,
\delta_{\bm k_1+\bm k_2,\,\bm k_3+\bm k_4+\bm G},\\
\Lambda(1,2;3,4)&=\sum_{\gamma}
u^{*}_\gamma(\bm k_1)\,u^{*}_\gamma(\bm k_2)\,u_\gamma(\bm k_3)\,u_\gamma(\bm k_4)\,
e^{-i\bm G\cdot\bm d_\gamma};
\end{aligned}
\label{eq:lambdadef}
\end{equation}
the Hubbard vertex couples opposite spins only.

Restricted to the four Fermi points, momentum conservation is very
restrictive. Along $x$ the possible transfers are
$\Delta k_x=(r_1+r_2-r_3-r_4)k_F\in\{0,\pm2k_F,\pm4k_F\}$ with
$G_x\in\pi\mathbb Z$; at $k_F=3\pi/8$ neither $2k_F$ nor $4k_F$ is a multiple of $\pi$, so only chirality-balanced processes
survive, with $\Delta k_x=0$ exactly and no umklapp at the four-fermion
level (higher-order multi-fermion umklapp operators are momentum-allowed
at this commensurate density, but they arise only at higher order in $U$
and we do not consider them). Along $y$ a process
transferring a single fermion between the pockets would carry $\Delta
k_y=\pi\notin2\pi\mathbb Z$, so only even numbers of legs may move between
$X$ and $M$. Every surviving $\bm G$ then has $e^{-i\bm G\cdot\bm d_\gamma}=1$, and
linearizing about the Fermi points
[$c_{(rk_F+q,\,k_y^{\mathcal P})\sigma}\to\psi_{\mathcal P,r,\sigma}(x)$,
with pocket label $\mathcal P=X,M$, i.e.\ $k_y^{X}=0$ and $k_y^{M}=\pi$,
and chirality $r=R,L=\pm$] gives the one-dimensional interaction
\begin{equation}
H_U=\sideset{}{'}\sum_{\{\mathcal P_\ell,r_\ell\}}
\frac{2U}{L_y}\,\Lambda(1,2;3,4)\int\!dx\;
\psi^{\dagger}_{\mathcal P_1,r_1,\up}\psi^{\dagger}_{\mathcal P_2,r_2,\dn}
\psi_{\mathcal P_3,r_3,\dn}\psi_{\mathcal P_4,r_4,\up},
\label{eq:H1D}
\end{equation}
where the primed sum runs over the momentum-conserving assignments and the
factor $2U/L_y$ is the transverse normalization ($N_c=L_xL_y/2$ magnetic
cells at cylinder length $L_x$).

\begin{figure}[t]
\centering
\includegraphics[width=0.98\textwidth]{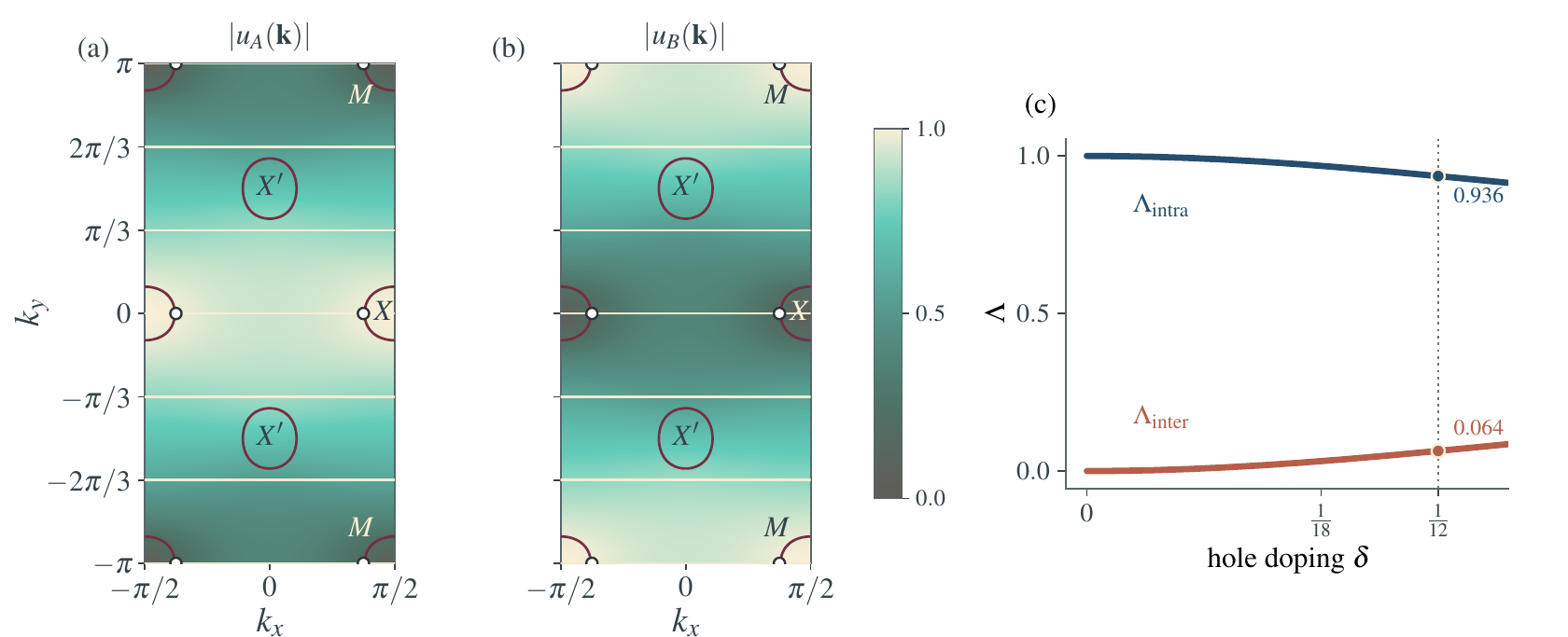}
\caption{Pocket polarization and near decoupling at $t'=0.75$ on the
six-leg cylinder. (a),(b) Sublattice weights $|u_A(\bm k)|$ and $|u_B(\bm k)|$ of
the lower band over one magnetic Brillouin zone.
Dark red contours: the 2D Fermi surface at $\delta=1/12$;
white lines: the six allowed cuts \eqnref{eq:kycuts}; open circles mark the
Fermi points on those cuts; labels identify the pocket centers $X$, $M$, $X'$.
The $X$ pocket
resides almost entirely on sublattice $A$ and the $M$ pocket on sublattice
$B$ (exactly so at the pocket centers), while the sublattice-balanced $X'$
pockets are the valleys removed by the six-leg momentum grid. (c) The two
Fermi-point form factors of \eqnref{eq:lambdavalues} versus hole doping:
intra-pocket $\Lambda_{\rm intra}$ and inter-pocket
$\Lambda_{\rm inter}$. The dashed vertical line marks the DMRG doping
$\delta=1/12$.}
\label{fig:pockets}
\end{figure}

In the reference gauge of \eqnref{eq:bloch} the four Fermi-point spinors
are real and inversion even; writing
$u_{\mathcal P,\gamma}\equiv u_\gamma(\pm k_F,k_y^{\mathcal P})$, the two
cuts are related by the sublattice flip
\begin{equation}
u_{M,A}=u_{X,B},
\qquad
u_{M,B}=u_{X,A}
\qquad
\bigl[h_{(k_x,k_y+\pi)}=\tau_xh_{(k_x,k_y)}\tau_x\bigr].
\label{eq:uflip}
\end{equation}
Consequently every intra-pocket assignment of the four legs takes one and
the same value, and so does every inter-pocket one:
\begin{equation}
\Lambda_{\rm intra}=\sum_\gamma|u_{\mathcal P,\gamma}|^{4},
\qquad
\Lambda_{\rm inter}
=\sum_\gamma|u_{X,\gamma}|^{2}|u_{M,\gamma}|^{2}
=\sum_\gamma u^{*2}_{X,\gamma}u^{2}_{M,\gamma}
=2\,u_{X,A}^{2}u_{X,B}^{2},
\label{eq:lambdavalues}
\end{equation}
where $\Lambda_{\rm intra}$, equal on the two pockets by
\eqnref{eq:uflip}, covers the four-legs-on-one-pocket processes ($XXXX$
and $MMMM$), while $\Lambda_{\rm inter}$ covers every two-plus-two process,
the first form being inter-pocket forward and backscattering ($XMXM$
type), the second pair tunneling ($XXMM$ type); they coincide because
the spinors are real. The hierarchy between the two is the sublattice
polarization at work. Exactly at the pocket centers the polarization is
complete and opposite on the two pockets, $w_-(X)=(1,0)^{T}$ and
$w_-(M)=(0,1)^{T}$, so at the band edge $\Lambda_{\rm inter}$ vanishes
identically; at small but finite doping the Fermi points sit slightly off
the band edge and pick up only a weak sublattice admixture, so
$\Lambda_{\rm inter}$ remains small while $\Lambda_{\rm intra}$ stays close to unity. This is seen directly in \figref{fig:pockets}, where the weights $|u_A(\bm k)|$ and $|u_B(\bm k)|$ partition the MBZ into complementary domains
(wherever $|u_A|$ is large, $|u_B|$ is small and vice versa), the $X$ and
$M$ pockets sit deep inside opposite domains, and any vertex connecting
them must pick up two powers of the small admixture; the $X'$ points sit
on the domain wall, which is why those valleys, when present, couple at
$O(1)$. At the DMRG doping $\delta=1/12$ the two form factors evaluate to
$\Lambda_{\rm intra}=0.936$ and $\Lambda_{\rm inter}=0.064$, so the
intra-pocket coupling dominates by a factor of ${\simeq}15$.

\subsection{Bosonization and inter-pocket couplings}
With the inter-pocket vertices
suppressed, each pocket $\mathcal P$ is a standard one-dimensional spinful
chain with Fermi velocity $v_F$, and the chirality-balanced intra-pocket
quartic terms allowed by the selection rules are listed in Table~\ref{tab:processes}, four density-density (forward) combinations and
two backscattering partners, all carrying the same bare coupling
$g=(2U/L_y)\Lambda_{\rm intra}$.

\begin{table}[t]
\caption{The six chirality-balanced intra-pocket interactions of pocket
$\mathcal P$, listed by the chirality assignment of the vertex
$\psi^{\dagger}_{\mathcal P,r_1,\up}\psi^{\dagger}_{\mathcal P,r_2,\dn}
\psi_{\mathcal P,r_3,\dn}\psi_{\mathcal P,r_4,\up}$; all six carry the
same bare coupling $g=(2U/L_y)\Lambda_{\rm intra}$. Here
$\rho_{\mathcal P,r,\sigma}=\,:\!\psi^{\dagger}_{\mathcal P,r,\sigma}
\psi_{\mathcal P,r,\sigma}\!:$\,.}
\label{tab:processes}
\begin{ruledtabular}
\begin{tabular}{lll}
$(r_1,r_2)$ & $(r_3,r_4)$ & operator \\
\colrule
$(R,R)$ & $(R,R)$ & $\rho_{\mathcal P,R,\up}\,\rho_{\mathcal P,R,\dn}$ \\
$(L,L)$ & $(L,L)$ & $\rho_{\mathcal P,L,\up}\,\rho_{\mathcal P,L,\dn}$ \\
$(R,L)$ & $(L,R)$ & $\rho_{\mathcal P,R,\up}\,\rho_{\mathcal P,L,\dn}$ \\
$(L,R)$ & $(R,L)$ & $\rho_{\mathcal P,L,\up}\,\rho_{\mathcal P,R,\dn}$ \\
$(R,L)$ & $(R,L)$ &
$\psi^{\dagger}_{\mathcal P,R,\up}\psi^{\dagger}_{\mathcal P,L,\dn}
\psi_{\mathcal P,R,\dn}\psi_{\mathcal P,L,\up}$ \\
$(L,R)$ & $(L,R)$ &
$\psi^{\dagger}_{\mathcal P,L,\up}\psi^{\dagger}_{\mathcal P,R,\dn}
\psi_{\mathcal P,L,\dn}\psi_{\mathcal P,R,\up}$ \\
\end{tabular}
\end{ruledtabular}
\end{table}

Bosonizing along standard lines~\cite{Giamarchi2004,Song2025moire}, we
introduce per pocket and spin the boson pair
$(\phi_{\mathcal P,\sigma},\theta_{\mathcal P,\sigma})$, defined by
$\partial_x\phi_{\mathcal P,\sigma}=-\pi\rho_{\mathcal P,\sigma}$ and
$\partial_x\theta_{\mathcal P,\sigma}=\pi j_{\mathcal P,\sigma}$, with
$\rho_{\mathcal P,\sigma}=\rho_{\mathcal P,R,\sigma}+\rho_{\mathcal
P,L,\sigma}$ the smooth density and $j_{\mathcal P,\sigma}=\rho_{\mathcal
P,R,\sigma}-\rho_{\mathcal P,L,\sigma}$ the chiral difference, obeying
$[\phi_{\mathcal P,\sigma}(x),\partial_{x'}\theta_{\mathcal
P',\sigma'}(x')]=i\pi\,\delta_{\mathcal P\mathcal P'}
\delta_{\sigma\sigma'}\delta(x-x')$, together with the charge and spin
combinations $\phi_{\mathcal P,\rho/s}=(\phi_{\mathcal
P,\up}\pm\phi_{\mathcal P,\dn})/\sqrt2$ (and likewise
$\theta_{\mathcal P,\rho/s}$). Following the standard steps, the four
density rows of Table~\ref{tab:processes} renormalize the quadratic form
while the two backscattering rows produce a spin cosine, and the pocket
Hamiltonian separates into commuting charge and spin sectors,
$H^{(\mathcal P)}=H^{(\mathcal P)}_{\rho}+H^{(\mathcal P)}_{s}$, of
Luttinger form
\begin{equation}
\begin{aligned}
H^{(\mathcal P)}_{\rho}
&=\frac{v_{\rho}}{2\pi}\int\!dx\,
\Bigl[K_{\rho}\bigl(\partial_x\theta_{\mathcal P,\rho}\bigr)^{2}
+K_{\rho}^{-1}\bigl(\partial_x\phi_{\mathcal P,\rho}\bigr)^{2}\Bigr],\\
H^{(\mathcal P)}_{s}
&=\frac{v_{s}}{2\pi}\int\!dx\,
\Bigl[K_{s}\bigl(\partial_x\theta_{\mathcal P,s}\bigr)^{2}
+K_{s}^{-1}\bigl(\partial_x\phi_{\mathcal P,s}\bigr)^{2}\Bigr]
+\frac{2g_{1}}{(2\pi a_0)^{2}}\int\!dx\,
\cos\bigl(2\sqrt2\,\phi_{\mathcal P,s}\bigr),
\end{aligned}
\label{eq:LLform}
\end{equation}
with $g_{1}(0)=g$, $a_0$ a short-distance cutoff, and bare parameters
\begin{equation}
K_{\rho}=\Bigl[1+\frac{g}{\pi v_F}\Bigr]^{-1/2}<1,
\qquad
K_{s}(0)=\Bigl[1-\frac{g}{\pi v_F}\Bigr]^{-1/2}
\simeq1+\frac{g}{2\pi v_F}>1,
\qquad
v_{\rho/s}=\frac{v_F}{K_{\rho/s}}.
\label{eq:KLL}
\end{equation}
The two sectors then behave oppositely: the charge sector contains no cosine at all (an umklapp $\cos(2\sqrt2\,\phi_{\mathcal P,\rho})$ would
require $4k_F\in\pi\mathbb Z$), so $H^{(\mathcal P)}_{\rho}$ is a pure
Gaussian model with $K_{\rho}<1$, and the charge mode stays gapless at
weak coupling. The spin sector is
the standard sine-Gordon model; with $y_{s}\equiv g_{1}/(\pi v_{s})$ its
one-loop Kosterlitz flow is~\cite{Giamarchi2004}
\begin{equation}
\frac{dK_{s}}{d\ell}=-\frac{1}{2}K_{s}^{2}y_{s}^{2},
\qquad
\frac{dy_{s}}{d\ell}=2\bigl(1-K_{s}\bigr)y_{s},
\label{eq:KTflow}
\end{equation}
and spin-rotation symmetry places the bare values \eqnref{eq:KLL} on the
separatrix $2[K_{s}(0)-1]=y_{s}(0)=g/(\pi v_F)$, along which both
equations collapse onto $dy_{s}/d\ell=-y_{s}^{2}$, i.e.
\begin{equation}
y_{s}(\ell)=\frac{y_{s}(0)}{1+y_{s}(0)\,\ell}\;\longrightarrow\;0^{+},
\qquad
g_{1}(\ell)\to0^{+},
\qquad
K_{s}(\ell)\to1.
\label{eq:KTsolution}
\end{equation}
The repulsive backscattering is marginally irrelevant, and the spin mode,
like the charge mode, stays gapless. Each pocket by itself is therefore a
gapless C1S1 spinful Luttinger liquid, and the direct product of the two
decoupled pockets is C2S2, two gapless charge and two gapless spin modes,
total central charge $c=4$.

The bare hierarchy of \eqnref{eq:lambdavalues} is now essential, since every inter-pocket vertex (forward scattering, backscattering, or pair tunneling) starts with a bare coupling more than an order of magnitude smaller than its
intra-pocket counterpart. These operators are tree-level marginal at the
decoupled free-fermion fixed point. Consequently, in the weak-coupling
regime, such a small inter-pocket seed can remain weak over a long RG
interval, even if one of these channels ultimately grows and opens a
spin gap. On a finite cylinder, the RG flow is cut off at the infrared
scale set by $L_x$. For sufficiently weak $U$ and the numerically
accessible lengths, this cutoff can therefore be reached before the
inter-pocket couplings become strong, producing a long nearly decoupled
crossover before any eventual two-pocket instability is resolved.

We stress that this is a simple qualitative estimate; we have not
solved the complete coupled multiband
RG~\cite{BalentsFisher1996,LinBalentsFisher1997}. Thus the bare hierarchy
supplies a candidate preasymptotic interpretation of the DMRG chiral
metal, in which finite cylinders may retain the mode content of the decoupled
C2S2 theory over the accessible window, with no discernible pair binding [\figref{fig:model}(b)].

We do not repeat this analysis for
$t'=0.25$. On the one hand the problem is genuinely more involved,
four crossed cuts and eight Fermi points per spin, with a considerably
richer set of inter-cut couplings; on the other hand it is also less
necessary, because there the six-leg cuts essentially recover the
two-dimensional pocket structure, since the two $K_\pm$ pockets are sampled by
the four adjacent cuts and retain their sublattice-balanced character,
so the intervalley couplings enter at $O(1)$ and the two-valley
Kohn-Luttinger mechanism of \secref{app:KL} applies directly,
consistent with the superconductor found by DMRG at $t'=0.25$.

\section{Strong coupling: doping the chiral spin liquid}
\label{app:parton}

\subsection{Gauge-equivalent $d+id$ spinon pairing}
\label{app:parton_pairing}
In the Mott regime the CSL
spinon ansatz is the hopping Hamiltonian of \eqnref{eq:bloch} with effective
amplitudes $t^{\mathrm{eff}}_{\mathrm{NN}},t^{\mathrm{eff}}_{\mathrm{NNN}}$, written in real space (Landau gauge) as
\begin{equation}
H_f=\sum_{i,\sigma}\Big[-t^{\mathrm{eff}}_{\mathrm{NN}}\,f^{\dagger}_{i+\hat x,\sigma}f_{i\sigma}
-(-1)^{i_x}\,t^{\mathrm{eff}}_{\mathrm{NN}}\,f^{\dagger}_{i+\hat y,\sigma}f_{i\sigma}
-i(-1)^{i_x}\,t^{\mathrm{eff}}_{\mathrm{NNN}}\,f^{\dagger}_{i+\hat x+\hat y,\sigma}f_{i\sigma}
+i(-1)^{i_x}\,t^{\mathrm{eff}}_{\mathrm{NNN}}\,f^{\dagger}_{i-\hat x+\hat y,\sigma}f_{i\sigma}\Big]+\mathrm{H.c.}
\label{eq:spinonreal}
\end{equation}
In the Nambu doublet $\psi_i=(f_{i\up},f^{\dagger}_{i\dn})^T$ this ansatz is
encoded by $2\times2$ link matrices,
$H_f=\sum_{\langle ij\rangle}\psi^{\dagger}_iu_{ij}\psi_j+\mathrm{H.c.}$, with
\begin{equation}
\begin{aligned}
u_{i+\hat x,i}&=-t^{\mathrm{eff}}_{\mathrm{NN}}\,\tau_z, &
u_{i+\hat y,i}&=-(-1)^{i_x}\,t^{\mathrm{eff}}_{\mathrm{NN}}\,\tau_z,\\
u_{i+\hat x+\hat y,i}&=-i(-1)^{i_x}\,t^{\mathrm{eff}}_{\mathrm{NNN}}\,\tau_0, &
u_{i-\hat x+\hat y,i}&=+i(-1)^{i_x}\,t^{\mathrm{eff}}_{\mathrm{NNN}}\,\tau_0,
\end{aligned}
\label{eq:links}
\end{equation}
where $\tau_{x,y,z}$ act in the gauge (Nambu) space. Every closed Wilson loop of
these links is proportional to $\tau_0$, since closed square-lattice loops
contain an even number of NN steps, so the loop products lie in the center of
$SU(2)$ and the ansatz admits nontrivial gauge transformations. Two
successive site-dependent rotations,
\begin{equation}
\begin{aligned}
g^{(1)}_i&=e^{i\phi^{(1)}_i\tau_z}, &
\phi^{(1)}_i&=-\pi\,i_x-\frac{\pi}{2}\,i_y^{\,2}+\frac{\pi}{8}\,s_i,\\
g^{(2)}_i&=e^{i\phi^{(2)}_i\tau_y}, &
\phi^{(2)}_i&=-\frac{\pi}{4}\,s_i,
\end{aligned}
\label{eq:gaugerot}
\end{equation}
with $s_i=(-1)^{i_x+i_y}$, transform the links as $u_{ij}\to
g_iu_{ij}g_j^{\dagger}$ and remove all staggered phases. After
$g^{(1)}$ alone the four link types already take site-independent moduli with
sublattice-staggered phases,
\begin{equation}
u'_{i+\hat x,i}=\frac{t^{\mathrm{eff}}_{\mathrm{NN}}}{\sqrt2}\big(\tau_z-is_i\tau_0\big),
\qquad
u'_{i+\hat y,i}=\frac{t^{\mathrm{eff}}_{\mathrm{NN}}}{\sqrt2}\big(\tau_z+is_i\tau_0\big),
\qquad
u'_{i\pm\hat x+\hat y,i}=\pm t^{\mathrm{eff}}_{\mathrm{NNN}}\,s_i\tau_z,
\label{eq:midlinks}
\end{equation}
and the staggered $\tau_y$ rotation $g^{(2)}$ then removes the residual
$s_i$ dependence, yielding the uniform link matrices
\begin{equation}
u''_{i+\hat x,i}=\frac{t^{\mathrm{eff}}_{\mathrm{NN}}}{\sqrt2}\big(\tau_z+\tau_y\big),
\qquad
u''_{i+\hat y,i}=\frac{t^{\mathrm{eff}}_{\mathrm{NN}}}{\sqrt2}\big(\tau_z-\tau_y\big),
\qquad
u''_{i\pm\hat x+\hat y,i}=\pm t^{\mathrm{eff}}_{\mathrm{NNN}}\,\tau_x.
\label{eq:uniformlinks}
\end{equation}
Reading the transformed links in the standard Abrikosov form
$u_{ij}=\big(\begin{smallmatrix}\chi_{ij}&\Delta_{ij}\\
\Delta^{*}_{ij}&-\chi^{*}_{ij}\end{smallmatrix}\big)$, with $\chi_{ij}$ the
hopping and $\Delta_{ij}$ the singlet spinon pairing on the directed link,
gives
\begin{equation}
\Delta_{i+\hat x,i}=-\frac{it^{\mathrm{eff}}_{\mathrm{NN}}}{\sqrt2},\quad
\Delta_{i+\hat y,i}=+\frac{it^{\mathrm{eff}}_{\mathrm{NN}}}{\sqrt2},\quad
\Delta_{i\pm\hat x+\hat y,i}=\pm t^{\mathrm{eff}}_{\mathrm{NNN}},
\label{eq:did}
\end{equation}
a chiral $d+id$ pattern, showing that the hopping-only CSL ansatz is gauge
equivalent to a chirally paired spinon state. The uniform pattern of \eqnref{eq:did} is
independent of the gauge chosen for the electron hopping, as it must be, since the
electron gauge choice only reshuffles the site-dependent rotations in
\eqnref{eq:gaugerot}, not the $SU(2)$ gauge-equivalence class of the physical
spinon state.
If bosonic holons condense upon doping, this spinon pairing induces physical
electron pairing; the final electron pairing phase also contains the holon
condensate phase and therefore cannot be read off directly from
\eqnref{eq:did}. The holon condensate, however, carries charge but neither
spin nor band topology, so it cannot alter the quantized spin response, and the
superconductor obtained by condensing holons inherits the spin Chern number
of the parent CSL, $C_s=-2\,\sgn(t^{\mathrm{eff}}_{\mathrm{NNN}})$, as confirmed by the response
theory below and by the DMRG measurement in the main text.

\subsection{Topological response and edge theory}
The gauge transformation of the previous subsection exposes the $d+id$ spinon pairing hidden in the hopping-only CSL ansatz; for the topological response, however, it is more direct to return to the Chern-band form \eqnref{eq:spinonreal} of the ansatz. At finite hole doping the rotor charge sector of \secref{app:rotor} is then organized in terms of holons. Relative to the half-filled Mott background, the $L_i=-1$ rotor state of \eqnref{eq:constraint} is an empty electronic site and therefore represents one doped hole, and in the doped low-energy theory the rotor bond factor $e^{i(\theta_i-\theta_j)}$ of \eqnref{eq:rotorkinetic} is represented by the hopping of bosonic holons $h$, while the neutral fermionic spinons $f_\sigma$ remain in the Chern-band state. The result is a $U(1)$ slave-boson theory with the physical electron $c_\sigma\sim h^{\dagger}f_\sigma$, in which a compact internal gauge field $a_\mu$ implements the local constraint $n_h+n_f=1$, with $n_h\equiv h^{\dagger}h$ and $n_f\equiv\sum_\sigma f^{\dagger}_{\sigma}f_{\sigma}$. Both partons carry the same internal gauge charge, the holon in addition carries the electromagnetic charge, and the spinon $f_\sigma$ carries spin charge $\sigma=\pm1$ for $\up,\dn$; the corresponding external electromagnetic and spin probes are denoted $A^e_\mu$ and $A^s_\mu$. Below we use the differential-form notation $X\wedge dY\equiv\epsilon^{\mu\nu\rho}X_\mu\partial_\nu Y_\rho\,d^3x$ and $(da)^2\equiv(\epsilon^{\mu\nu\rho}\partial_\nu a_\rho)^2$ with $\epsilon^{0xy}=+1$, and $D_A$ denotes a covariant derivative with gauge field $A$. The holons see the same $\pi$ flux and, at the commensurate filling considered, form a bosonic IQH state, while the spinons remain in their Chern bands. The parton Lagrangian is then
\begin{equation}
\mathcal L[h,f]=\mathcal L_{\rm bIQH}\big[D_{A^e-a}h\big]+\mathcal L_{\rm CSL}\big[D_{\sigma A^s-a}f_\sigma\big].
\label{eq:partonL}
\end{equation}
Integrating out the spinon Chern bands and then the bosonic IQH holons gives
the effective response
\begin{equation}
\mathcal L_{\rm eff}=\frac{2i\,\sgn(t^{\mathrm{eff}}_{\mathrm{NNN}})}{4\pi}A^e\!\wedge\! dA^e
-\frac{2i\,\sgn(t^{\mathrm{eff}}_{\mathrm{NNN}})}{2\pi}A^e\!\wedge\! da
-\frac{2i\,\sgn(t^{\mathrm{eff}}_{\mathrm{NNN}})}{4\pi}A^s\!\wedge\! dA^s
+\frac{1}{2g^2}(da)^2.
\label{eq:Leff}
\end{equation}
Three conclusions follow. First, the $A^s\wedge dA^s$ coefficient gives the
spin Chern number $C_s=-2\,\sgn(t^{\mathrm{eff}}_{\mathrm{NNN}})$ and the spin Hall conductivity
$\sigma^s_{xy}=C_s(\hbar/2)^2/h$, in agreement with the weak-coupling BdG
result and with DMRG. Second, no $a\wedge da$ term survives, so the phase has
no residual intrinsic topological order, and integrating out $a_\mu$
generates a Higgs mass for $A^e$, so the state is a superconductor. Third,
varying \eqnref{eq:Leff} with respect to $A^e_0$ gives the charge density
$\rho_e=-\frac{1}{\pi}\sgn(t^{\mathrm{eff}}_{\mathrm{NNN}})\,\nabla\times\bm a$, so a $2\pi$ flux of
the internal gauge field binds electromagnetic charge $2e$; a standard
compact-$U(1)$ duality then identifies the monopole operators $\mathcal
M_\pm\sim e^{\pm i\phi}$ of $a_\mu$ as the Cooper-pair creation and
annihilation operators of the superconductor.

The same conclusions follow from the edge, which also
makes explicit why the superconductor carries chiral edge modes but no
protected quantized charge Hall response. In \eqnref{eq:Leff} the
electromagnetic probe enters both through $A^e\wedge dA^e$ and through the
mixed $A^e\wedge da$ term, so the charge response cannot be read off
directly. On the boundary we take $\mu,\nu\in\{0,x\}$ with $\epsilon_{0x}=+1$
and omit nonuniversal velocity terms.

For the undoped CSL with $t^{\mathrm{eff}}_{\mathrm{NNN}}>0$, the two spinon Chern bands contribute two chiral bosons $\phi_{f\sigma}$ coupling to
$\sigma A^s_\mu-a_\mu$,
\begin{equation}
\mathcal L_{\rm CSL}
=-\frac{i}{4\pi}\sum_{\sigma}\partial_0\phi_{f\sigma}\partial_x\phi_{f\sigma}
-\frac{i}{2\pi}\sum_{\sigma}\epsilon_{\mu\nu}a_\mu\partial_\nu\phi_{f\sigma}
+\frac{i}{2\pi}\epsilon_{\mu\nu}A^s_\mu\partial_\nu\big(\phi_{f\up}-\phi_{f\dn}\big).
\label{eq:CSLedge}
\end{equation}
Integrating out the internal gauge field imposes
$\phi_{f\up}+\phi_{f\dn}=\mathrm{const}$, leaving a single chiral field
$\phi_f\equiv\phi_{f\up}=-\phi_{f\dn}+\mathrm{const}$ with
\begin{equation}
\mathcal L_{\rm CSL}
=-\frac{2i}{4\pi}\partial_0\phi_f\partial_x\phi_f
+\frac{2i}{2\pi}\epsilon_{\mu\nu}A^s_\mu\partial_\nu\phi_f,
\label{eq:CSLedgefinal}
\end{equation}
i.e., one chiral edge mode ($c_-=1$ in this convention) carrying the
quantized spin response $C_s=-2$. The bosonic IQH holon sector instead
contributes two counterpropagating bosons $\phi_{h,1},\phi_{h,2}$ with $K$
matrix $K^h=\big(\begin{smallmatrix}0&1\\1&0\end{smallmatrix}\big)$, whose
eigenvalues $\pm1$ give no net chirality.

In the superconductor the two sectors combine, with the holon modes coupling
to $A^e_\mu-a_\mu$:
\begin{equation}
\begin{aligned}
\mathcal L_{\rm SC}
={}&-\frac{i}{4\pi}\sum_{\sigma}\partial_0\phi_{f\sigma}\partial_x\phi_{f\sigma}
-\frac{i}{2\pi}\sum_{\sigma}\epsilon_{\mu\nu}a_\mu\partial_\nu\phi_{f\sigma}
+\frac{i}{2\pi}\epsilon_{\mu\nu}A^s_\mu\partial_\nu\big(\phi_{f\up}-\phi_{f\dn}\big)\\
&+\frac{i}{4\pi}\big(\partial_0\phi_{h,1}\partial_x\phi_{h,2}+\partial_0\phi_{h,2}\partial_x\phi_{h,1}\big)
+\frac{i}{2\pi}\sum_{j=1,2}\epsilon_{\mu\nu}\big(A^e_\mu-a_\mu\big)\partial_\nu\phi_{h,j}.
\end{aligned}
\label{eq:SCedgefull}
\end{equation}
Integrating out $a_\mu$ now imposes the constraint
\begin{equation}
\phi_{h,2}=-\sum_{\sigma}\phi_{f\sigma}-\phi_{h,1}+\mathrm{const}.
\label{eq:SCedgeconstraint}
\end{equation}
Renaming $\phi_h\equiv\phi_{h,1}$ and introducing the neutral, spin, and
charge combinations $\phi_0=\frac12(\phi_{f\up}+\phi_{f\dn})+\phi_h$,
$\phi_s=\frac12(\phi_{f\up}-\phi_{f\dn})$, and
$\phi_c=\frac12(\phi_{f\up}+\phi_{f\dn})$, the edge action reduces to
\begin{equation}
\mathcal L_{\rm SC}=-\frac{2i}{4\pi}\big(\partial_0\phi_0\partial_x\phi_0
+\partial_0\phi_s\partial_x\phi_s\big)
-\frac{2i}{2\pi}\epsilon_{\mu\nu}A^e_\mu\partial_\nu\phi_c
+\frac{2i}{2\pi}\epsilon_{\mu\nu}A^s_\mu\partial_\nu\phi_s.
\label{eq:SCedge}
\end{equation}
Only $\phi_0$ and $\phi_s$ carry chiral kinetic terms, and they propagate in
the same direction, giving the chiral central charge $|c_-|=2$. The spin mode
$\phi_s$ couples to $A^s$ and reproduces $C_s=-2$. The field $\phi_c$ labels
the combination that couples to $A^e$, but it has no
$\partial_0\phi_c\partial_x\phi_c$ term of its own in \eqnref{eq:SCedge}; the
superconductor therefore has no quantized charge Hall response protected by a
gapped charge sector. These are the edge signatures quoted in the main text.

\section{Additional DQMC results and pairing formalism}
\label{app:dqmc_supp}

In this section, we present additional DQMC results for the fermion sign, spin correlations, and pairing response, together with the definitions and formalism used for the pairing analysis in the main text. Except for the average sign itself, all DQMC expectation values are sign reweighted, and statistical uncertainties are estimated using jackknife and bootstrap resampling.
The DQMC simulations use a Trotter step $\Delta\tau t=0.05$ and are performed in the grand-canonical ensemble, with the chemical potential tuned to the target density for the doped calculations.

\subsection{Sign and spin correlations at half filling}

\begin{figure}[t]
\centering
\includegraphics[width=0.8\columnwidth]{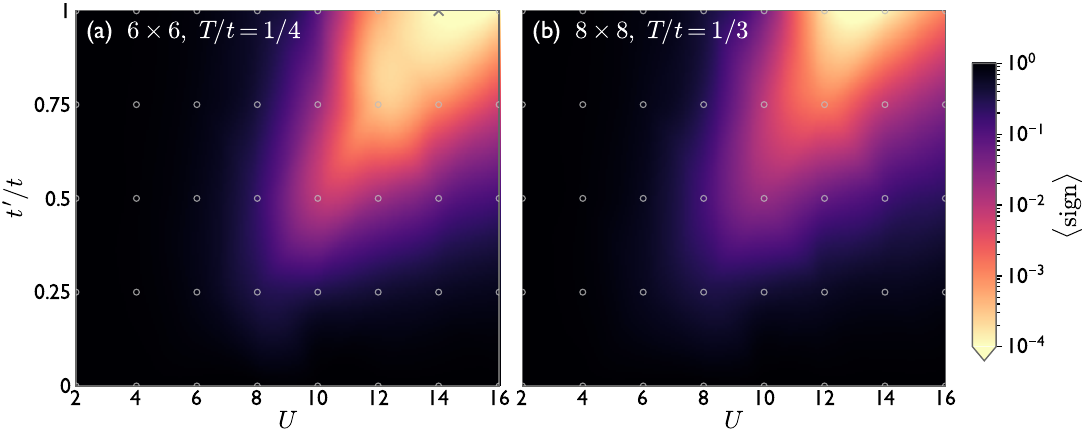}
\caption{Average DQMC fermion sign at half filling as a function of $U/t$ and $t'/t$ for (a) a $6\times6$ lattice at $T/t=1/4$ and (b) an $8\times8$ lattice at $T/t=1/3$.
Open circles mark measured parameter points included in the interpolation, while the cross in (a) marks the point excluded because of its large statistical uncertainty.
The smooth background is obtained by interpolating the reliable measurements and serves as a guide to the eye.
The logarithmic color scale is clipped to
$10^{-4}\leq\langle\mathrm{sign}\rangle\leq1$; values below the lower limit are saturated at the lower end of the scale.}
\label{fig:sign}
\end{figure}

\begin{figure}[t]
\centering
\includegraphics[width=0.8\columnwidth]{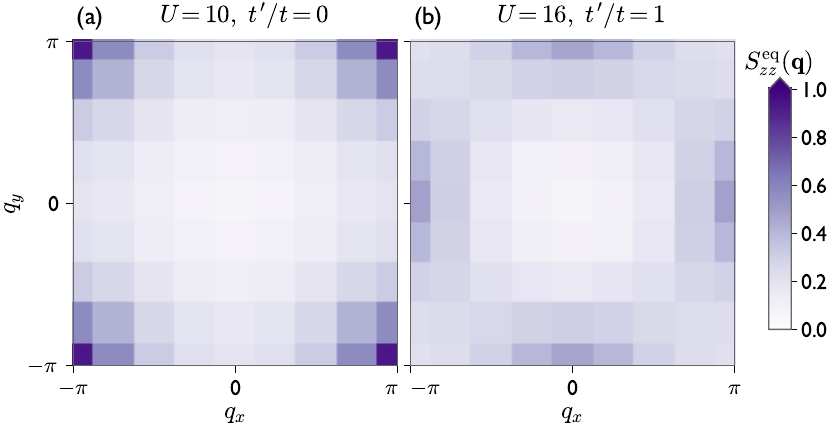}
\caption{Momentum-space equal-time spin correlations,
$S_{zz}^{\mathrm{eq}}(\mathbf q)
=\sum_{\mathbf r}
e^{-i\mathbf q\cdot\mathbf r}C_{zz}(\mathbf r)$,
where
$C_{zz}(\mathbf r)
=\frac{1}{4N}\sum_{\mathbf R}
\left\langle
\bigl(n_{\mathbf R+\mathbf r,\uparrow}
 -n_{\mathbf R+\mathbf r,\downarrow}\bigr)
\bigl(n_{\mathbf R,\uparrow}
 -n_{\mathbf R,\downarrow}\bigr)
\right\rangle$.
Results are shown at half filling on an $8\times8$ lattice at $T/t=1/3$.
Panel (a) shows $(U/t,t'/t)=(10,0)$ in the AFM region, whereas panel (b) shows $(U/t,t'/t)=(16,1)$ in the CSL region, as identified by the DMRG ground-state phase diagram.}
\label{fig:spin}
\end{figure}

Figure~\ref{fig:sign} shows the average DQMC fermion sign at half filling.
The strongest suppression of the sign broadly follows the IQH-CSL transition region and becomes particularly severe toward the large-$t'$, large-$U$ corner of the phase diagram.

Figure~\ref{fig:spin} shows the momentum-resolved equal-time spin correlations, complementing Fig.~\ref{fig:undoped}(d).
In the AFM region identified by DMRG, a representative point exhibits a clear peak at $\mathbf{q}=(\pi,\pi)$, as shown in Fig.~\ref{fig:spin}(a).
By contrast, at a representative point in the low-compressibility CSL region identified in Figs.~\ref{fig:undoped}(a)-(c), the dominant spin-correlation peak shifts away from $\mathbf{q}=(\pi,\pi)$ to $\mathbf{q}=(\pi,0)$, as shown in Fig.~\ref{fig:spin}(b).

\subsection{Pairing correlations, susceptibility, and vertex in the doped case}
\label{app:dqmc_pair}

Using the singlet bond pair operator introduced in the main text,
we label an oriented bond as $b=(i_b,j_b)$ and write
\begin{equation}
\hat\Delta_b=\frac{1}{\sqrt{2}}
\left(c_{i_b\up}c_{j_b\dn}
-c_{i_b\dn}c_{j_b\up}\right).
\label{eq:dqmc_bond_pair}
\end{equation}
The pairing channel is specified by
\begin{equation}
\hat\Delta_{\mathrm{pair}}=
\frac{1}{\sqrt{N}}\sum_{\bm r,\ell}
\eta_{\bm r\ell}\,
\hat\Delta_{\bm r\ell},
\qquad
\eta_{\bm r\ell}=e^{-i\phi_{\bm r\ell}}.
\label{eq:dqmc_pair_channel}
\end{equation}
where $N$ is the number of lattice sites, $\bm r=(r_x,r_y)$ labels a lattice coordinate, and $\ell=0,\ldots,3$ labels the four bonds
\begin{align}
b_{\bm r,0}&=(\bm r,\bm r+\hat x),&
b_{\bm r,1}&=(\bm r,\bm r+\hat y),\nonumber\\
b_{\bm r,2}&=(\bm r+\hat x,\bm r+\hat y),&
b_{\bm r,3}&=(\bm r,\bm r+\hat x+\hat y).
\end{align}
The corresponding bond phases are
\begin{equation}
\left(
\phi_{\bm r,0},
\phi_{\bm r,1},
\phi_{\bm r,2},
\phi_{\bm r,3}
\right)
=
\left(
\frac{\pi}{2}+\pi(r_x+r_y),
\pi+\pi r_y,
\frac{3\pi}{4}+\pi r_y,
\frac{5\pi}{4}+\pi r_y
\right)
\quad (\mathrm{mod}\ 2\pi).
\label{eq:dqmc_pair_form_factor}
\end{equation}
These phases specify the form-factor coefficients used in the DQMC projection. They are the bond phases $\phi_\alpha(\bm r)$ of the main text, with each bond labeled by $b_{\bm r,\ell}$ instead of $(\bm r,\alpha)$.

The imaginary-time pair correlation projected onto this channel is
\begin{align}
C_{\mathrm{pair}}(\tau)
&=
\left\langle
T_\tau\hat\Delta_{\mathrm{pair}}^\dagger(\tau)
\hat\Delta_{\mathrm{pair}}(0)
\right\rangle \nonumber\\
&=
\frac{1}{N}
\sum_{b,c}
\eta_b^*\eta_c
\left\langle
T_\tau\hat\Delta_b^\dagger(\tau)
\hat\Delta_c(0)
\right\rangle,
\label{eq:dqmc_pair_correlation}
\end{align}
where $b$ and $c$ run over all four bond types associated with every lattice site.

For the finite-temperature analysis, we use the midpoint pair correlator to define
\begin{equation}
\chi_{\mathrm{pair}}(\beta/2)
\equiv
\chi_{\mathrm{pair}}
\equiv
\beta C_{\mathrm{pair}}(\beta/2),
\label{eq:pair_midpoint}
\end{equation}
where $\beta=1/T$ is the inverse temperature.
This midpoint quantity preferentially probes low-energy pairing correlations~\cite{wang2025finite,Chen2026TriangularHofstadterSC}.
We suppress the explicit argument $\beta/2$ below when no confusion can arise.
The midpoint quantity should be distinguished from the conventional static susceptibility
\begin{equation}
\chi_{\mathrm{pair}}^{\mathrm{stat}}
=
\int_0^\beta d\tau\,C_{\mathrm{pair}}(\tau),
\end{equation}
which is not the quantity used in this work.

To isolate the contribution beyond the particle-particle bubble, we also calculate
\begin{equation}
\bar{\chi}_{\mathrm{pair}}(\beta/2)
\equiv
\bar{\chi}_{\mathrm{pair}}
\equiv
\beta\bar C_{\mathrm{pair}}(\beta/2),
\end{equation}
where
\begin{equation}
\bar C_{\mathrm{pair}}(\beta/2)
=
\frac{1}{N}
\sum_{b,c}\eta_b^*\eta_c\,
\bar C_{bc}(\beta/2).
\end{equation}
For $b=(i_b,j_b)$ and $c=(i_c,j_c)$, the Wick-factorized midpoint bond correlator is
\begin{align}
\bar C_{bc}(\beta/2)
=
\frac{1}{2}\Big[
&\mathcal G^\up_{i_ci_b}(\beta/2)
 \mathcal G^\dn_{j_cj_b}(\beta/2)
+
 \mathcal G^\up_{j_ci_b}(\beta/2)
 \mathcal G^\dn_{i_cj_b}(\beta/2)
\nonumber\\
&+
 \mathcal G^\up_{i_cj_b}(\beta/2)
 \mathcal G^\dn_{j_ci_b}(\beta/2)
+
 \mathcal G^\up_{j_cj_b}(\beta/2)
 \mathcal G^\dn_{i_ci_b}(\beta/2)
\Big],
\label{eq:pair_bubble}
\end{align}
with
\begin{equation}
\mathcal G^\sigma_{ij}(\tau)
=
\left\langle
c_{i\sigma}(\tau)c_{j\sigma}^\dagger(0)
\right\rangle
\end{equation}
the fully interacting, spin-resolved single-particle propagator.
We use $\mathcal G$ here to distinguish this unequal-time propagator from the equal-time, spin-summed $G(r)$ defined in the main text.
Thus, $\bar{\chi}_{\mathrm{pair}}$ is the no-vertex, dressed particle-particle bubble.
Operationally, the full DQMC correlator forms the Wick products on each auxiliary-field configuration before the Monte Carlo average,
whereas $\bar C_{\mathrm{pair}}$ is constructed from products of the separately averaged propagators.
We then define the dimensionless pairing-vertex indicator,
\begin{align}
\Gamma_{\mathrm{pair}}\bar{\chi}_{\mathrm{pair}}
&\equiv\left[\chi_{\mathrm{pair}}^{-1}(\beta/2) - \bar{\chi}_{\mathrm{pair}}^{-1}(\beta/2) \right] \bar{\chi}_{\mathrm{pair}}(\beta/2) \nonumber\\
&= \frac{\bar{\chi}_{\mathrm{pair}}(\beta/2)}{\chi_{\mathrm{pair}}(\beta/2)} -1.
\label{eq:pair_vertex}
\end{align}
Accordingly, $\Gamma_{\mathrm{pair}}\bar{\chi}_{\mathrm{pair}}<0$ indicates an enhancement of the pairing response relative to the dressed bubble.

\begin{figure}[t]
\centering
\includegraphics[width=\columnwidth]{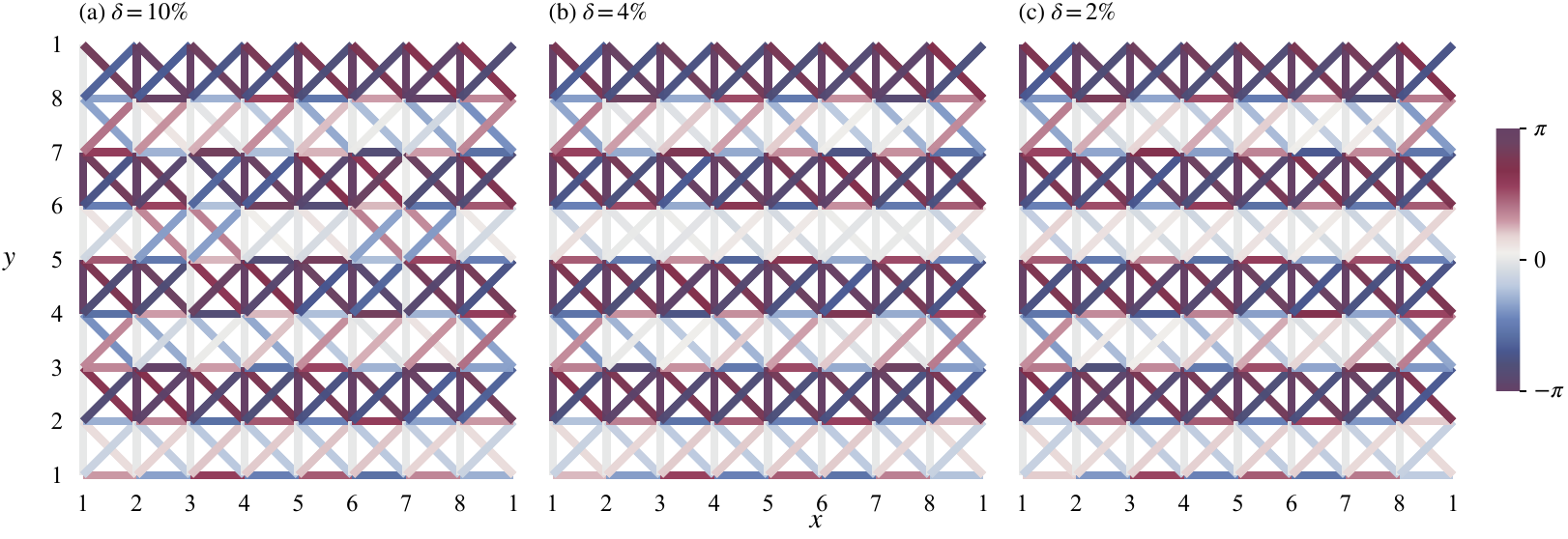}
\caption{Phase patterns of the bond-resolved DQMC midpoint correlator $C_{b_0,b}(\beta/2)$ on $8\times8$ lattices for $U/t=6$, $t'/t=0.75$, and $T/t=1/3$, at (a) $\delta=10\%$, (b) $\delta=4\%$, and (c) $\delta=2\%$. Bond color represents $\arg C_{b_0,b}(\beta/2)$, with $b_0=b_{\bm r_0,1}$ and $\bm r_0=(1,1)$.
A common linewidth is used for all bonds in all three panels.}
\label{fig:8x8phase}
\end{figure}

\begin{figure}[t]
\centering
\includegraphics[width=0.9\columnwidth]{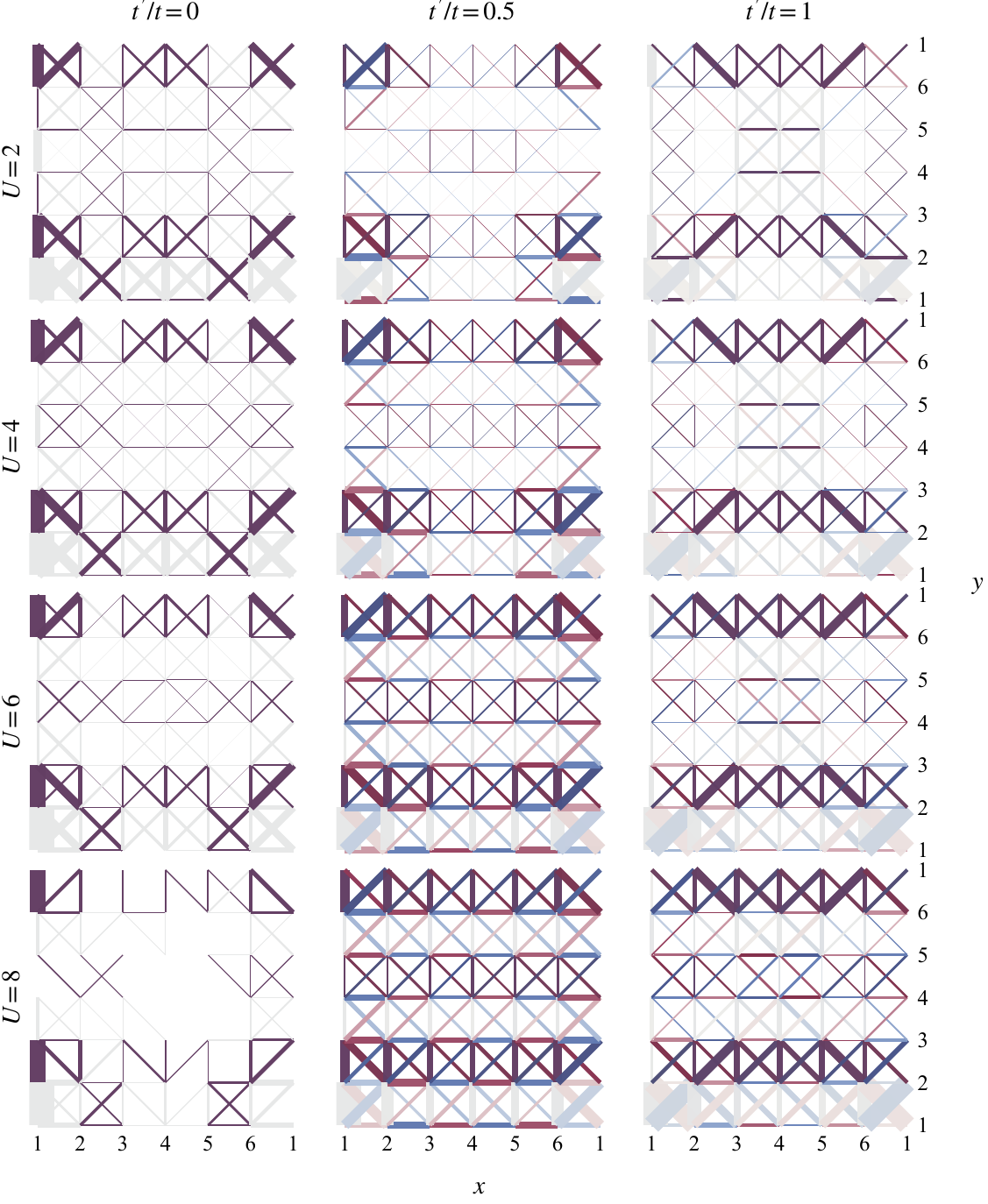}
\caption{Phase and magnitude patterns of the bond-resolved DQMC midpoint correlator
$C_{b_0,b}(\beta/2)$ on $6\times6$ lattices at $\delta=4\%$ and $T/t=1/4$.
Rows correspond to $U/t=2$, $4$, $6$, and $8$, while columns correspond to $t'/t=0$, $0.5$, and $1$.
The same reference bond $b_0$ and phase scale as in Fig.~\ref{fig:8x8phase} are used in every panel.
Line thickness is proportional to $\left|C_{b_0,b}(\beta/2)\right|$, normalized by the largest bond magnitude separately in each panel, and therefore encodes relative magnitude within a panel.
Bonds with $\left|C_{b_0,b}(\beta/2)\right|\leq2\sigma_C$ are omitted, where $\sigma_C$ is the jackknife standard error.}
\label{fig:phases}
\end{figure}

\begin{figure}[t]
\centering
\includegraphics[width=0.9\columnwidth]{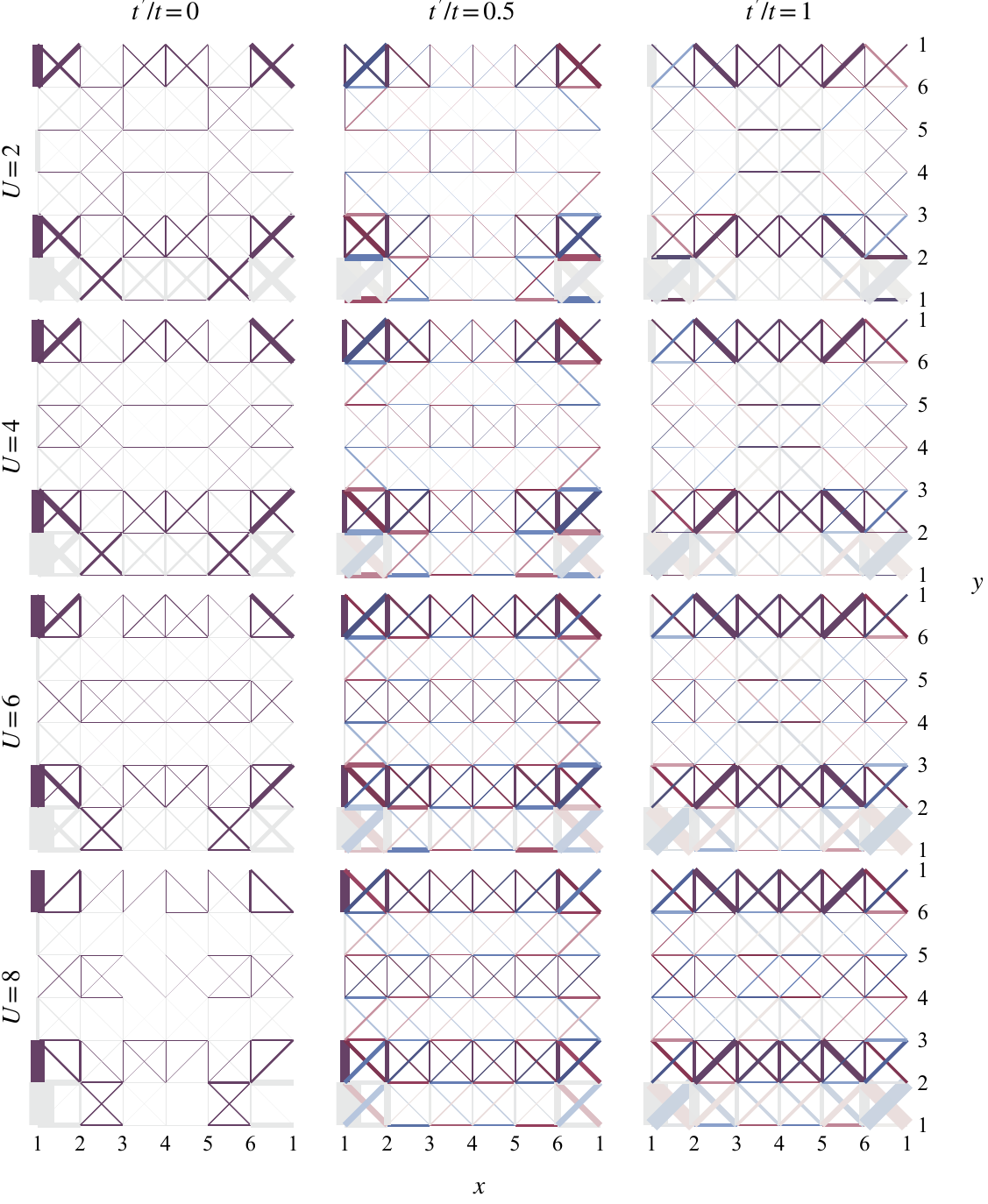}
\caption{The same analysis as in Fig.~\ref{fig:phases} but for $T/t=1/3$.}
\label{fig:phases3}
\end{figure}

We first examine in real space the bond-resolved midpoint correlator relative to a fixed reference bond $b_0$,
\begin{equation}
C_{b_0,b}(\beta/2)
=
\left\langle
T_\tau
\hat\Delta_{b_0}^\dagger(\beta/2)
\hat\Delta_b(0)
\right\rangle,
\label{eq:dqmc_bond_phase}
\end{equation}
as shown in Figs.~\ref{fig:8x8phase}, \ref{fig:phases}, and \ref{fig:phases3}.
Figure~\ref{fig:8x8phase} compares the DQMC pairing-phase pattern at $U/t=6$, $t'/t=0.75$, and $T/t=1/3$ for dopings $\delta=10\%$, $4\%$, and $2\%$. The characteristic pattern is retained throughout this doping range and remains consistent with the DMRG result at $\delta=1/12$ in Fig.~\ref{fig:doped}(a), although localized phase variations become more visible at the largest doping. Because the average fermion sign decreases rapidly with increasing doping, we use $\delta=4\%$ for the broader $U$-$t'$ and temperature scans.
This choice allows us to access the $U/t=6$ and $8$ cases down to $T/t=1/4$ on the $6\times6$ lattice at the $t'$ values considered below.

Figures~\ref{fig:phases} and \ref{fig:phases3} show how the phase pattern evolves with $U$ and $t'$ at fixed doping $\delta=4\%$ on a $6\times6$ lattice. The characteristic pattern is best developed at larger $U$ and nonzero $t'$, where the correlations also extend over longer distances. Comparing the two figures shows that lowering the temperature from $T/t=1/3$ to $1/4$ makes the correlation pattern more spatially extended and better resolved.

\begin{figure*}[t]
 \centering
 \includegraphics[width=\textwidth]{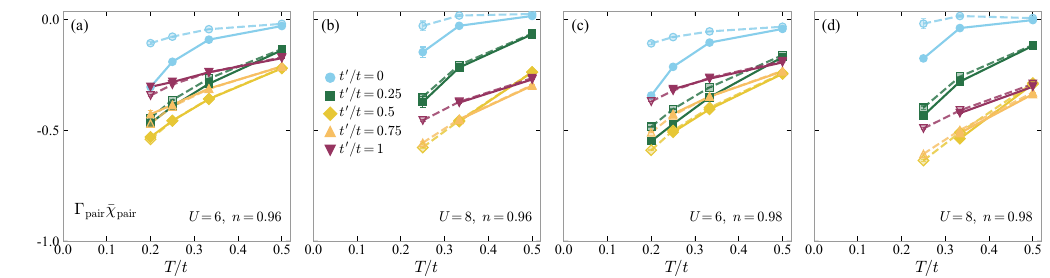}
 \caption{
 The dimensionless pairing-vertex indicator
 $\bar{\chi}_{\mathrm{pair}}(\beta/2)/\chi_{\mathrm{pair}}(\beta/2)-1$
 for (a) $U/t=6$, $\delta=4\%$;
 (b) $U/t=8$, $\delta=4\%$;
 (c) $U/t=6$, $\delta=2\%$; and
 (d) $U/t=8$, $\delta=2\%$.
 Colors and symbols distinguish the values of $t'/t$. Solid lines with filled markers denote the $8\times8$
 lattice, whereas dashed lines with semitransparent markers denote the $6\times6$ lattice. Error bars are jackknife standard errors.}
 \label{fig:Tdependence_vertex}
\end{figure*}

\begin{figure*}[t]
 \centering
 \includegraphics[width=\textwidth]{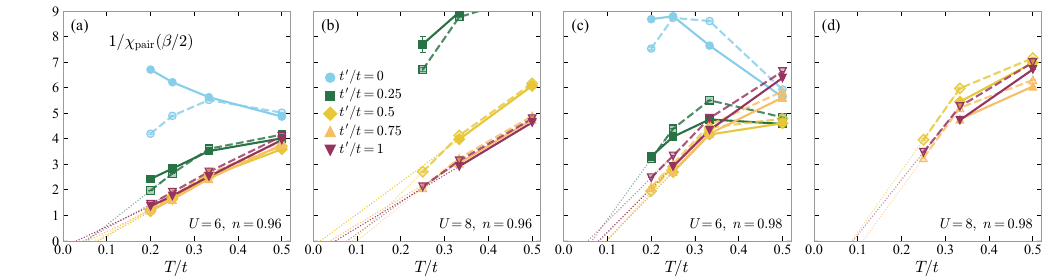}
 \caption{
 The inverse pair susceptibility, $1/[\chi_{\mathrm{pair}}(\beta/2)]$, is shown for
 (a) $U/t=6$, $\delta=4\%$;
 (b) $U/t=8$, $\delta=4\%$;
 (c) $U/t=6$, $\delta=2\%$; and
 (d) $U/t=8$, $\delta=2\%$.
 Colors and symbols distinguish $t'/t$. Solid lines with filled markers correspond to the $8\times8$ lattice, while dashed lines with semitransparent markers correspond to the $6\times6$ lattice.
 Dotted lines show linear extrapolations based on the two lowest available temperatures.}
 \label{fig:Tdependence_invchi}
\end{figure*}

We next examine the temperature dependence of the phase-projected pairing response.
Figures~\ref{fig:Tdependence_vertex} and \ref{fig:Tdependence_invchi} show that, at nonzero $t'$, $\Gamma_{\mathrm{pair}}\bar{\chi}_{\mathrm{pair}}$ generally becomes more negative and $1/\chi_{\mathrm{pair}}$ decreases upon cooling, consistent with an enhanced interaction-induced pairing tendency and an approach to a superconducting instability.
The $6\times6$ and $8\times8$ lattices exhibit the same qualitative trends.
At fixed $U$ and $t'$, $\chi_{\mathrm{pair}}$ is generally smaller at $\delta=2\%$ than at $4\%$.
Where a two-point extrapolation of $1/\chi_{\mathrm{pair}}$ crosses zero below the fitted temperature window, we use the root as an estimated pairing scale $T_0$ that is shown in Fig.~\ref{fig:dqmcmain}(b) in the main text.

\clearpage
\section{Additional DMRG results}
\label{app:dmrg_supp}

In this section we present additional DMRG results that complement the phase diagrams of the main text. All simulations use the $SU(2)$-symmetric cylinder setup described there, with $L_y=6$ legs except in \secref{app:dmrg_ly34}, and all doped results are taken at hole doping $\delta=1/12$. We first show the momentum-resolved spin structure factor of the three half-filled parent phases and the spin gap along a cut through the IQH-CSL transition. We then characterize the doped phases through the spin structure factor, the charge density profiles, the scalar spin chirality, the spin Chern number, and the electron and spin correlation lengths across the $U$-$t'$ plane, check the pair binding on three- and four-leg cylinders, and finally examine the correlation functions at two representative points, a chiral metal at weak coupling and the time-reversal-symmetric point $t'=0$.

All correlation functions labeled $M\to\infty$ are extrapolated in the bond dimension. At each parameter point we take the longest available cylinder and compute the correlator at three to five bond dimensions between $M=4000$ and $16000$, typically $M=8000$ to $14000$. At each separation $r$ we fit the magnitude of the correlator, for example $|P_{yy}(r)|$ or $|G(r)|^2$, to a quadratic polynomial in $1/M$, $O_M(r)=O_\infty(r)+a(r)/M+b(r)/M^2$, by least squares, and take the intercept $O_\infty(r)$ as the extrapolated value. The power-law exponents $K$ and correlation lengths $\xi$ quoted in the main text and below are obtained from fits to these extrapolated curves. A pair correlation $|P_{yy}(r)|\sim r^{-K_P}$ implies a pairing susceptibility $\chi_{\rm SC}(T)\sim T^{-(2-K_P)}$ that diverges as $T\to0$ for $K_P<2$~\cite{Giamarchi2004}, the criterion adopted in earlier DMRG studies~\cite{Jiang2021npjDopedQSL} and used here to identify the SC phase.

\subsection{Spin structure factor at half filling and at $\delta=1/12$}
\label{app:dmrg_half}

\begin{figure}[t]
\centering
\includegraphics[width=0.86\textwidth]{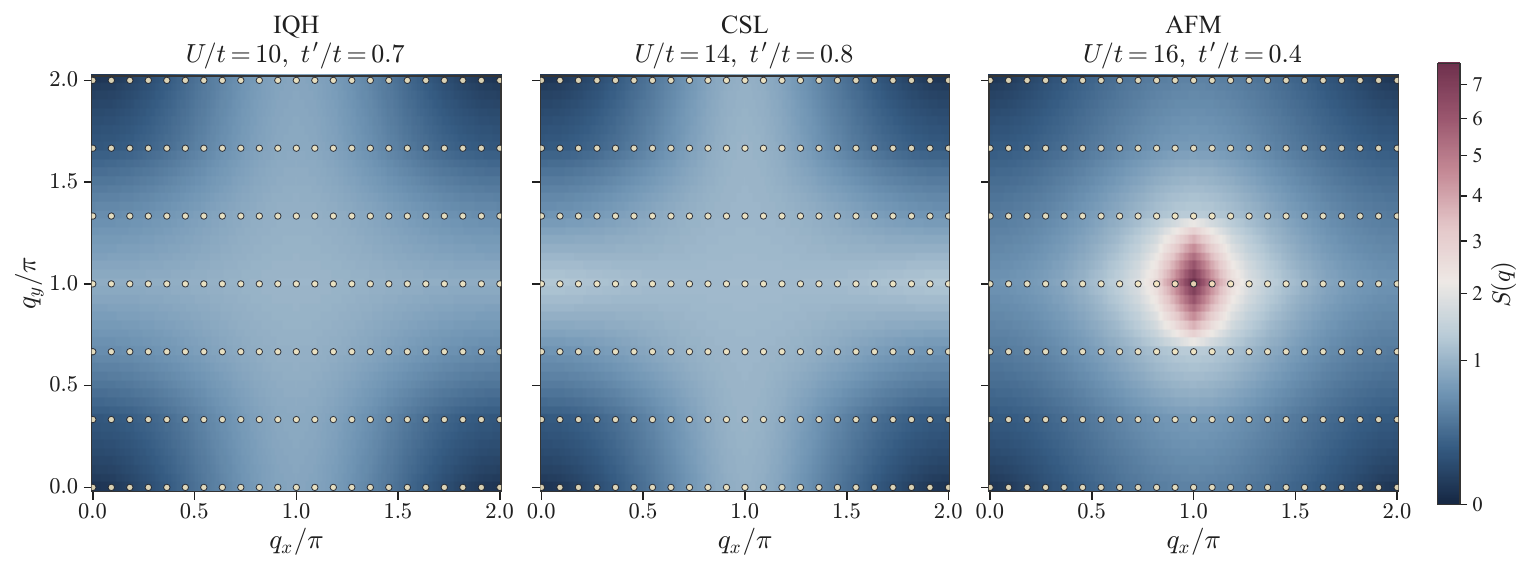}
\caption{Spin structure factor $S(\bm q)$ at half filling, measured by DMRG on $24\times6$ cylinders at representative points of the three phases of \figref{fig:undoped}(a), the IQH phase at $U=10$, $t'/t=0.7$, the CSL at $U=14$, $t'/t=0.8$, and the AFM at $U=16$, $t'/t=0.4$. The three panels share the color scale on the right, running from dark blue at small $S(\bm q)$ to dark red at large $S(\bm q)$. White circles mark the discrete cylinder momenta, between which the maps are interpolated.}
\label{fig:halfsq}
\end{figure}

At half filling the phase boundaries of \figref{fig:undoped}(a) are located by the charge and spin Chern numbers, and the spin structure factor $S(\bm q)$ defined in the main text provides an independent check of the resulting phase assignment. Figure~\ref{fig:halfsq} shows $S(\bm q)$ at a representative point of each phase, measured on $24\times6$ cylinders. In the IQH phase and in the CSL the response is broad and featureless, with no dominant peak anywhere in the Brillouin zone, consistent with short-ranged spin correlations in both phases. In the AFM region, by contrast, $S(\bm q)$ develops a pronounced peak at the N\'eel wave vector $\bm q=(\pi,\pi)$. These ground-state maps complement the finite-temperature DQMC spin correlations of \figref{fig:spin} and support the phase assignment used in the main text.

\begin{figure}[t]
\centering
\includegraphics[width=0.78\textwidth]{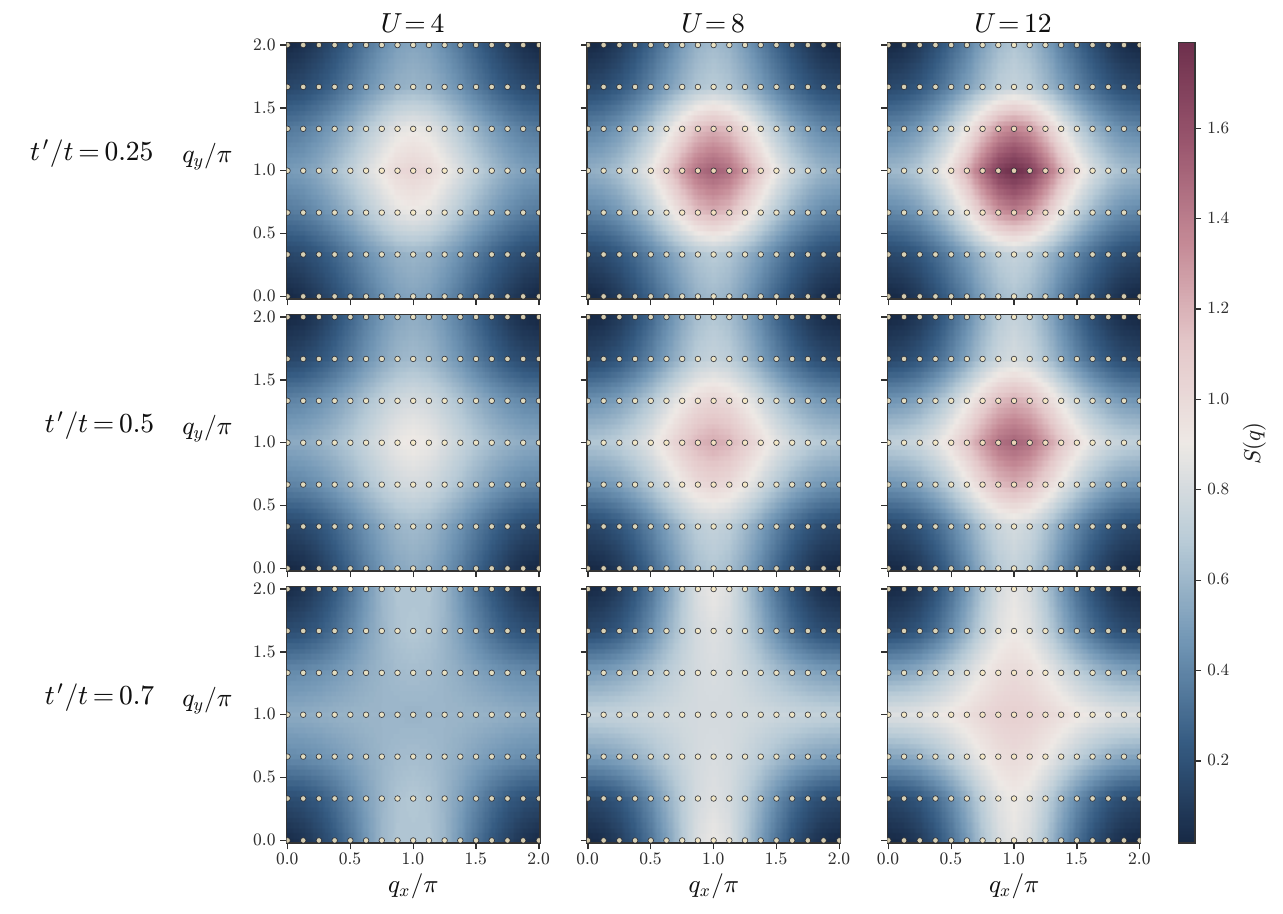}
\caption{Spin structure factor $S(\bm q)$ at $\delta=1/12$ on $32\times6$ cylinders, for $t'/t=0.25$, $0.5$, and $0.7$ (rows) and $U=4$, $8$, and $12$ (columns). All nine panels share the color scale on the right. White circles mark the discrete cylinder momenta, between which the maps are interpolated.}
\label{fig:dopedsq}
\end{figure}

Figure~\ref{fig:dopedsq} shows the spin structure factor of the doped ground states on $32\times6$ cylinders, for $t'/t=0.25$, $0.5$, and $0.7$ and $U=4$, $8$, and $12$. A broad maximum at $\bm q=(\pi,\pi)$ is present throughout and strengthens monotonically as $U$ grows and as $t'$ decreases, following the proximity of the underlying half-filled state to the AFM phase. Even where it is strongest, at $U=12$ and $t'/t=0.25$, the peak remains broad and its height stays far below the N\'eel peak of the undoped AFM in \figref{fig:halfsq}, consistent with the short spin correlation lengths mapped in \secref{app:dmrg_xi}.

\subsection{Spin gap at half filling}
\label{app:dmrg_gap}

\begin{figure}[t]
\centering
\includegraphics[width=0.42\textwidth]{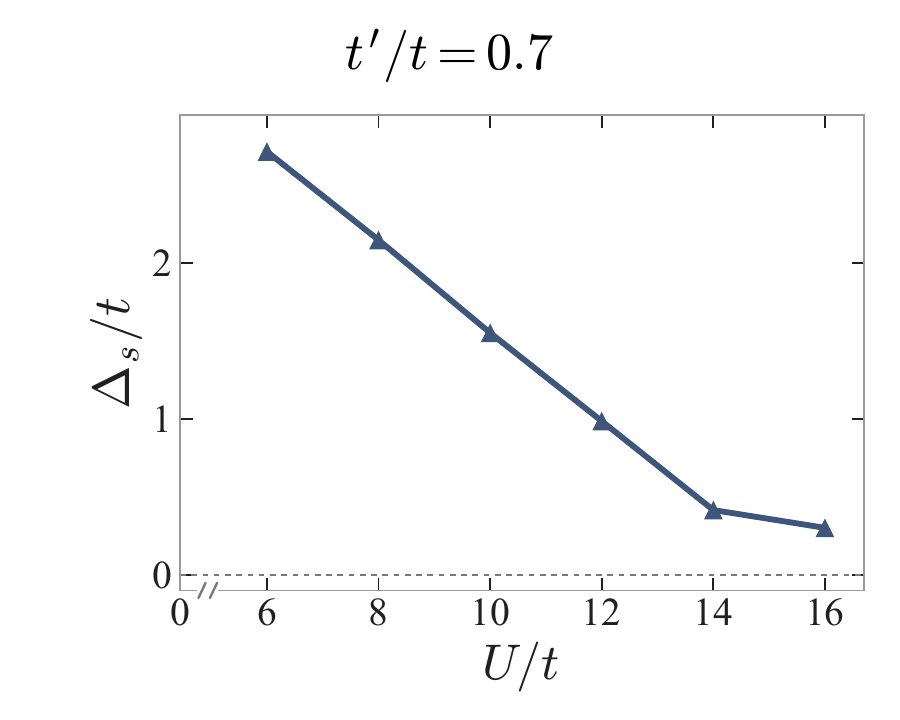}
\caption{Spin gap at half filling, $\Delta_s=E_0(N,S=1)-E_0(N,S=0)$, from DMRG at $t'/t=0.7$ as a function of $U$.}
\label{fig:spingap}
\end{figure}

The DQMC compressibility and staggered spin structure factor of \figref{fig:undoped}(c),(d) probe the charge and spin gaps through their finite-temperature fluctuations. DMRG gives direct ground-state access to the same two gaps along the cut $t'/t=0.7$, which crosses the IQH-CSL boundary of \figref{fig:undoped}(a) near $U\approx12.5$. The charge sector is characterized by the two-electron binding energy $E_b$ defined in the main text and shown in the inset of \figref{fig:undoped}(a). For the spin sector we compute the spin gap at half filling, $\Delta_s=E_0(N,S=1)-E_0(N,S=0)$, the difference between the lowest energies in the total-spin $S=1$ and $S=0$ sectors at fixed electron number $N_e=N$, where $E_0(N,S=0)=E_0(N)$ is the half-filled ground-state energy entering $E_b$. Figure~\ref{fig:spingap} shows $\Delta_s$ for $U=6$--$16$. The spin gap is large at weak coupling, decreases monotonically with $U$, and remains finite over the whole range, including across the IQH-CSL transition where $E_b$ is most negative. The two gaps thus behave differently at the transition: the charge sector softens, favoring the injection of charge as Cooper pairs, while the spin sector stays gapped, consistent with the DQMC trends in the main text and with the critical theory of \secref{app:su2k}.

\subsection{Charge density profiles and scalar spin chirality at $\delta=1/12$}
\label{app:dmrg_density}

\begin{figure}[t]
\centering
\includegraphics[width=0.72\textwidth]{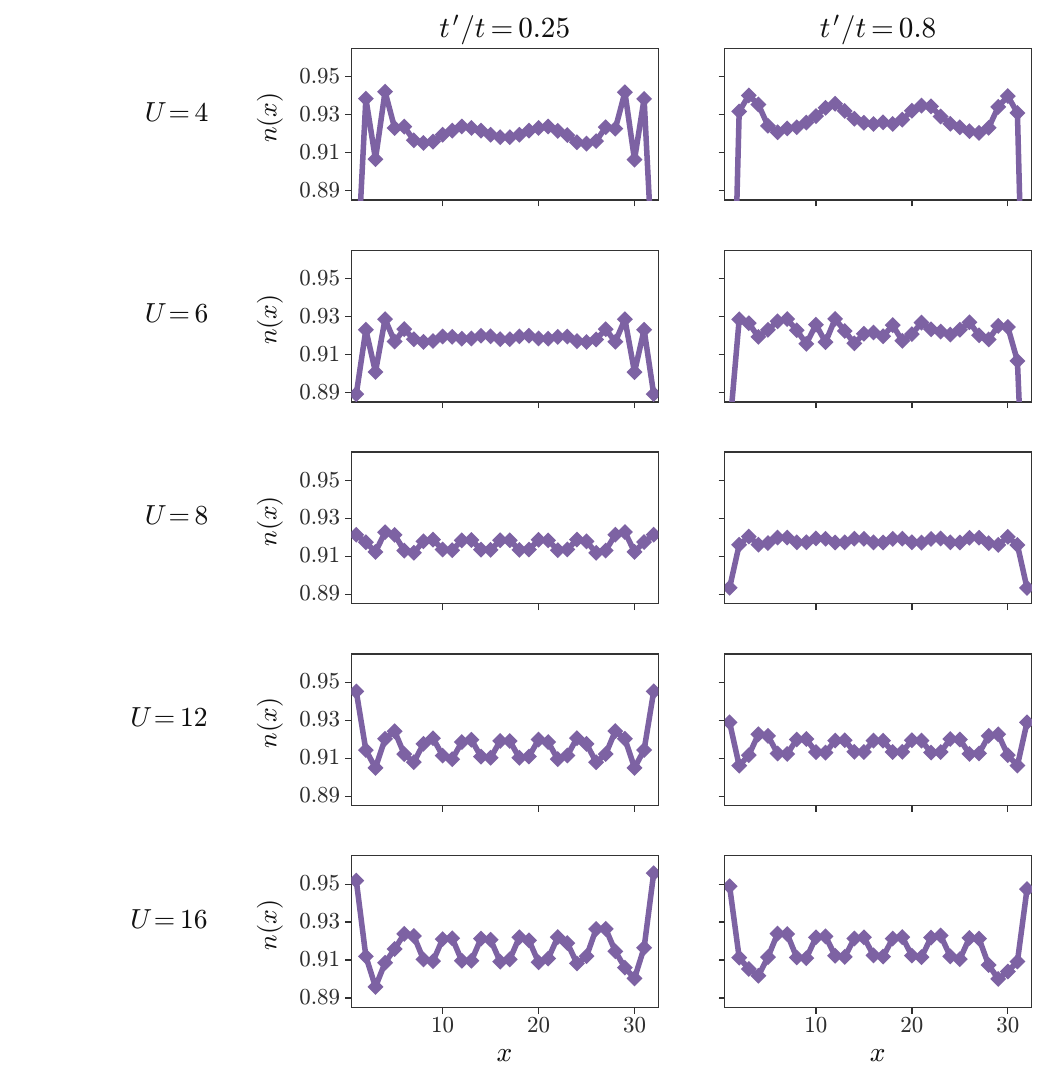}
\caption{Charge density profiles $n(x)$ along a fixed row of the $32\times6$ cylinder at $\delta=1/12$ (purple diamonds), for $t'/t=0.25$ (left column) and $t'/t=0.8$ (right column) and, from top to bottom, $U=4$, $6$, $8$, $12$, and $16$, each at the largest bond dimension available. All panels share the same axes, so the modulation amplitudes can be compared directly across panels.}
\label{fig:densityprof}
\end{figure}

Figure~\ref{fig:densityprof} shows the site density along a fixed row of the cylinder, $n(x)=\sum_\sigma\langle n_{(x,y_0)\sigma}\rangle$. We compare $t'/t=0.25$ and $0.8$ for $U$ between $4$ and $16$ on $L_x=32$ cylinders. At weak coupling the profile is essentially flat away from the open ends. As $U$ increases, a periodic modulation of the density develops around the mean value $\bar n=1-\delta$ and becomes increasingly well defined, in the same way for both values of $t'$. The modulation amplitude remains small, of order $0.01$, throughout the accessible parameter range.

\begin{figure}[t]
\centering
\includegraphics[width=0.45\textwidth]{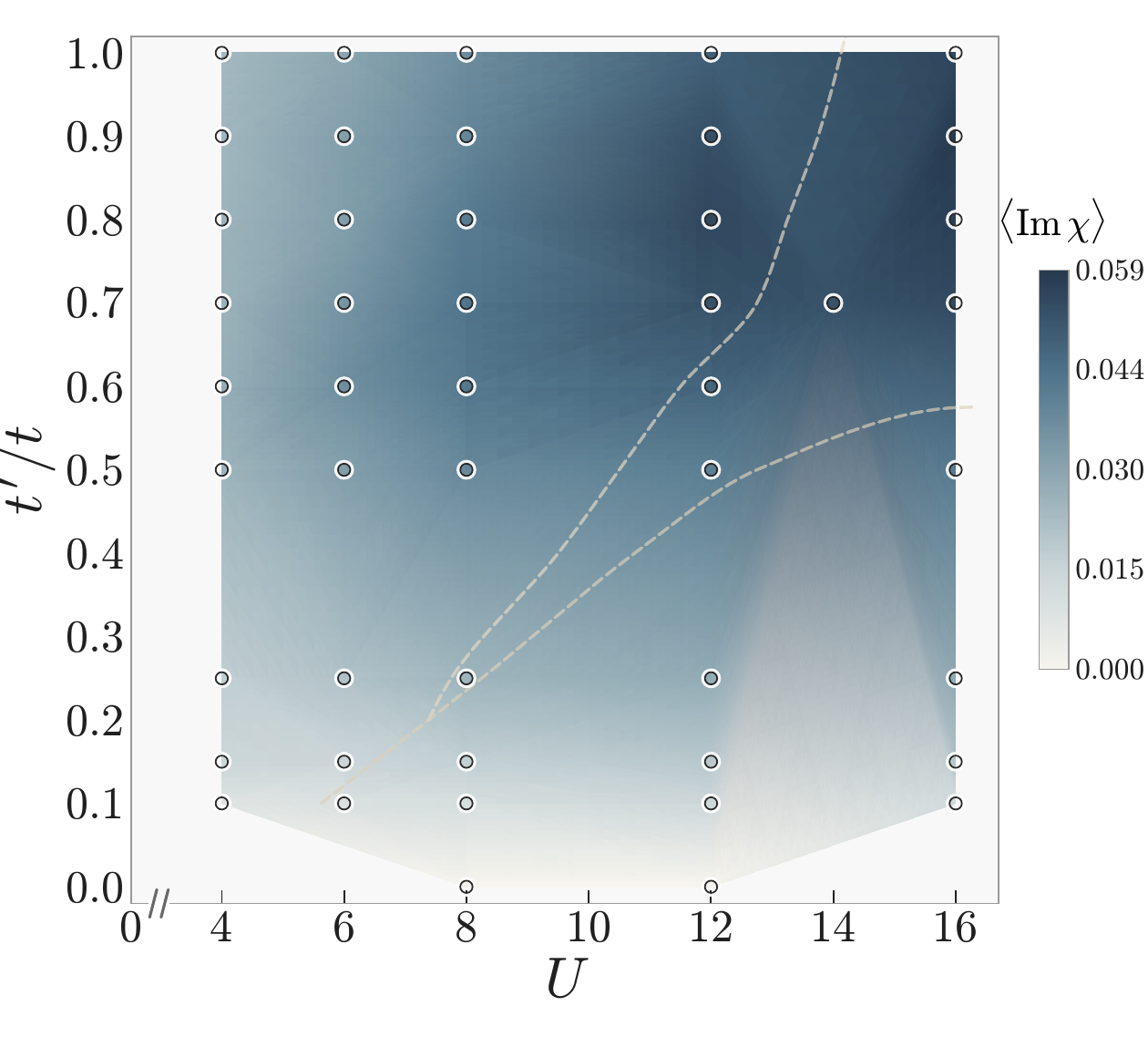}
\caption{Triangle-averaged scalar spin chirality $\langle\mathrm{Im}\,\chi\rangle$ of \eqnref{eq:chirality} at $\delta=1/12$ as a function of $U$ and $t'/t$, measured on the longest available cylinder at each simulated point (circles) and interpolated between them, with darker blue marking larger chirality. The dashed white curves reproduce the half-filling phase boundaries of \figref{fig:undoped}(a).}
\label{fig:chiralorder}
\end{figure}

The imaginary NNN hopping threads flux $\pm\pi/2$ through the elementary triangles of the lattice (\secref{app:model}), so for $t'\neq0$ time reversal is broken explicitly and the doped ground states can sustain a nonzero scalar spin chirality. On every elementary triangle, formed by two NN bonds and one NNN bond, we measure
\begin{equation}
\chi_{ijk}=\big\langle\bm S_i\cdot\big(\bm S_j\times\bm S_k\big)\big\rangle,
\label{eq:chirality}
\end{equation}
with the three sites $(i,j,k)$ ordered anticlockwise, and average the result over all triangles of the cylinder. Figure~\ref{fig:chiralorder} maps $\langle\mathrm{Im}\,\chi\rangle$ over the doped phase diagram. The chirality is nonzero at every simulated point with $t'\neq0$ and grows with $t'$, reaching its largest values toward the large-$t'$ side of the phase diagram. It diminishes with decreasing $t'$, and at $t'=0$, where time reversal is restored, the measured chirality is consistent with zero.

\subsection{\texorpdfstring{Spin Chern number at $\delta=1/12$}{Spin Chern number at delta=1/12}}
\label{app:dmrg_chern}

\begin{figure}[t]
\centering
\includegraphics[width=0.45\textwidth]{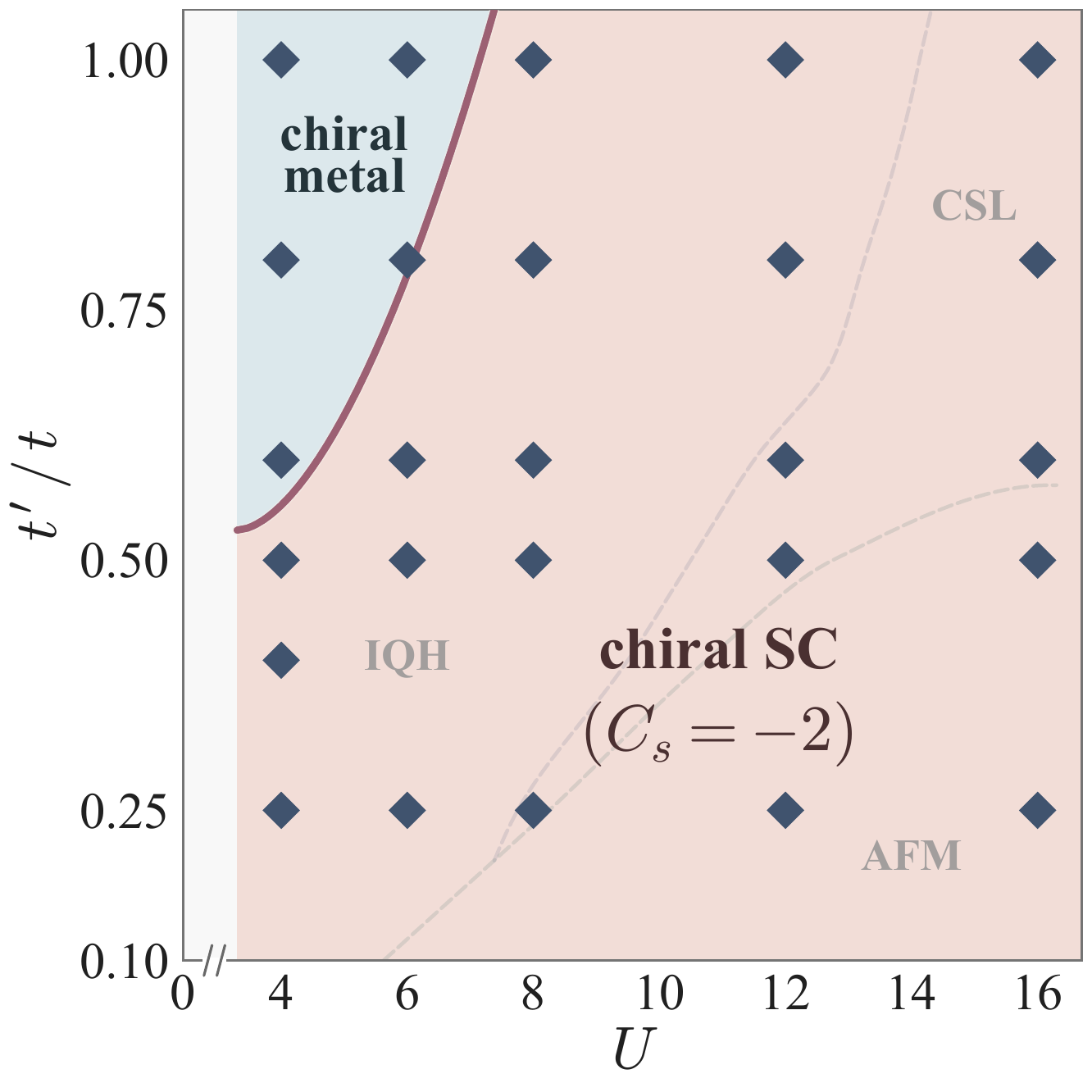}
\caption{Parameter points at which the spin Chern number is measured by spin-flux insertion at $\delta=1/12$ (filled diamonds), shown on the doped DMRG phase diagram of \figref{fig:model}(b). The dashed curves reproduce the half-filling phase boundaries of \figref{fig:undoped}(a).}
\label{fig:dopedchern}
\end{figure}

We measure the spin Chern number of the doped states by the same spin-flux insertion as at half filling, threading $\theta_F$ from $0$ to $2\pi$ and recording the spin $\Delta Q_s$ pumped to the cylinder edge. Figure~\ref{fig:dopedchern} marks the 26 parameter points at which we carry out this measurement. They cover $U=4$ to $16$ and $t'/t=0.25$ to $1$, and hence the doped IQH, CSL, and AFM regions alike.

\subsection{Electron and spin correlation lengths}
\label{app:dmrg_xi}

\begin{figure}[t]
\centering
\includegraphics[width=0.8\textwidth]{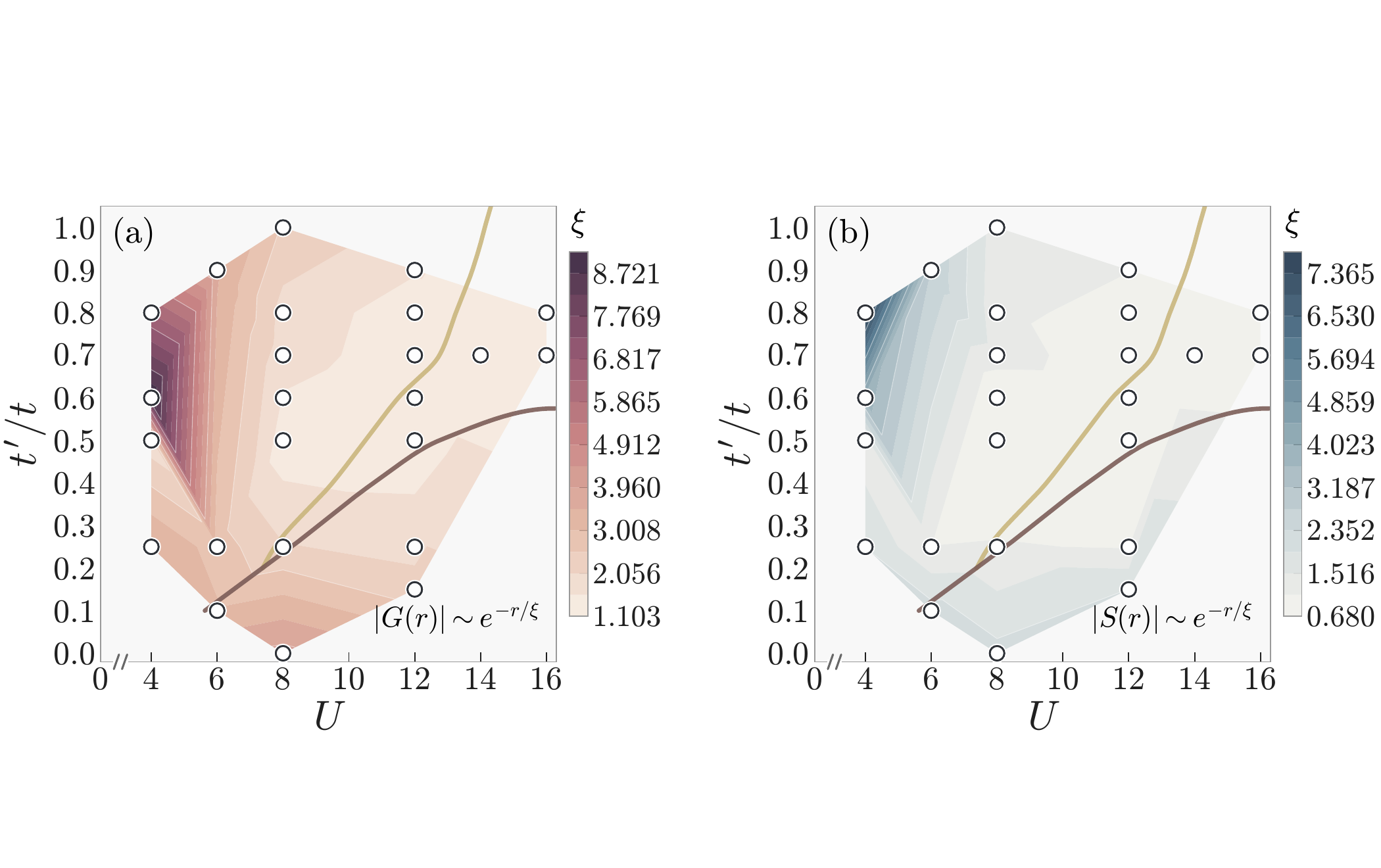}
\caption{Correlation lengths from exponential fits to the $M\to\infty$ extrapolated correlations at $\delta=1/12$, for (a) the single-particle Green's function, $|G(r)|\sim e^{-r/\xi}$, and (b) the spin correlation of \eqnref{eq:spincorr}, $|S(r)|\sim e^{-r/\xi}$. Open circles mark the simulated points, between which the maps are interpolated, with darker colors marking longer correlation lengths. The gold and brown curves reproduce the half-filling phase boundaries of \figref{fig:undoped}(a).}
\label{fig:xiphase}
\end{figure}

To follow the single-electron and spin sectors across the doped phase diagram we consider, in addition to $G(r)$ defined in the main text, the real-space spin correlation along the cylinder,
\begin{equation}
S(r)=\big\langle\bm S(x_0,y_0)\cdot\bm S(x_0+r,y_0)\big\rangle.
\label{eq:spincorr}
\end{equation}
At every simulated point we extrapolate both correlators to $M\to\infty$ quadratically in $1/M$, and fit the extrapolated curves to the exponential forms $|G(r)|\sim e^{-r/\xi}$ and $|S(r)|\sim e^{-r/\xi}$. Figure~\ref{fig:xiphase} shows the two correlation lengths across the $U$-$t'$ plane. The electron correlation length in \figref{fig:xiphase}(a) is short, a few lattice spacings, over most of the phase diagram, consistent with gapped single-electron excitations in the chiral superconductor. Only at small $U$ and $t'/t\gtrsim0.6$ does it grow markedly; there the correlations decay slowly and the fitted $\xi$ becomes large, signaling behavior closer to a power law and matching the chiral-metal region of \figref{fig:model}(b), which is analyzed in \secref{app:dmrg_metal}. The spin correlation length in \figref{fig:xiphase}(b) is short essentially everywhere, so at $\delta=1/12$ the doping has destroyed the N\'eel order of the undoped AFM region, and only in the same weak-coupling, large-$t'$ corner is it somewhat enhanced.

\subsection{Pair binding on three- and four-leg cylinders}
\label{app:dmrg_ly34}

\begin{figure}[t]
\centering
\includegraphics[width=0.95\textwidth]{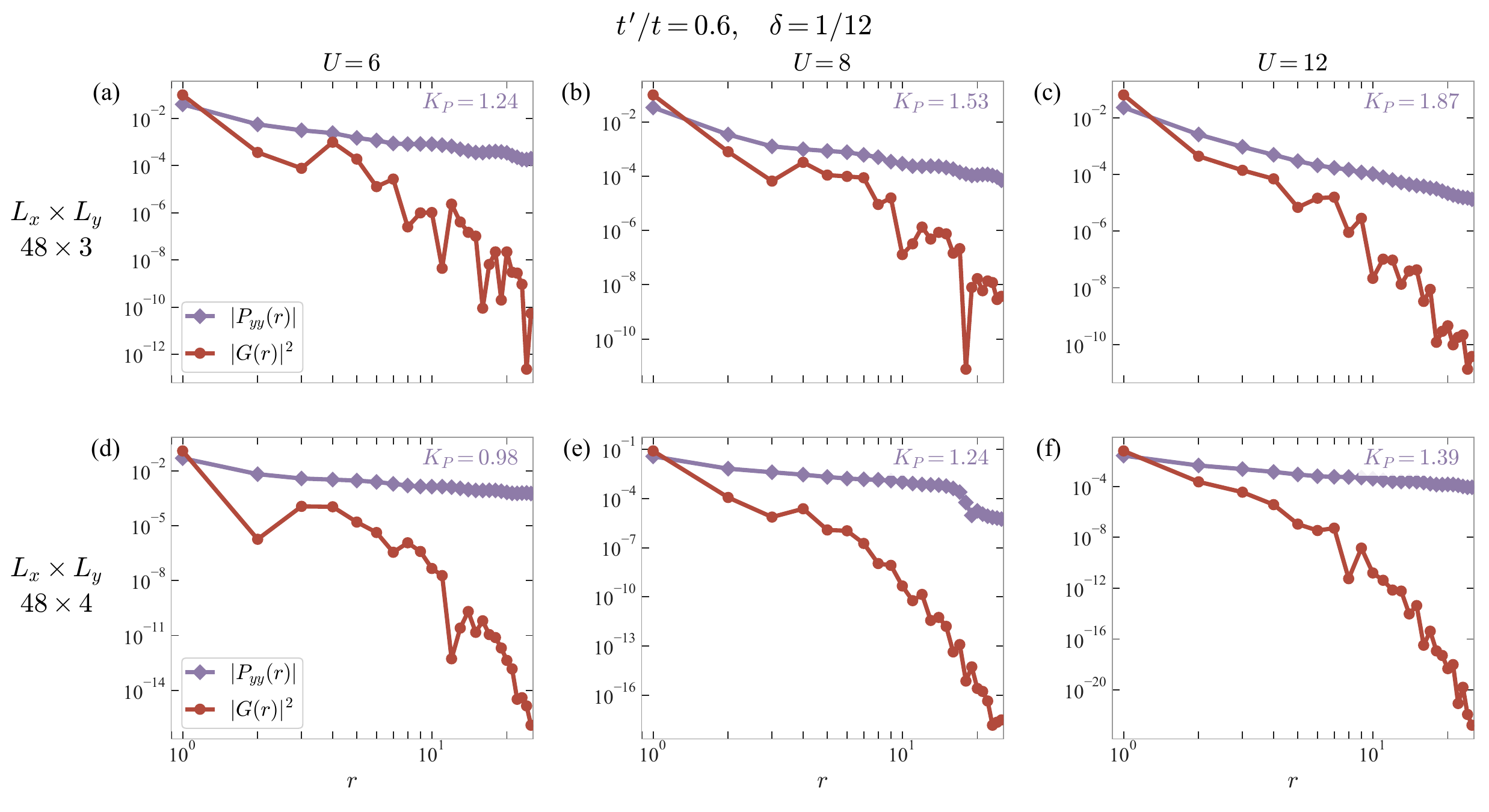}
\caption{Pair-field correlation $|P_{yy}(r)|$ (purple diamonds) and squared single-particle Green's function $|G(r)|^{2}$ (red circles) at $t'/t=0.6$ and $\delta=1/12$ on (a)--(c) $48\times3$ and (d)--(f) $48\times4$ cylinders, for $U=6$, $8$, and $12$ (columns), at the largest bond dimension reached on each cylinder, $M=4000$ for $L_y=3$ and $M=8000$ for $L_y=4$, without extrapolation in $M$. $K_P$ is the exponent of a power-law fit $|P_{yy}(r)|\sim r^{-K_P}$ over intermediate distances.}
\label{fig:ly34pair}
\end{figure}

To check that the pair binding does not rely on the six-leg cylinder width used above, we repeat the comparison between the pair-field correlation $|P_{yy}(r)|$ and the squared single-particle Green's function $|G(r)|^{2}$ of \figref{fig:doped}(b) on $L_y=3$ and $L_y=4$ cylinders of length $L_x=48$, at $t'/t=0.6$, $\delta=1/12$, and $U=6$, $8$, and $12$. Figure~\ref{fig:ly34pair} shows the data at the largest bond dimension reached on each cylinder, $M=4000$ for $L_y=3$ and $M=8000$ for $L_y=4$, without extrapolation in $M$. In all six cases $|P_{yy}(r)|$ follows a slow power law with $K_P<2$, while $|G(r)|^{2}$ falls off by many orders of magnitude over the same distances. The pair correlation therefore dominates the single-electron channel on the narrower cylinders just as on $L_y=6$, so the pair binding and the superconducting quasi-long-range order are robust against the cylinder width.

\subsection{Correlation functions in the chiral metal}
\label{app:dmrg_metal}

\begin{figure}[t]
\centering
\includegraphics[width=\textwidth]{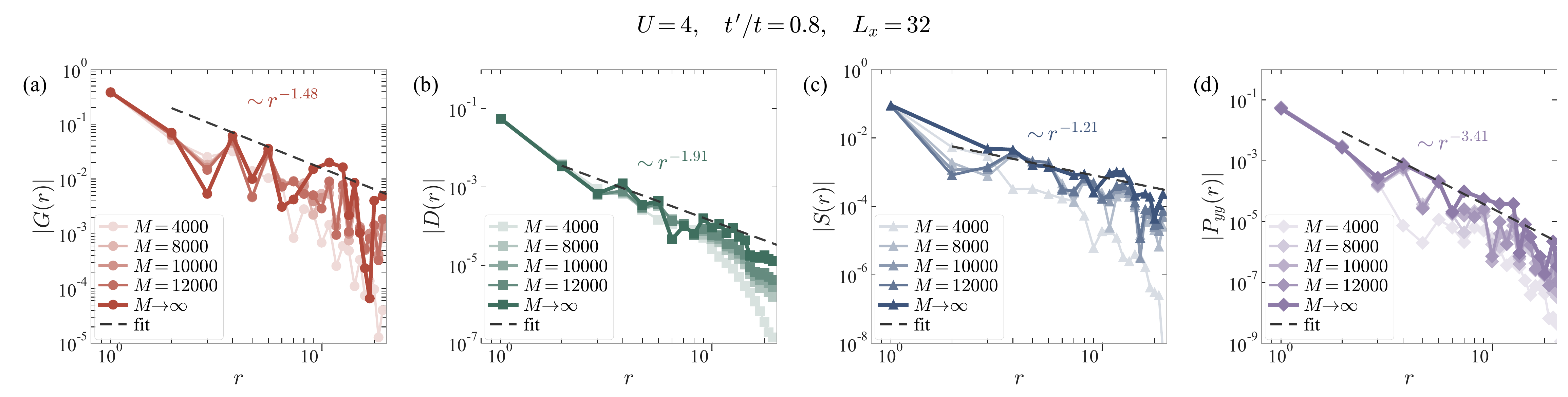}
\caption{Correlation functions at $U=4$, $t'/t=0.8$, and $L_x=32$ on a log-log scale, showing (a) $|G(r)|$ (red circles), (b) $|D(r)|$ (green squares), (c) $|S(r)|$ (blue triangles), and (d) $|P_{yy}(r)|$ (purple diamonds). Light to dark colors show increasing bond dimension $M$, thick curves the $M\to\infty$ extrapolation, quadratic in $1/M$, and dashed lines power-law fits to the extrapolated curves with the quoted exponents.}
\label{fig:metalcorr}
\end{figure}

Figure~\ref{fig:metalcorr} collects the correlation functions at $U=4$ and $t'/t=0.8$, a representative point of the chiral metal of \figref{fig:model}(b). Besides $G(r)$, $S(r)$, and the pair-field correlation $P_{yy}(r)$ of the main text, we show the connected density correlation
\begin{equation}
D(r)=\big\langle n(x_0,y_0)\,n(x_0+r,y_0)\big\rangle-\big\langle n(x_0,y_0)\big\rangle\big\langle n(x_0+r,y_0)\big\rangle,
\label{eq:densitycorr}
\end{equation}
where $n(x,y)=\sum_\sigma n_{(x,y)\sigma}$ is the local density. Each panel shows the finite-$M$ data, the $M\to\infty$ extrapolation, and a power-law fit $\sim r^{-K}$ to the extrapolated curve. All four correlators are consistent with power-law decay, with exponents $K=1.48$ for $|G(r)|$, $1.91$ for $|D(r)|$, $1.21$ for $|S(r)|$, and $K_P=3.41$ for $|P_{yy}(r)|$, the last value being the one quoted in the inset of \figref{fig:doped}(b). Gapless charge, spin, and single-electron excitations thus coexist at this point, and the pair correlation decays faster than $|G(r)|^{2}$, so there is no pair binding. This behavior identifies a metal rather than a superconductor and is consistent with the two nearly decoupled spinful chains found for this regime in \secref{app:sixleg}.

\subsection{Correlation functions at the time-reversal-symmetric point $t'=0$}
\label{app:dmrg_t0}

\begin{figure}[t]
\centering
\includegraphics[width=\textwidth]{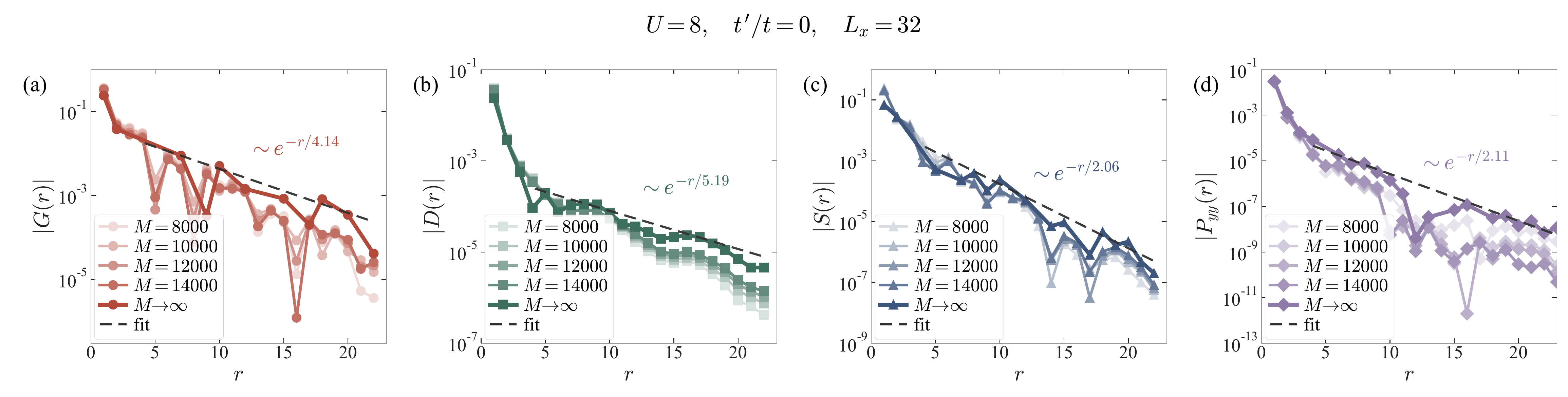}
\caption{Correlation functions at $U=8$, $t'=0$, and $L_x=32$ on a log-linear scale, showing (a) $|G(r)|$ (red circles), (b) $|D(r)|$ (green squares), (c) $|S(r)|$ (blue triangles), and (d) $|P_{yy}(r)|$ (purple diamonds). Light to dark colors show increasing bond dimension $M$, thick curves the $M\to\infty$ extrapolation, quadratic in $1/M$, and dashed lines exponential fits $\sim e^{-r/\xi}$ to the extrapolated curves with the quoted correlation lengths. Convergence at this time-reversal-symmetric point is not assured, and the data represent the current status at the largest bond dimensions reached.}
\label{fig:t0corr}
\end{figure}

At $t'=0$ the NNN bonds are absent and the Hamiltonian is time-reversal symmetric, so a chiral ground state could only arise by breaking this symmetry spontaneously, selecting one of two degenerate chirality sectors. As noted in the main text, DMRG converges far more slowly in this situation than at any finite $t'$, and we do not regard the results at this point as converged; we nevertheless report the current status. Figure~\ref{fig:t0corr} shows the correlation functions at $U=8$ and $t'=0$ on $L_x=32$ cylinders, with bond dimensions up to $M=14000$ and the $M\to\infty$ extrapolation taken quadratically in $1/M$. Over the accessible distances all four correlators are well described by exponential decay, with fitted correlation lengths $\xi=4.14$ for $|G(r)|$, $5.19$ for $|D(r)|$, $2.06$ for $|S(r)|$, and $2.11$ for $|P_{yy}(r)|$. In particular the pair-field correlation decays exponentially with a short correlation length and shows no trace of the slow power law that characterizes the superconductor at $t'>0$. Within the bond dimensions reached we therefore find no indication of pairing or superconductivity at this point, although we cannot rule out a different behavior once the calculations converge, which motivates the Hofstadter $t$-$J$ study proposed in the main text.

\end{document}